\documentclass[12pt]{article}
\usepackage[english]{babel}
\usepackage{graphicx}
\usepackage{amsmath,bm}
\usepackage{graphicx}% Include figure files
\usepackage{bm}% bold math
\usepackage[normalem]{ulem}
\usepackage{hyperref}
\hypersetup{
    colorlinks=true,
    linkcolor=blue,
    }
\usepackage{natbib}
\usepackage{amssymb}
\usepackage{overpic}
\usepackage{color}
\usepackage{tikz}

\providecommand\bnabla{\boldsymbol{\nabla}}
\providecommand\bcdot{\boldsymbol{\cdot}}

\newcommand{\tcr}[1]{\textcolor{red}{#1}}

\newcommand*\circled[1]{\tikz[baseline=(char.base)]{ \node[shape=circle, draw, inner sep=2pt] (char) {#1};} }
     
\title{ \Large From thin plates to Ahmed bodies: linear and weakly non-linear stability of rectangular prisms}

\author{G. A. Zampogna, E. Boujo}

\date{\today}

\begin{document}

\maketitle
%\begin{comment}

\begin{abstract}
We study the stability of laminar wakes past three-dimensional rectangular prisms. The width-to-height ratio is set to $W/H=1.2$, while the length-to-height ratio $1/6<L/H<3$ covers a wide range of geometries from thin plates to elongated Ahmed bodies. First, global linear stability analysis yields a series of pitchfork and Hopf bifurcations: (i)~at lower Reynolds numbers $Re$, two stationary modes, $A$ and $B$, become unstable, breaking the top/bottom and left/right planar symmetries, respectively; (ii)~at larger $Re$, two oscillatory modes become unstable and, again, each mode breaks one of the two symmetries. The critical $Re$ of these four modes increase with $L/H$, qualitatively reproducing the trend of stationary and oscillatory bifurcations in axisymmetric wakes (e.g. thin disk, sphere and bullet-shaped bodies). Next, a weakly non-linear analysis based on the two stationary modes $A$ and $B$ yields coupled amplitude equations. For Ahmed bodies, as $Re$ increases state $(A,0)$ appears first, followed by state $(0,B)$. While there is a range of bistability of those two states, only $(0,B)$ remains stable at larger $Re$, similar to the static wake deflection (across the larger base dimension) observed in the turbulent regime. The bifurcation sequence, including bistability and hysteresis, is validated with fully non-linear direct numerical simulations, and is shown to be robust to variations in $W$ and $L$ in the range of common Ahmed bodies.
\end{abstract}


%-----------------------------------------------------
%-----------------------------------------------------
%-----------------------------------------------------
\section{Introduction}
\label{sec:intro}

Flows past two-dimensional (2D) bluff bodies, steady at low Reynolds number, generally become unstable to oscillatory perturbations, leading to vortex shedding. 
For example, the wake of a 2D circular cylinder undergoes a Hopf bifurcation at the critical Reynolds number $Re_c=47$, which breaks time invariance and results in the famous von K\'arm\'an vortex street. 
Other examples include flows past rectangular cylinders of various aspect ratios, including the square cylinder, but also ellipses, wedges, etc.
This first Hopf bifurcation seems to be generic to all flows past isolated 2D bluff bodies,
although other types of 2D flows can first become unstable to stationary perturbations via a pitchfork bifurcation, like the planar sudden expansion.

Flows past three-dimensional (3D) bluff bodies exhibit a richer series of bifurcations, whose sequence depends on the specific geometry.
Often, wakes past axisymmetric bodies (e.g. thin disk, sphere, elongated bullet-shape bodies) first become unstable to stationary perturbations of azimuthal wavenumber $m=1$. 
The pitchfork bifurcation breaks the axisymmetry, leading to a steady wake deflected in some azimuthal direction, selected in practice by noise or imperfections.
At larger Reynolds number, the flow  undergoes a Hopf bifurcation, still with $m=1$, and the wake oscillates. For example, the stationary and oscillating critical Reynolds numbers
are $Re_c^s \simeq 115$ and $Re_c^o \simeq 125$ for the wake of a thin disk 
and $Re_c^s \simeq 210$ and $Re_c^o \simeq 275$ for the wake of a sphere \citep{natarajan_acrivos_1993, Gumowski08, Fabre08, meliga_global_2009}.
Axisymmetric rings also undergo an $m=1$ pitchfork bifurcation followed by an $m=1$ Hopf bifurcation for sufficiently small ratios of the torus diameter to the cross-section diameter \citep{sheard2003}.

Rectangular prisms are some of the simplest non-axisymmetric 3D bluff bodies.
In a systematic numerical study, \citet{MARQUET2015}  investigated the linear stability of relatively thin rectangular plates (length-to-height ratio $L/H=1/6$), and found that the nature of the first bifurcation depends on the frontal aspect ratio (width-to-height aspect ratio $W/H$). For large aspect ratios ($W/H>2.5$), the wake becomes unstable via the Hopf bifurcation of an oscillatory eigenmode that breaks the planar symmetry across the smaller dimension while preserving the planar  symmetry in the larger dimension. This is  fully consistent with the limit of infinite aspect ratios, i.e. 2D cylinders.
For intermediate aspect ratios ($2<W/H<2.5$), the wake still becomes unstable via a Hopf bifurcation but, remarkably, the oscillatory eigenmode breaks the planar symmetry across the larger dimension.
For small aspect ratios ($W/H<2$), the wake becomes unstable via a pitchfork bifurcation, which is reminiscent of the first bifurcation of the flow past a thin axisymmetric disk, and the stationary eigenmode breaks the planar symmetry across the smaller dimension.

A particular example of longer rectangular prisms is the cube ($W=L=H$). Direct numerical simulations (DNS) by \citet{Saha2004} and \citet{meng_an_cheng_kimiaei_2021} identified a pitchfork bifurcation at $Re_c \simeq 205-220$, leading to a steady wake with one planar symmetry. 
It is worth mentioning that, since $W=H$, the two cross-flow directions (say, top/down and left/right) are equivalent, so there are actually two simultaneous bifurcations, each breaking one of the two planar symmetries. The flow then becomes unstable to oscillatory perturbations at $Re_c \simeq 250-270$, leading to oscillations that preserve one of the two planar symmetries. 
Those regimes have also been observed in the experiments of \citet{klotz2014}.

Even longer rectangular prisms ($L>W,H$) include Ahmed bodies and simplified ground vehicles, of strong interest in the automotive industry. 
Motivated by practical applications, many studies have been conducted  at large $Re$ and in the presence of a horizontal ground 
\citep[etc.]{Grandemange_PoF2013, Cadot2015, brackston2016, EVRARD2016, barros2017, Varon2017, bonnavion_cadot_2018, Pavia2018, Legeai20}.
Most of the time, the turbulent wake is not aligned with the body but rather deflected in one of two preferred states, and randomly switches between the two states, leading to a bimodal probability density function for the wake deflection.
Interestingly, the direction of that deflection is not dictated by the ground but by the body geometry.
For instance,  bodies that are wider than tall  have a wake deflected to the left or the right, i.e. in the direction parallel  to the ground, and random switches restore the left/right planar symmetry in the long-term mean flow;
conversely, bodies that are taller than wide have a wake deflected upwards or downwards, i.e. in the direction perpendicular  to the ground \citep{Grandemange_PoF2013}. 
A recent turbulent study with no ground proximity \citep{Legeai20} confirmed that the static deflection is in the larger direction.
A similar scenario was observed in laminar experiments \citep{Grandemange12PRE} and laminar simulations \citep{Evstafyeva17} for a wide Ahmed square-back body in ground proximity: the wake, initially symmetric, bifurcates  to a steady deflected state at $Re_c \simeq 340$ and becomes oscillatory at $Re_c \simeq 410-430$, both bifurcations breaking the left/right planar symmetry.

To the best of our  knowledge, there is no systematic study on rectangular prisms  in the laminar regime, and several questions remain unanswered.
For instance, what are the critical Reynolds numbers and the properties of the first linear instabilities? 
Are those instabilities oscillatory or stationary? 
Which spatial symmetry do they break?
What kind of flow is expected in the non-linear regime?
In the present study, we address those questions by investigating the stability of rectangular prisms of fixed width-to-height aspect ratio $W/H=1.2$, varying length-to-height ratio $L/H$ and varying front fillet radius $R$.
After describing the flow configuration in section~\ref{sec:problem} and the numerical methods in  section~\ref{sec:num}, we characterise the steady symmetric base flow in section~\ref{sec:BF}. 
We then perform a 3D linear stability analysis in section~\ref{sec:LSA}.
Anticipating on the results, we find that  the flow always become unstable via two pitchfork bifurcations, at two critical Reynolds numbers close to one another. 
In almost all cases, the first mode breaks the top/down symmetry (across the smaller dimension), and the second mode breaks the left/right symmetry (across the longer dimension). 
In section~\ref{sec:WNL}, we then study the non-linear regime with a weakly non-linear (WNL) analysis incorporating the two stationary modes for a reference Ahmed body. 
We show that, as $Re$ increases, a top/down symmetry-breaking state appears first, but is eventually replaced by a left/right symmetry-breaking state.
A series of fully non-linear DNS confirms the results.
Finally, we observe that the WNL bifurcation sequence is robust to geometric variations for values of $W$ and $L$ typical of common Ahmed bodies.

%-----------------------------------------------------
%-----------------------------------------------------
%-----------------------------------------------------
\section{Flow configuration \label{sec:problem}}

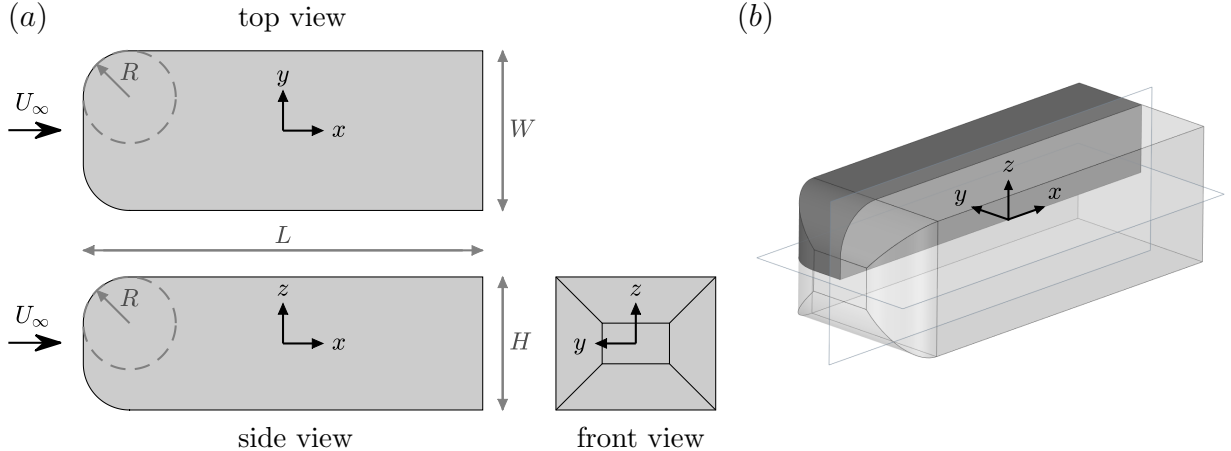
\begin{figure}
\centerline{   
 	\begin{overpic}[width=17cm, trim=50mm 45mm 70mm 57mm, clip=true]{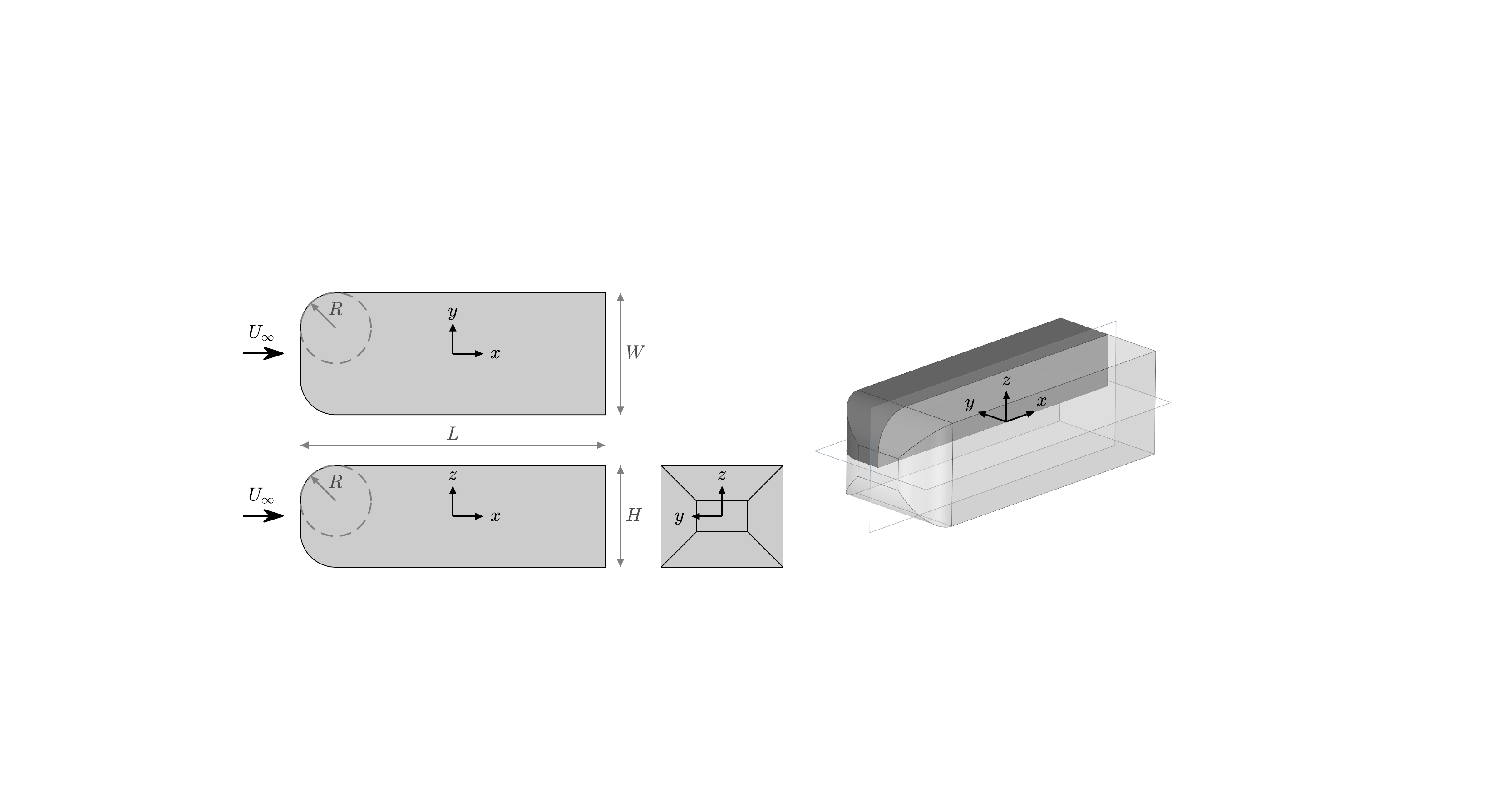}
      \put(2,34){$(a)$}
      \put(20,34){\small top view}
      \put(20,1){\small side view}
      \put(46.5,1){\small front view}
      \put(59,34){$(b)$}
 	\end{overpic}
}
\caption{
Sketch of the flow configuration. 
A rectangular prism of height $H$, width $W$, length $L$ and front edge fillet radius $R$ is aligned with the incoming flow $\bm{U}_\infty = (U_\infty,0,0)^T$.
$(a)$~Top, side and front views.
$(b)$~Three-dimensional view, highlighting the symmetry planes $y=0$ and $z=0$, and the quarter-body (darker) used in our linear and weakly non-linear calculations.
In this sketch, the specific geometry ($W/H=1.2$, $L/H=3$, $R/H=0.3472$) is shown, a reference Ahmed body that we also investigate with direct numerical simulations around the full body.
}
\label{fig:geometry}
\end{figure}

We consider the incompressible flow of a Newtonian fluid past a three-dimensional 
%bluff body.
rectangular prism.
The planar faces of the bluff body define a Cartesian coordinate system.
The incoming flow is $\bm{U}_\infty = (U_\infty,0,0)^T$, i.e. the body's roll, pitch and yaw are zero.
For convenience, we call the $x$, $y$ and $z$ directions the streamwise, lateral and vertical directions, respectively. 
We can therefore say that the front/rear faces of the body are normal to the $x$ direction, left/right faces  to the $y$ direction, and upper/lower faces  to the $z$ direction.
Similarly, we define the body dimensions as length $L$, width $W$ and  height $H$ (cf. figure~\ref{fig:geometry}).
Bodies with rounded front edges are also considered, with a fillet radius $R$.
Hereafter, all quantities are made dimensionless using the body height $H$ as reference length and free-stream velocity $U_\infty$ as reference velocity.

In this study, we fix the body width, $W=1.2$, except in the final part of the weakly non-linear analysis (section~\ref{sec:WNL-effect_W_L}).
The length is varied between $L=1/6 \simeq 0.167$ (thin plate orthogonal to the flow) to $L=3.8$ (similar to some Ahmed bodies).
The fillet radius of the front edges is varied between $R=0$ (sharp edges) and $R = \min(0.5, L)$ (fully rounded edges, with either upper/lower fillets meeting tangentially and the  front face vanishing, or all fillets reaching the rear face and all lateral faces vanishing). 
In between, the value $R= 100/288 \simeq 0.347$ is typical of Ahmed bodies.

The velocity field $\bm{u}(\bm{x},t)=(u,v,w)^T$ and pressure field $p(\bm{x},t)$ are solution of the incompressible  Navier-Stokes (NS) equations, expressing the conservation of mass and momentum and written in dimensionless form 
\begin{align}
\bnabla \cdot \bm u = 0,
\quad
\partial_t \bm{u}  + (\bm{u} \cdot \bnabla) \bm{u} = -\bnabla p + \frac{1}{Re} \bnabla^2 \bm{u},
\label{eq:NS}
\end{align}
where the Reynolds number $Re=U_\infty H/\nu$ is based on the  free-stream velocity, the body height and the fluid kinematic viscosity.
%We write (\ref{eq:NS}) in compact form as
%\begin{align}
%\mathcal B \partial_t  \bm{q} + 
%\mathcal N (\bm q) = \bm 0,
%\label{eq:NS-compact}
%\end{align}
%where $\bm{q}(\bm x,t)=(\bm u, p)^T$.
Throughout this paper, all lengths and velocities are made dimensionless using  $H$ and  $U_\infty$, respectively.

%-----------------------------------------------------
%-----------------------------------------------------
%-----------------------------------------------------
\section{Numerical methods \label{sec:num}}

In this study, two different numerical methods are used.
The non-linear base flow calculation, linear stability analysis and weakly non-linear stability analysis  are performed with the finite element software FreeFEM \citep{hecht_new_2012}, while direct numerical simulations are performed with the finite volume software OpenFOAM \citep{greenshields2022}.

For finite element calculations, we first build a tetrahedral mesh using the three-dimensional finite-element mesh generator Gmsh \citep{Geuzaine09}, with mesh nodes strongly clustered near the body surface. 
See Appendix A for details about the domain size, mesh characteristics and convergence study.
We use the same mesh to solve the base flow problem (\ref{eq:BF}), the direct and adjoint eigenvalue problems (\ref{eq:EVP_compact}) and (\ref{eq:ADJ_compact}), the linear problems (\ref{eq:epsilon2-detail_alpha})-(\ref{eq:epsilon2-detail_AB}) appearing in the weakly non-linear analysis and the coefficients (\ref{eq:lambdaA})-(\ref{eq:chiB}) of the amplitude equations (\ref{eq:A})-(\ref{eq:B}). 
We discretise the weak form of the equations to be solved with the FreeFEM, using a basis of Arnold-Brezzi-Fortin MINI-elements, i.e. P1 (linear) elements for pressure and P1b (linear enriched with a cubic bubble function) elements for each velocity component.
We solve for the non-linear base flow using an iterative Newton method. 
Linear systems involved in each Newton iteration and in the weakly non-linear analysis are inverted with the PETSc library. 
Eigenvalue problems are solved with the SLEPc library, using a Krylov-Schur method to obtain a set of eigenvalues closest to a given complex shift, together with the associated eigenmodes. Calculations are repeated with a series of shift chosen so as to obtain all leading eigenvalues in a range of frequencies of interest (typically $|\omega| \leq 1$).
All problems involve several millions of degrees of freedom, and are  solved in parallel  using a domain-decomposition method on 24 to 48 processes.  
Typical calculation times are of the order of one hour for the base flow at a few $Re$ values, and one hour for a few eigenvalues.

The DNS have been carried out on the same domain as the linear and WNL analyses, without assuming a priori symmetry conditions. 
The mesh was generated exploiting the routine snappyHexMesh.
As starting point, an initial Cartesian mesh generated with blockMesh has been employed, with 5 cells per unit length. 
Four nested  regions of increasing refinements have been introduced around the solid body, to guarantee 0.5$\times10^6$ cells per unit volume in the vicinity of the body. 
To test convergence, the same mesh generation strategy has been used, employing 2.5 and 10 cells per unit length as initial mesh; the drag coefficient has been used as a convergence criterion, showing differences of less than 1.3\% between the two finest meshes. 
A time-marching strategy has been employed to calculate the steady flow solution, based on a Crank-Nicolson scheme with initial uniform solution equal to $\bm{U}=(U_{\infty},0,0)^T$. 
The NS equations have been solved by exploiting the PIMPLE method, which decouples velocity from pressure. 
The spatial discretisation of each equation is based on the finite volume method with Gauss linear integration scheme, assuring second order precision in both time and space. 
Typical calculation times are of the order of one week for one non-linear flow simulated over  $10^3$ convective times.

%-----------------------------------------------------
%-----------------------------------------------------
%-----------------------------------------------------
\section{Base flow \label{sec:BF}}

%In this section, we compute steady base flows 
In this section we characterise the steady base flow 
$\bm q_0(\bm x) = (\bm u_0 ,p_0)^T$  past rectangular prisms of
various lengths $L$ and fillet radii $R$, for a fixed width $W = 1.2$. 
The base flow is a solution of the steady Navier-Stokes equations
\begin{align}
\bnabla \cdot \bm u_0 = 0,
\quad
(\bm{u}_0 \cdot \bnabla) \bm{u}_0 = -\bnabla p_0 + \frac{1}{Re} \bnabla^2 \bm{u}_0
\label{eq:BF}
\end{align}
in the fluid domain $\Omega$.
We look for base flows that have the same symmetries as the body: reflectional symmetries in the $y$ and $z$ directions (i.e. symmetry with respect to the vertical $xz$ plane and the horizontal $xy$ plane, respectively).
We take advantage of these symmetries to reduce the computational cost. 
Namely, we solve the stationary NS equations (\ref{eq:BF}) on the quarter-space $\Omega' = \{ x,y,z \, | \, y\geq0, z\geq0 \}$ instead of the full space $\Omega$,  
%$\Omega={}$
and we impose the following symmetry conditions on the symmetry planes:
%$\bm q_0 \cdot \bm{n} = 0$ and $\partial_{\bm n} (\bm  q_0 \cdot \bm{t}_i) = \bm{0}$ (where $\bm n$ denotes the unit normal vector and $\bm t_i$, $i=1,2$, two unit tangential vectors), i.e.
\begin{align}
\partial_y u_0 = v_0 = \partial_y w_0 = 0 \quad \mbox{on the } y=0 \mbox{ plane},
\nonumber 
\\
\partial_z u_0 = \partial_z v_0 = w_0 = 0 \quad \mbox{on the } z=0 \mbox{ plane}.
\label{eq:BC_BF}
\end{align}
Additionally, we impose a
free-stream velocity $\bm{u}_0 = (1,0,0)^T$ on the inlet plane,
a no-slip condition $\bm u_0 = \bm 0$ on the body surface,
stress-free condition $-p_0 \bm n + Re^{-1}  \bnabla \bm u_0 \cdot \bm n = \bm 0$ on the outlet plane (with $\bm n$ the outward unit normal vector),
and symmetry conditions identical to (\ref{eq:BC_BF}) on the remaining (lateral and upper) boundaries.

%%-----------------------------------------------------
%\subsection{Results}
%
%In this section we characterise the  steady, symmetric base flow past rectangular prisms of various lengths $L$ and fillet radii $R$, for a fixed width $W=1.2$.

%-----------------------------------------------------
\subsection{Effect of $Re$ \label{sec:BF-Re}}

\begin{figure}
\centerline{   
    \begin{overpic}[width=8cm, trim=32mm 105mm 39mm 118mm, clip=true]{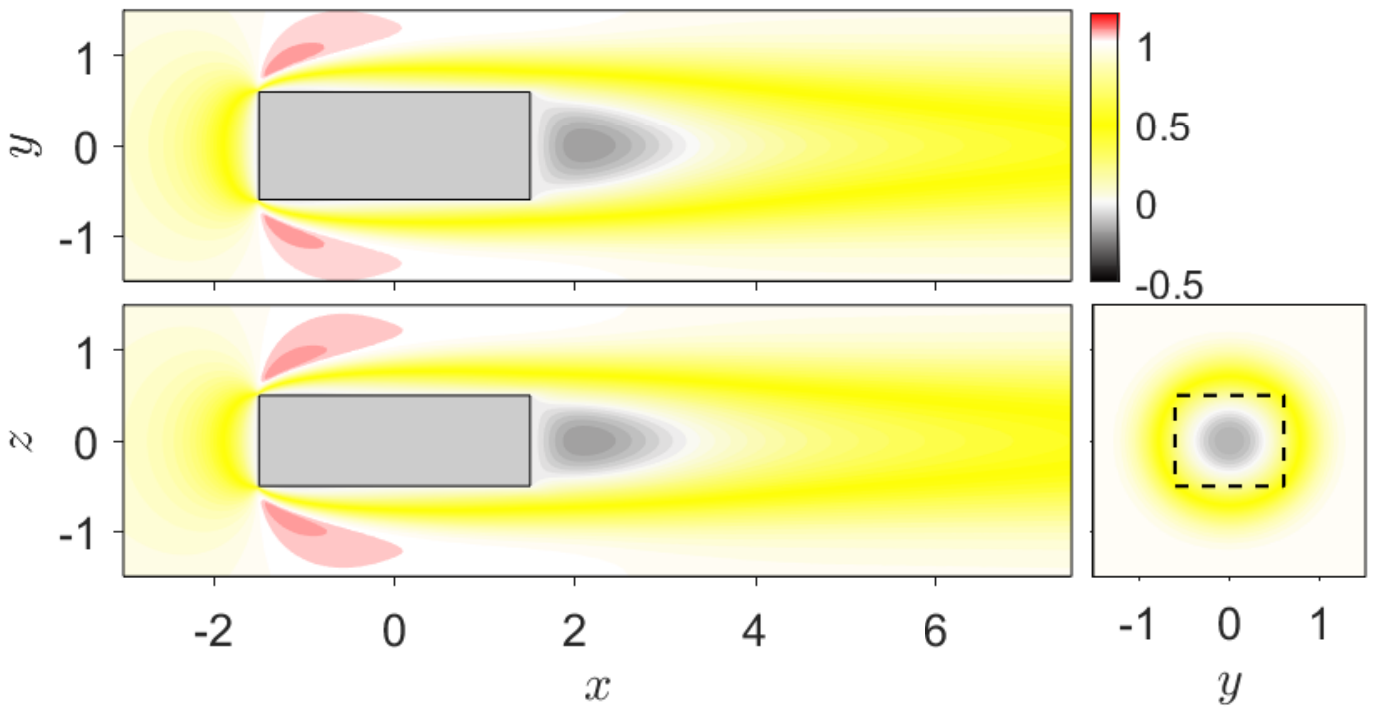}
      \put(-3,48){$(a)$}   
 	\end{overpic}
    \begin{overpic}[width=8cm, trim=32mm 105mm 39mm 118mm, clip=true]{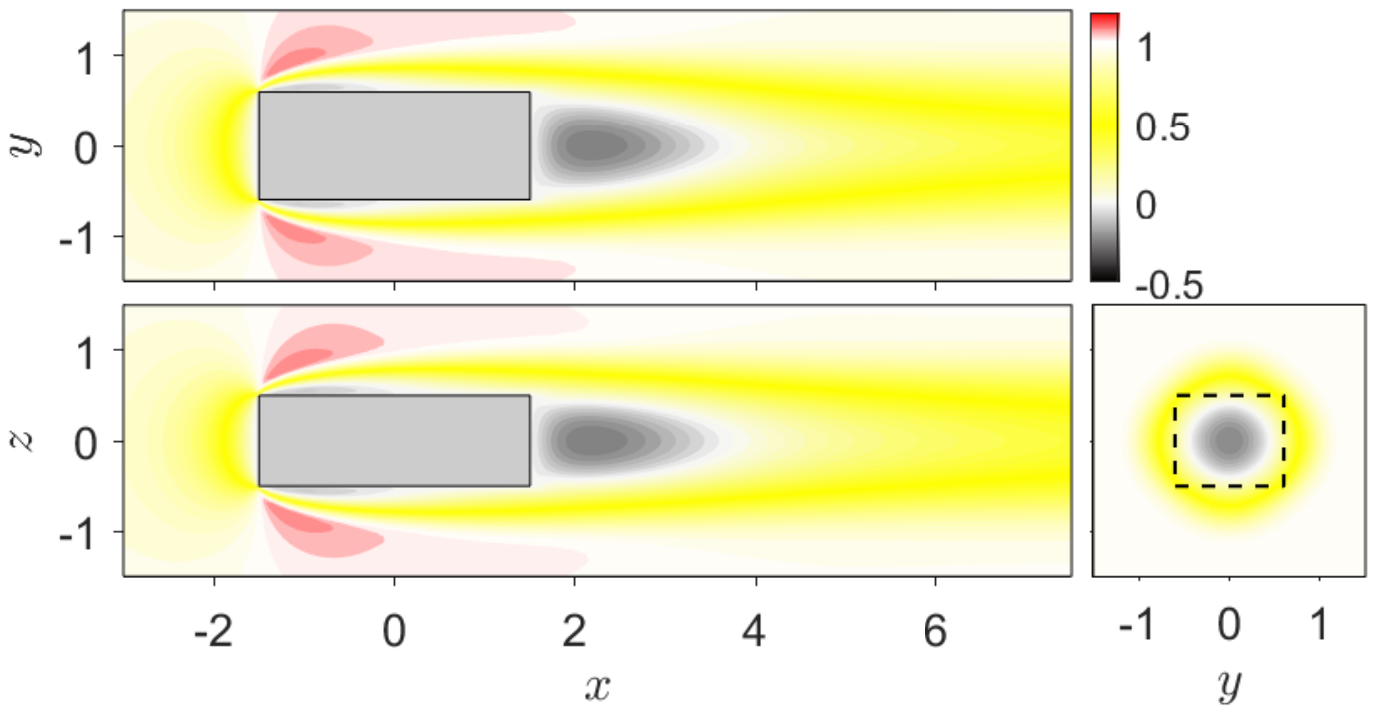}
      \put(-3,48){$(b)$}   
 	\end{overpic}
}
\centerline{   
    \begin{overpic}[width=8cm, trim=32mm 105mm 39mm 118mm, clip=true]{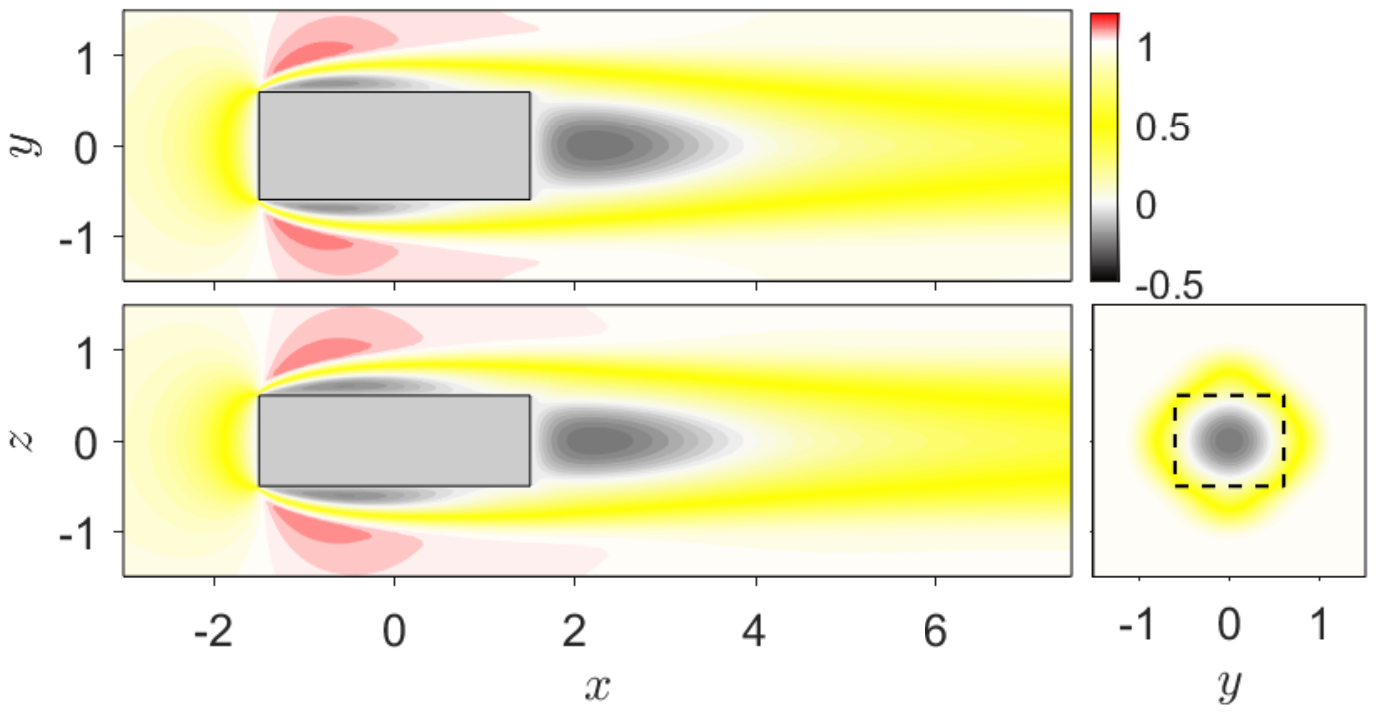}
      \put(-3,48){$(c)$}   
 	\end{overpic}
    \begin{overpic}[width=8cm, trim=32mm 105mm 39mm 118mm, clip=true]{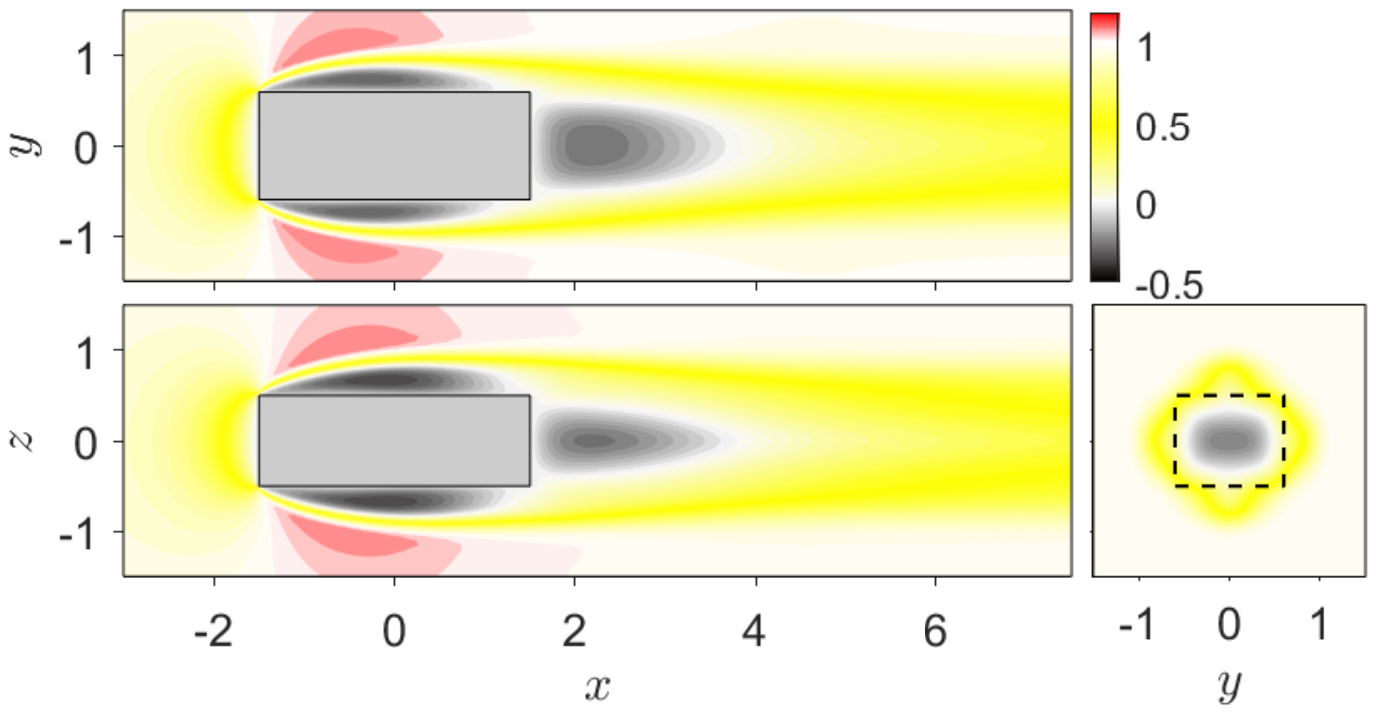}
      \put(-3,48){$(d)$}   
 	\end{overpic}
}
\caption{
Streamwise velocity $u_0$ of the base flow past a rectangular prism,  $W=1.2$, $L=3$, $R=0$, in the planes $z=0$ (top view), $y=0$ (side view) and $x=2.5$ (rear view):
$(a)$~$Re=150$,
$(b)$~$Re=250$,
$(c)$~$Re=350$,
$(d)$~$Re=450$.
}
\label{fig:BF_Re_effect}
\end{figure}

\begin{figure}
%\centerline{   
%    \begin{overpic}[height=6.5cm, trim=34mm 92mm 40mm 90mm, clip=true]{C:/Users/boujo/Documents/_Projects_MSc/2021_Adrien_Gimonnet/results/quarter-1-1.2-3/all_recirc_lengths_vs_Re_L3.pdf}
%      \put(0,75){$(a)$}
% 	\end{overpic}
%    \begin{overpic}[height=6.5cm, trim=30mm 92mm 40mm 90mm, clip=true]{C:/Users/boujo/Documents/_Projects_MSc/2021_Adrien_Gimonnet/results/quarter-1-1.2-3/Cx-n30_8_2-xout60.pdf}
%      \put(0,74){$(b)$}
% 	\end{overpic}
%}
%\centerline{      
%\begin{overpic}[height=6.5cm, trim=34mm 92mm 40mm 90mm, clip=true]{C:/Users/boujo/Documents/_Projects_MSc/2021_Adrien_Gimonnet/results/quarter-1-1.2-3/all_recirc_lengths_vs_Re_L3_NEW.pdf}
%      \put(0,75){$(a)$}
% 	\end{overpic} 
%    \begin{overpic}[height=6.5cm, trim=30mm 92mm 40mm 90mm, clip=true]{C:/Users/boujo/Documents/_Projects_MSc/2021_Adrien_Gimonnet/results/quarter-1-1.2-3/Cx-n60_10_1_NEW_smaller.pdf}
%      \put(0,73){$(b)$}
% 	\end{overpic}
%}
\centerline{      
\begin{overpic}[height=6.5cm, trim=34mm 92mm 40mm 90mm, clip=true]{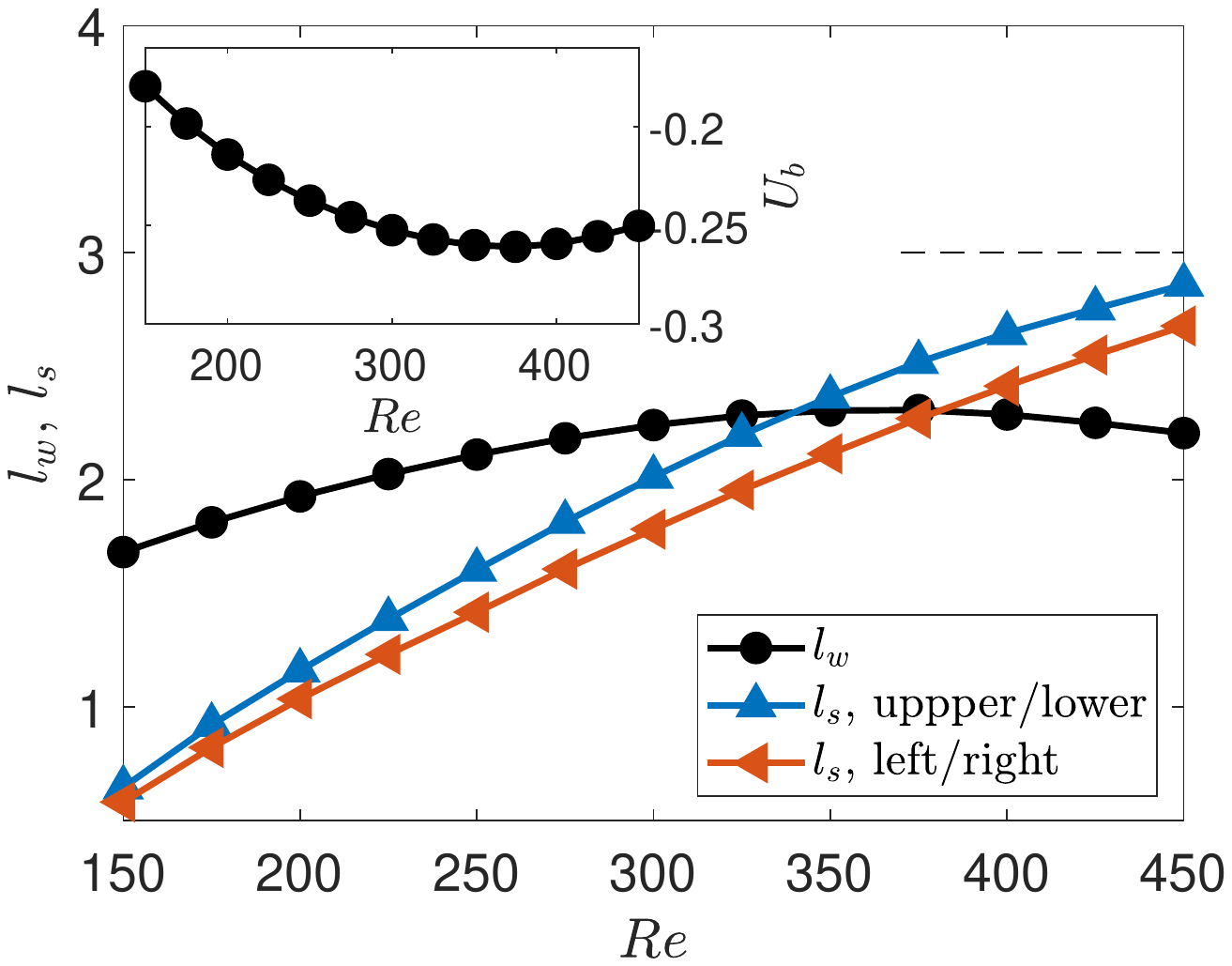}
      \put(0,75){$(a)$}
 	\end{overpic} 
    \begin{overpic}[height=6.5cm, trim=30mm 92mm 40mm 90mm, clip=true]{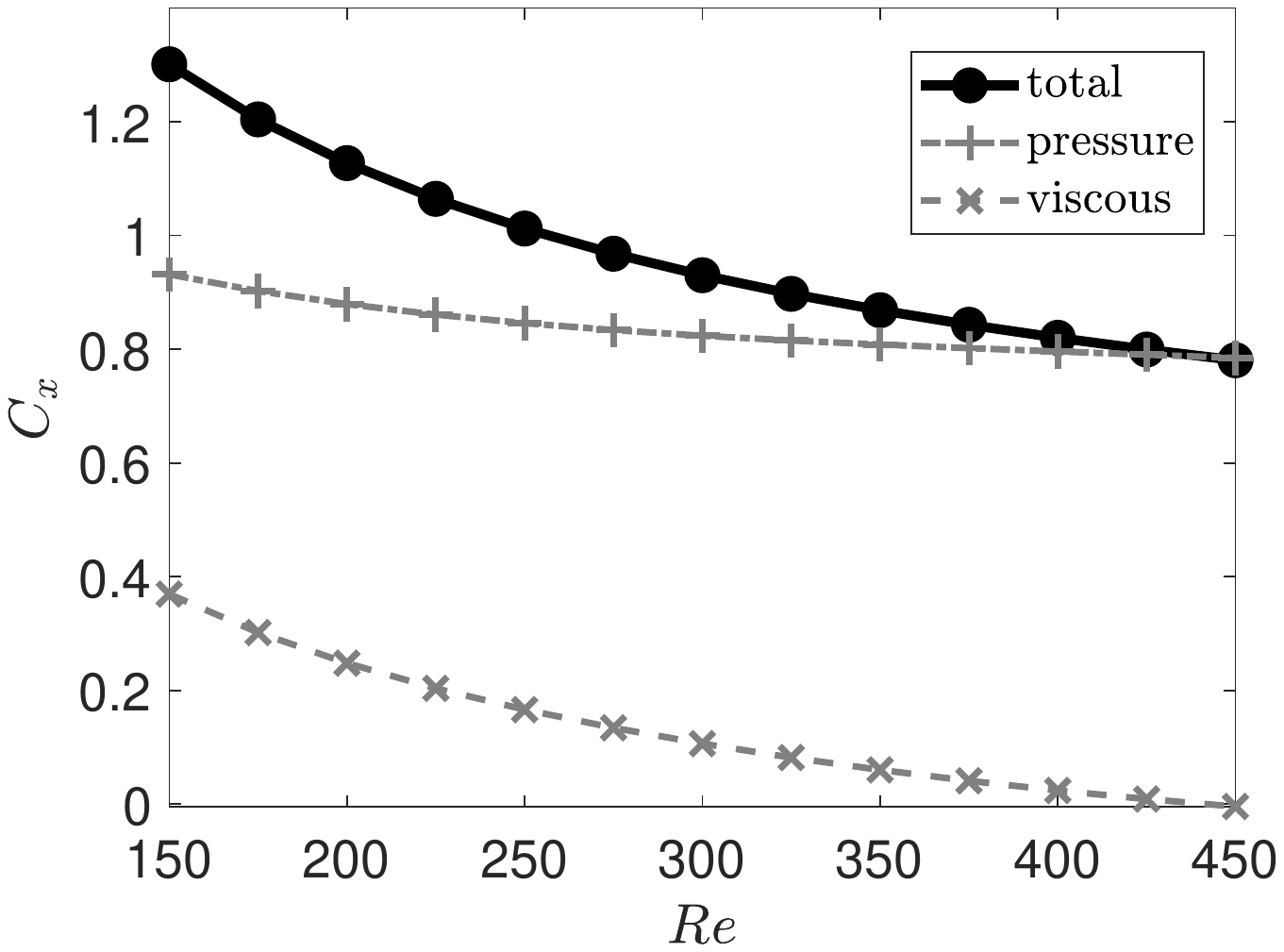}
      \put(0,73){$(b)$}
 	\end{overpic}
}
\caption{
Base flow past a rectangular prism,  $W=1.2$, $L=3$, $R=0$.
$(a)$~Length of the wake recirculation  and side recirculations.
Dashed line: $l_s=L=3$, when the side recirculations reach the end of the body and connect with the wake recirculation.
$(b)$~Drag coefficient.
}
\label{fig:recirc_length_L3_R0}
\end{figure}

We start by illustrating the effect of the Reynolds number for a selected geometry, namely  $L=3$ and $R=0$.
Figure~\ref{fig:BF_Re_effect} shows the streamwise velocity field for several values of $Re$, and figure~\ref{fig:recirc_length_L3_R0}$(a)$ shows the variation in length of the different recirculation regions found around the prism: $l_w$, the length of the ``wake recirculation'' originating from the trailing edges, measured along the symmetry axis $y=z=0$;
and $l_s$, the length of the ``side recirculations'' originating from the leading edges, measured in the symmetry planes on the left/right faces ($|y|=W/2, z=0$) and on the upper/lower faces ($y=0, |z|=H/2$).
Since $W>H$, side recirculations on the upper/lower faces are slightly longer than those on the left/right faces.
Their length is seen to increase with $Re$ in the investigated range of Reynolds number, until they extend all the way down to the trailing edges ($l_s = L = 3$) and connect with the wake recirculation, at $Re$ slightly below 500.
The length $l_w$ of the wake recirculation   increases more weakly than $l_s$, reaches a maximum at $Re \simeq 350$ before decreasing. 
The backflow, i.e. the minimum streamwise velocity along the symmetry axis, $U_b = \min u_0(x,0,0)$, decreases in the investigated range of $Re$ from -0.18 to -0.26 (inset).
At larger $Re$, the wake recirculation becomes substantially longer, wider and taller (not shown) as it merges with the side recirculations, and the backflow becomes stronger.
We do not, however, focus on this topological change of the base flow because bifurcations of interest occur at smaller $Re$, as will become clear in sections~\ref{sec:LSA}-\ref{sec:WNL}.

%\begin{figure}
%\centerline{
%  \fbox{    
%    \begin{overpic}[width=8cm, trim=28mm 90mm 40mm 90mm, clip=true]{C:/Users/boujo/Documents/_Projects_MSc/2021_Adrien_Gimonnet/results/quarter-1-1.2-3/Cx-n30_8_2-xout60.pdf}
% 	\end{overpic}
%  }
%}
%\caption{
%Drag coefficient of  a rectangular prism, $W=1.2$, $L=3$, $R=0$.
%}
%\label{fig:drag_L3_R0}
%\end{figure}

Figure~\ref{fig:recirc_length_L3_R0}$(b)$ shows the evolution of the drag coefficient,
\begin{align}
C_x = \frac{F_x}{\frac{1}{2} \rho U_\infty^2 W H}
\quad
\mbox{where }
F_x = - \oint_{\Gamma_{b}} \left( \bm{\sigma}_0 \cdot \bm{n} \right) \cdot \bm{e}_x \, \mathrm{d}\Gamma,
\end{align}
where $\Gamma_b$ is the surface of the full body,
and where the stress tensor $\bm{\sigma}_0   = -p_0 \bm{I} + Re^{-1}  \bnabla \bm u_0 $ includes  pressure and viscous effects.
The drag coefficient decreases with $Re$, as do both  pressure and viscous contributions, which is  typically of  bluff bodies in the laminar regime (see e.g. \citet{henderson_details_1995} for the two-dimensional circular cylinder wake). 
As $Re$ increases  the viscous contribution becomes much smaller than its pressure counterpart, because  the side recirculations extend over a larger body surface area while the wake recirculation is not substantially modified.

%real Fxp =int2d(Thbkp,1)( uuD*N.x );
%
% real Fxv = nu * int2d(Thbkp,1)(        -2.*dx(uu)*N.x -(dx(uuB)+dy(uu))*N.y -(dx(uuC)+dz(uu))*N.z );

%-----------------------------------------------------
\subsection{Effect of $L$ \label{sec:BF-L}}

\begin{figure}
\centerline{   
    \begin{overpic}[width=8cm, trim=32mm 105mm 39mm 118mm, clip=true]{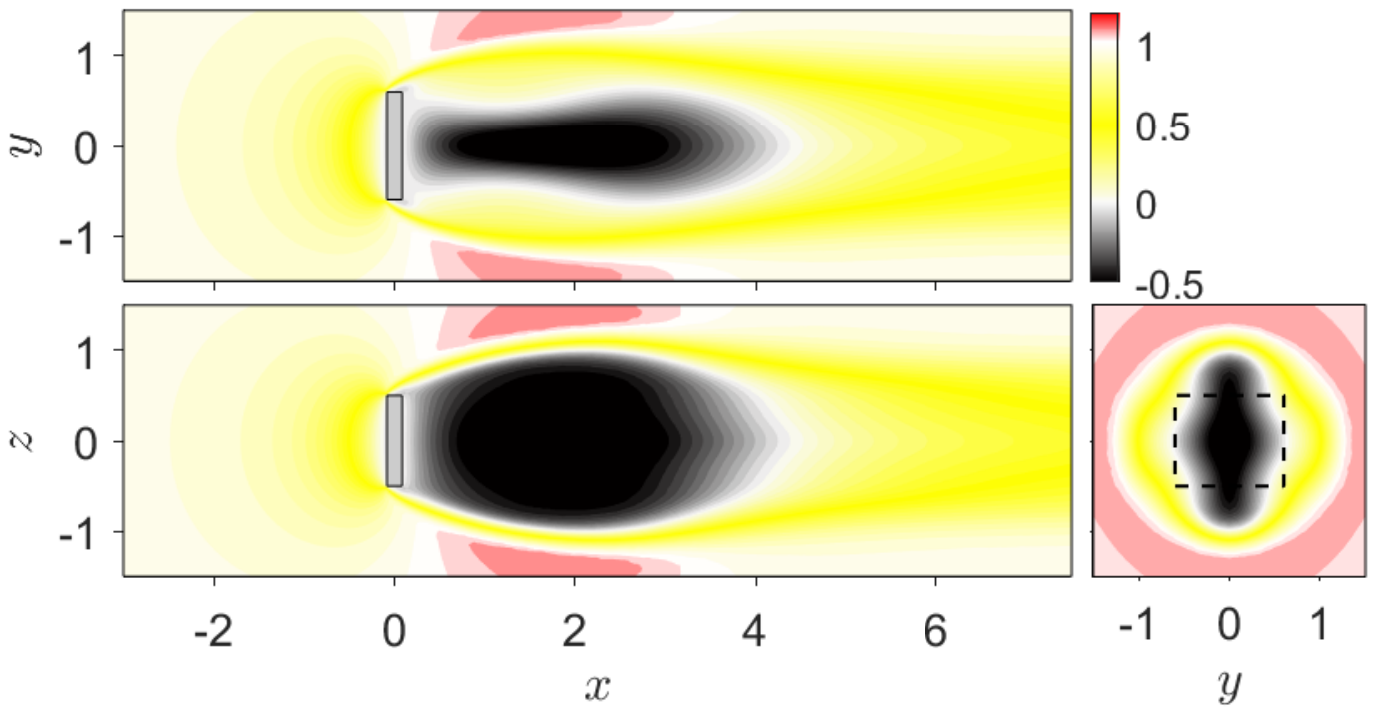}
      \put(-3,48){$(a)$}   
 	\end{overpic}
    \begin{overpic}[width=8cm, trim=32mm 105mm 39mm 118mm, clip=true]{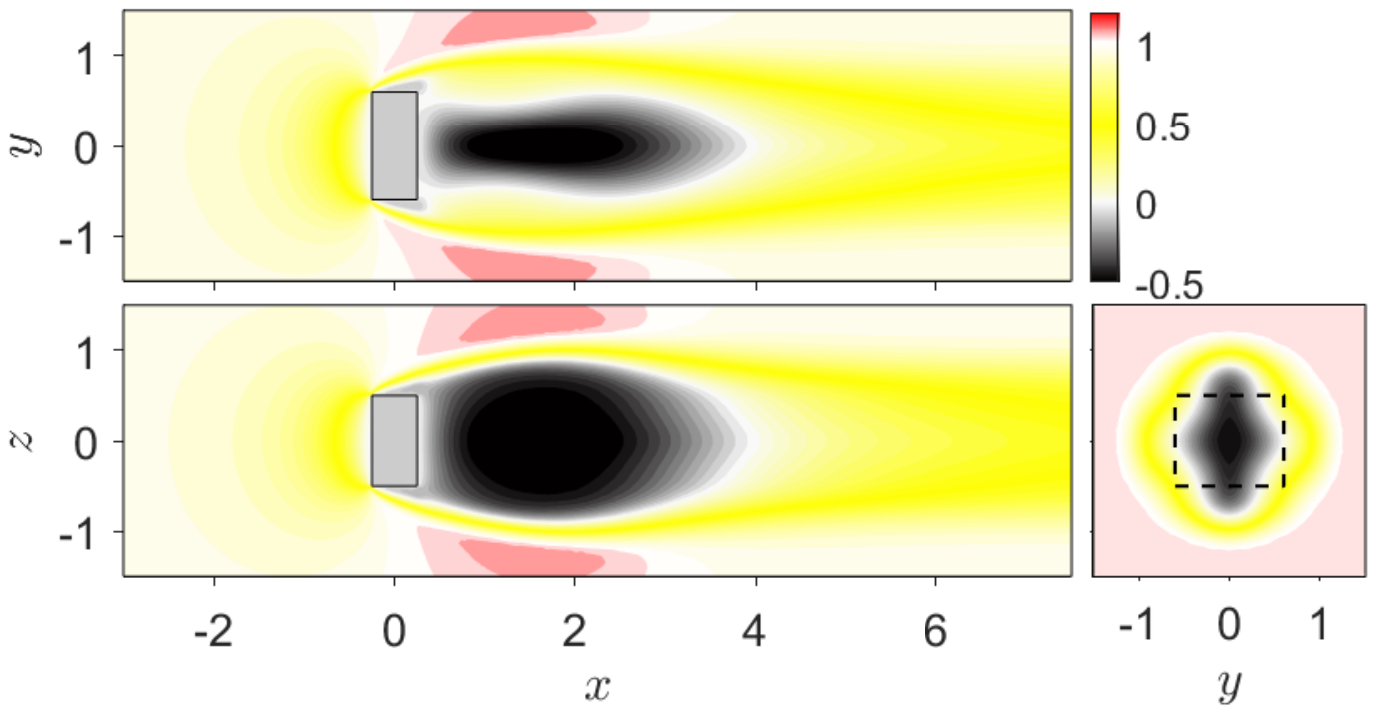}
      \put(-3,48){$(b)$}   
 	\end{overpic}
}
\centerline{   
    \begin{overpic}[width=8cm, trim=32mm 105mm 39mm 118mm, clip=true]{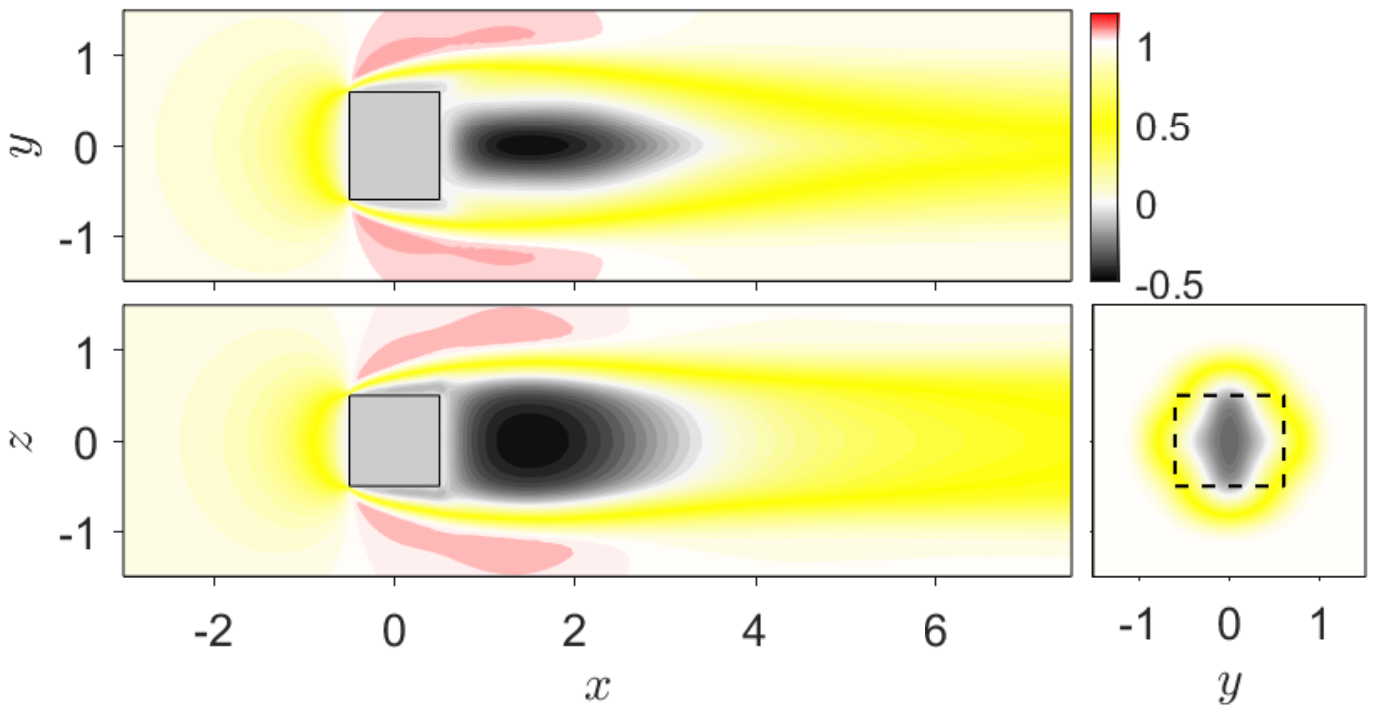}
      \put(-3,48){$(c)$}   
 	\end{overpic}
    \begin{overpic}[width=8cm, trim=32mm 105mm 39mm 118mm, clip=true]{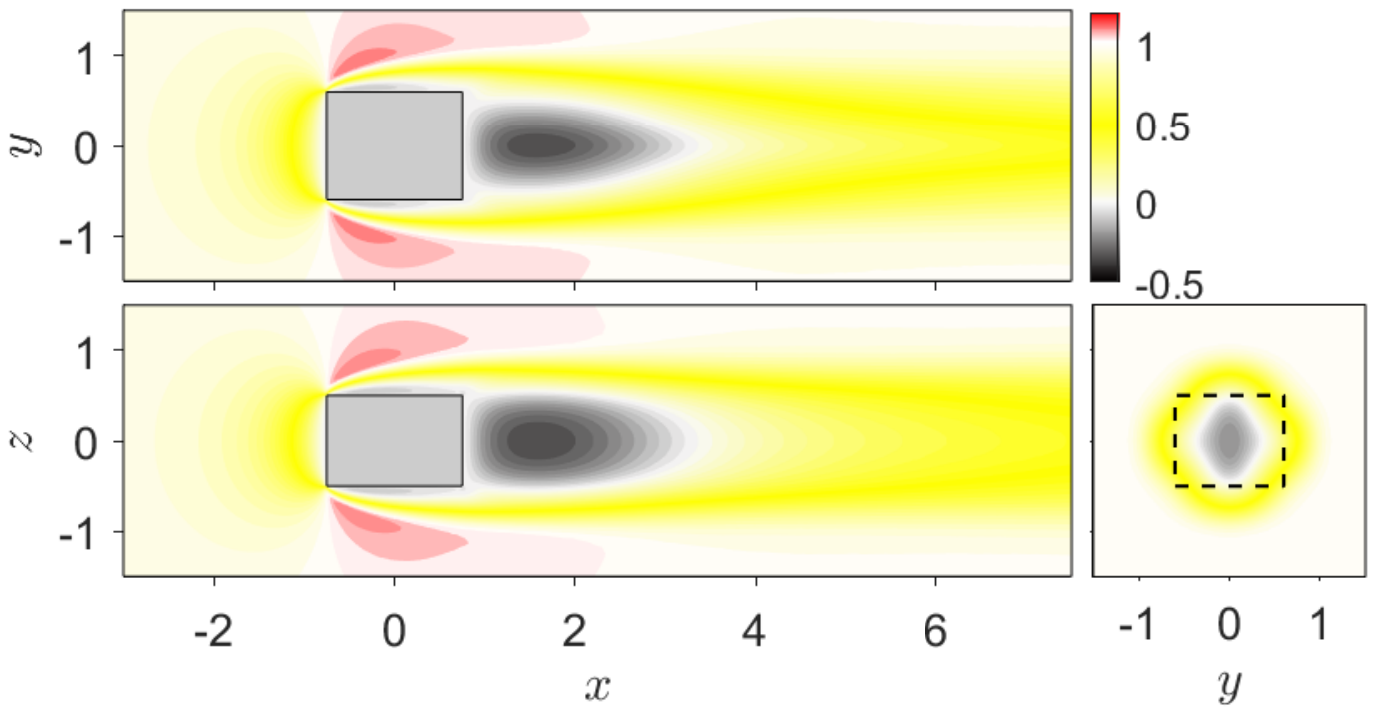}
      \put(-3,48){$(d)$}   
 	\end{overpic}
}
\centerline{   
    \begin{overpic}[width=8cm, trim=32mm 105mm 39mm 118mm, clip=true]{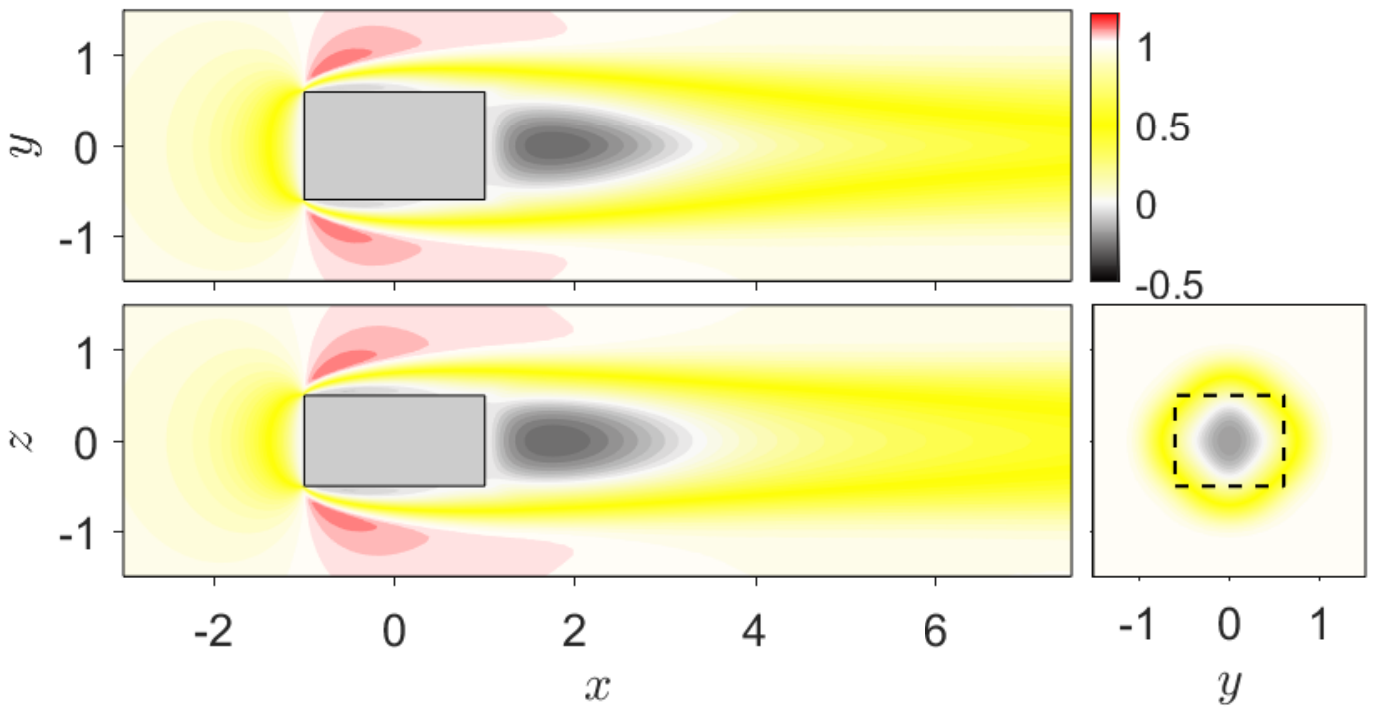}
      \put(-3,48){$(e)$}   
 	\end{overpic}
    \begin{overpic}[width=8cm, trim=32mm 105mm 39mm 118mm, clip=true]{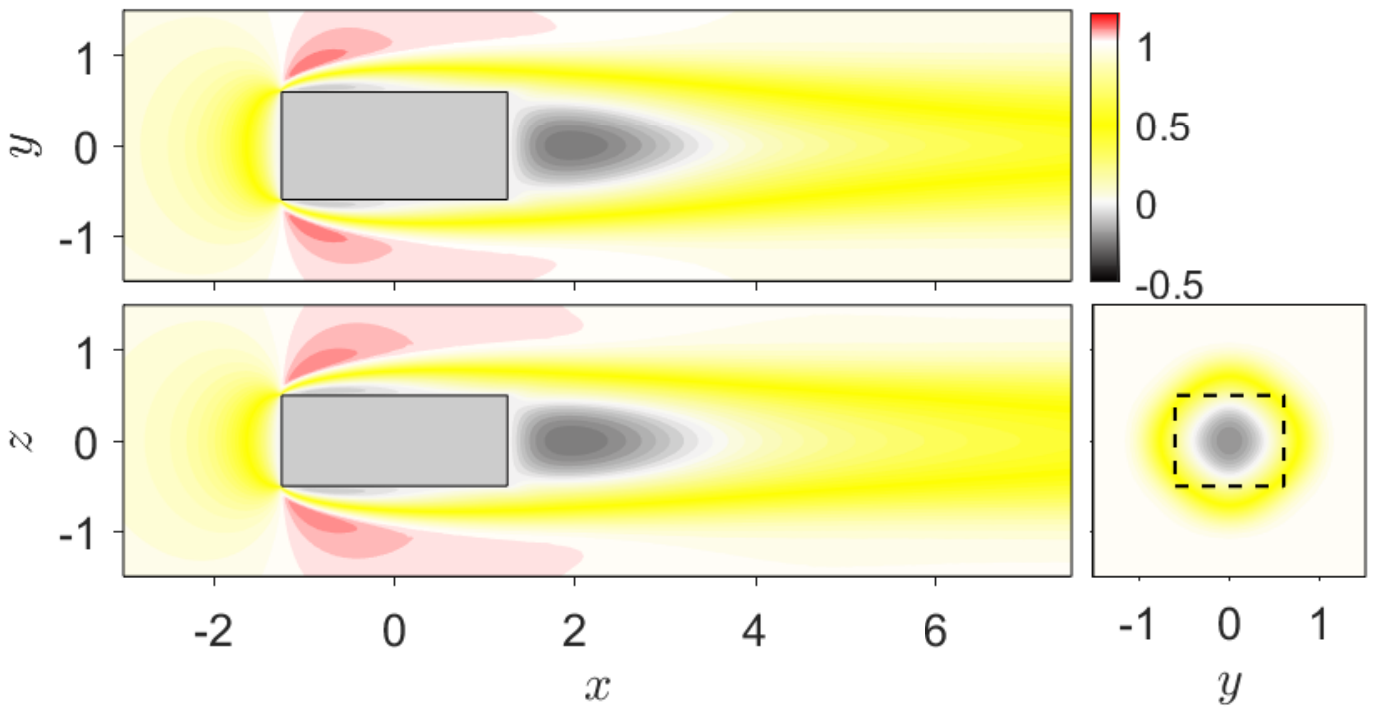}
      \put(-3,48){$(f)$}   
 	\end{overpic}
}
\caption{
Streamwise velocity $u_0$ of the base flow past rectangular prisms,  $W=1.2$, $R=0$, at $Re=250$, in the planes $z=0$ (top view), $y=0$ (side view) and $x=2.5$ (rear view):
$(a)$~$L=1/6$,
$(b)$~$L=0.5$,
$(c)$~$L=1$,
$(d)$~$L=1.5$,
$(e)$~$L=2$,
$(f)$~$L=2.5$.
}
\label{fig:BF_L_effect}
\end{figure}

We now investigate the effect of the length $L$ on the steady base flow past bodies with sharp edges ($R=0$). 
For the sake of illustration, we choose $Re=250$ (which will prove to be close to the critical Reynolds number in section~\ref{sec:LSA}).
Figure~\ref{fig:BF_L_effect} shows the streamwise velocity field for $L$ between $1/6$ and 3.
For large $L$, the flow separates at the leading edges and reattaches onto the body, as already described previously;
%in section~\ref{sec:BF-Re}; 
the flow then re-separates  at the trailing edges with a small angle with respect to the streamwise direction, which leads to a rather narrow and short wake recirculation.
As $L$ decreases, the side recirculations and the wake merge, the wake recirculation becomes longer and the backflow  stronger. 
The overall flow also becomes increasingly different in the two symmetry planes $y=0$ and $z=0$:  separation occurs with a larger angle from the upper/lower leading edges and the wake becomes significantly taller, but not wider.

\begin{figure}
%\centerline{  
%  \begin{overpic}[height=6.5cm, trim=30mm 90mm 40mm 90mm, clip=true]{C:/Users/boujo/Documents/_Projects_MSc/2021_Adrien_Gimonnet/results/quarter-other-ARs/all_recirc_lengths_other_ARs_R0_Re250.pdf}
%      \put(0,74){$(a)$}
%  \end{overpic}
%  \begin{overpic}[height=6.5cm, trim=28mm 90mm 40mm 90mm, clip=true]{C:/Users/boujo/Documents/_Projects_MSc/2021_Adrien_Gimonnet/results/quarter-other-ARs/Cx_vs_L-Re250.pdf}
%      \put(0,74){$(b)$}
%  \end{overpic}
%}
%\centerline{  
%  \begin{overpic}[height=6.5cm, trim=30mm 90mm 40mm 90mm, clip=true]{C:/Users/boujo/Documents/_Projects_MSc/2021_Adrien_Gimonnet/results/quarter-other-ARs/all_recirc_lengths_other_ARs_R0_Re250-inset_backflow.pdf}
%      \put(0,74){$(a)$}
%  \end{overpic}
%  \begin{overpic}[height=6.5cm, trim=28mm 90mm 40mm 90mm, clip=true]{C:/Users/boujo/Documents/_Projects_MSc/2021_Adrien_Gimonnet/results/quarter-other-ARs/Cx_vs_L-Re250.pdf}
%      \put(0,74){$(b)$}
%  \end{overpic}
%}
%\centerline{  
%  \begin{overpic}[height=6.5cm, trim=30mm 90mm 40mm 90mm, clip=true]{C:/Users/boujo/Documents/_Projects_MSc/2021_Adrien_Gimonnet/results/quarter-other-ARs/all_recirc_lengths_other_ARs_R0_Re250_NEW.pdf}
%      \put(0,74){$(a)$}
%  \end{overpic}
%  \begin{overpic}[height=6.5cm, trim=28mm 90mm 40mm 90mm, clip=true]{C:/Users/boujo/Documents/_Projects_MSc/2021_Adrien_Gimonnet/results/quarter-other-ARs/Cx_vs_L-Re250_NEW.pdf}
%      \put(0,74){$(b)$}
%  \end{overpic}
%}
\centerline{  
  \begin{overpic}[height=6.5cm, trim=30mm 90mm 40mm 90mm, clip=true]{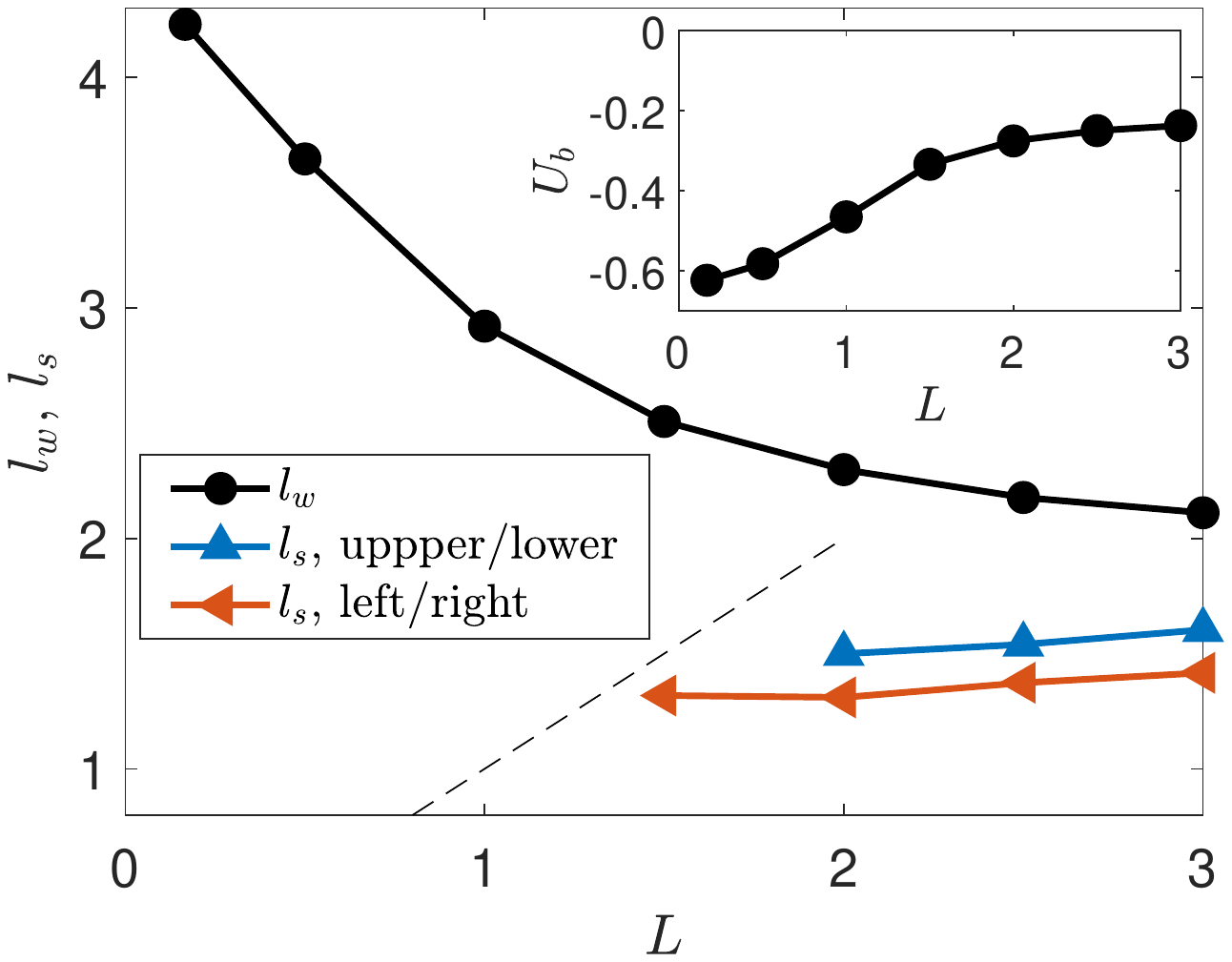}
      \put(0,74){$(a)$}
  \end{overpic}
  \begin{overpic}[height=6.5cm, trim=28mm 90mm 40mm 90mm, clip=true]{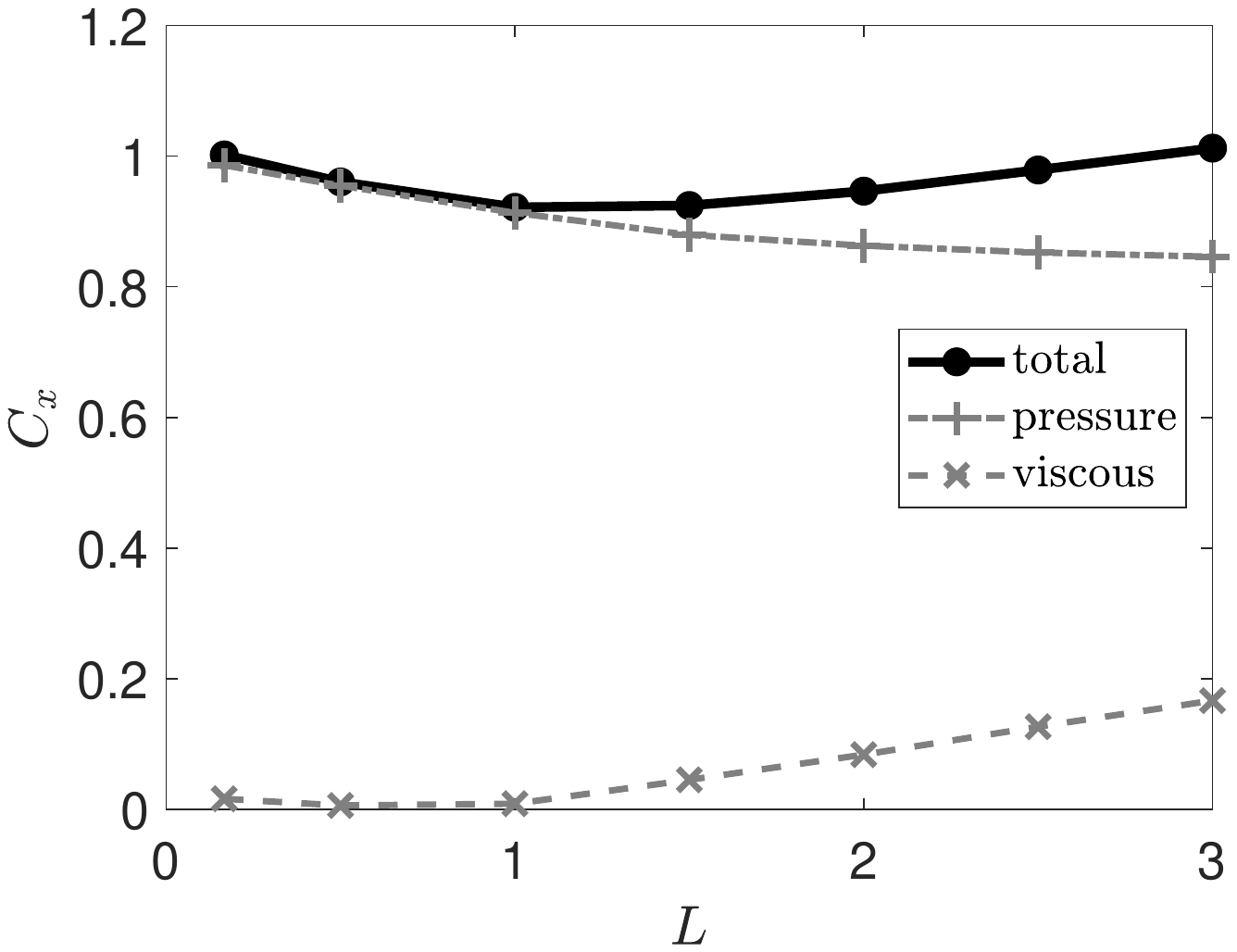}
      \put(0,73){$(b)$}
  \end{overpic}
}
\caption{
Base flow past rectangular prisms, $W=1.2$, $R=0$, $Re=250$.
$(a)$~Length of the wake recirculation and side recirculations.
Dashed line: $l_s=L$, when the side recirculations reach the end of the body  and connect with the wake recirculation.
$(b)$~Drag coefficient.
}
\label{fig:recirc_length_R0_Re250}
\end{figure}

These qualitative observations are confirmed in
figure~\ref{fig:recirc_length_R0_Re250}$(a)$: the length of the wake recirculation decreases with $L$, while the  side recirculations (disconnected from the wake only when $L > l_w$) keep a fairly constant length. 
The backflow $U_b$ (inset) is much stronger when $L\lesssim 1.5$, i.e. when the side and wake  recirculations are connected.

%\begin{figure}
%\centerline{
%  \fbox{    
%    \begin{overpic}[width=8cm, trim=28mm 90mm 40mm 90mm, clip=true]{C:/Users/boujo/Documents/_Projects_MSc/2021_Adrien_Gimonnet/results/quarter-other-ARs/Cx_vs_L-Re250.pdf}
%      \put(0,74){$(b)$}
% 	\end{overpic}
%  }
%}
%\caption{
%Drag coefficient of rectangular prisms, $W=1.2$, $R=0$, $Re=250$.
%}
%\label{fig:drag_R0_Re250}
%\end{figure}

Figure~\ref{fig:recirc_length_R0_Re250}$(b)$ shows that the drag coefficient  varies non-monotonically with $L$, reaching a minimum for $L \simeq 1.5$.
This is  the result of a competition between pressure and viscous effects.
Pressure drag decreases with $L$, as longer bodies have narrower wakes, associated with a weaker front/rear pressure difference.
Conversely, viscous drag increases with $L$, as longer bodies are subjected to positive wall shear stress over a wider surface area.

%-----------------------------------------------------
\subsection{Effect of $R$ \label{sec:BF-R}}

\begin{figure}
\centerline{   
    \begin{overpic}[width=8cm, trim=32mm 105mm 39mm 118mm, clip=true]{./figs/Ub-re250-UUU.pdf}
      \put(-3,48){$(a)$}   
 	\end{overpic}
    \begin{overpic}[width=8cm, trim=32mm 105mm 39mm 118mm, clip=true]{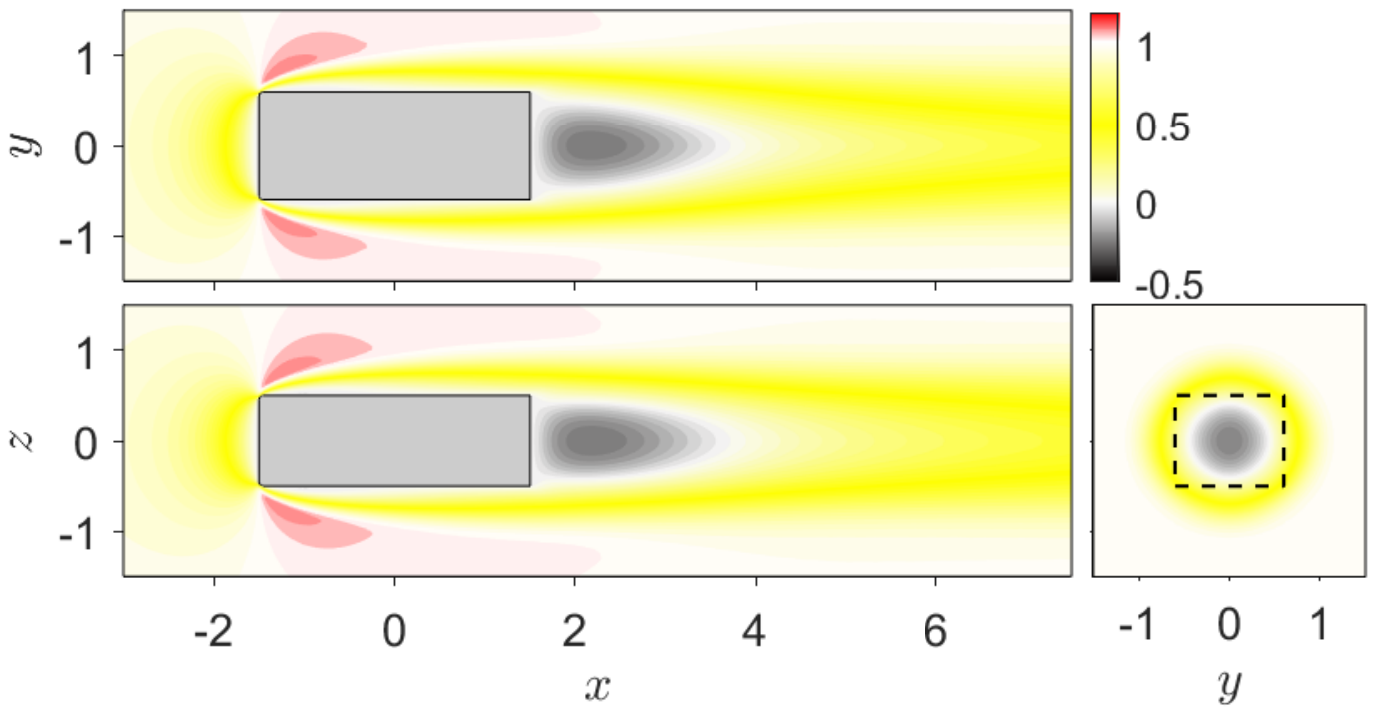}
      \put(-3,48){$(b)$}   
 	\end{overpic}
}
\centerline{   
    \begin{overpic}[width=8cm, trim=32mm 105mm 39mm 118mm, clip=true]{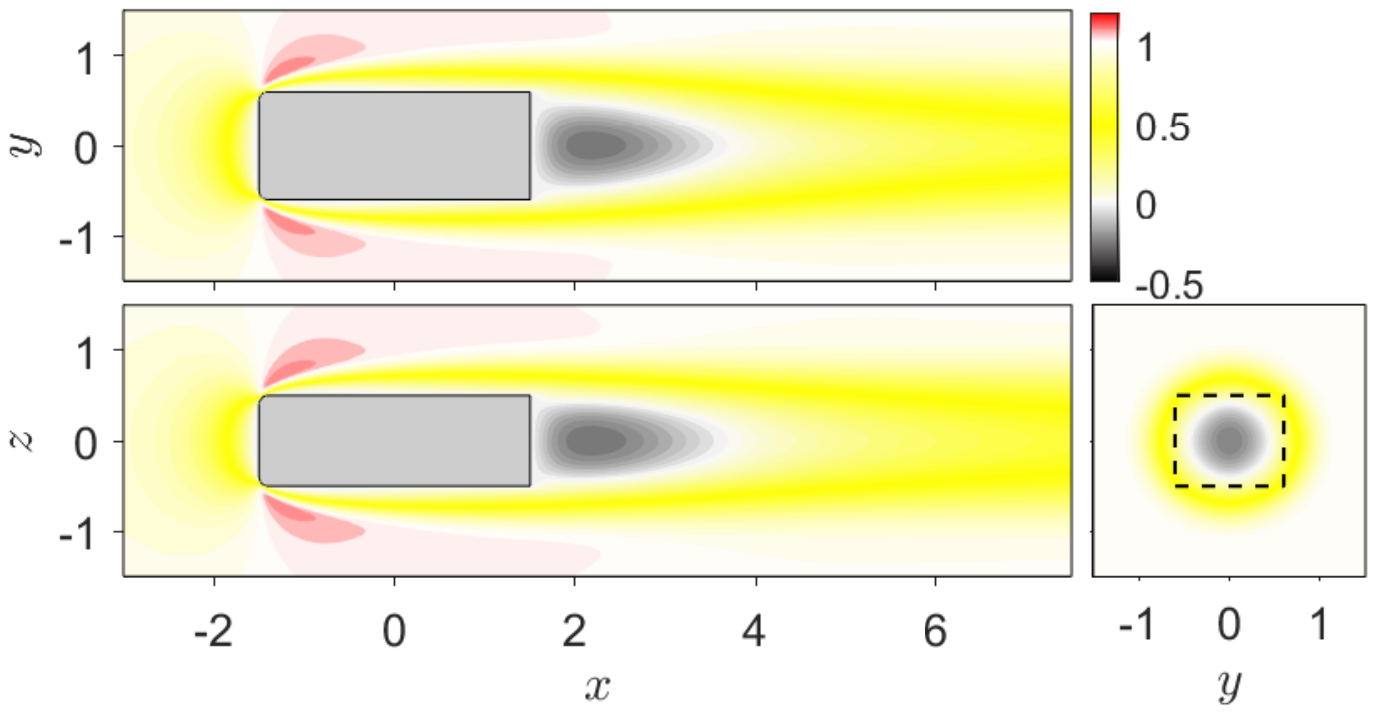}
      \put(-3,48){$(c)$}   
 	\end{overpic}
    \begin{overpic}[width=8cm, trim=32mm 105mm 39mm 118mm, clip=true]{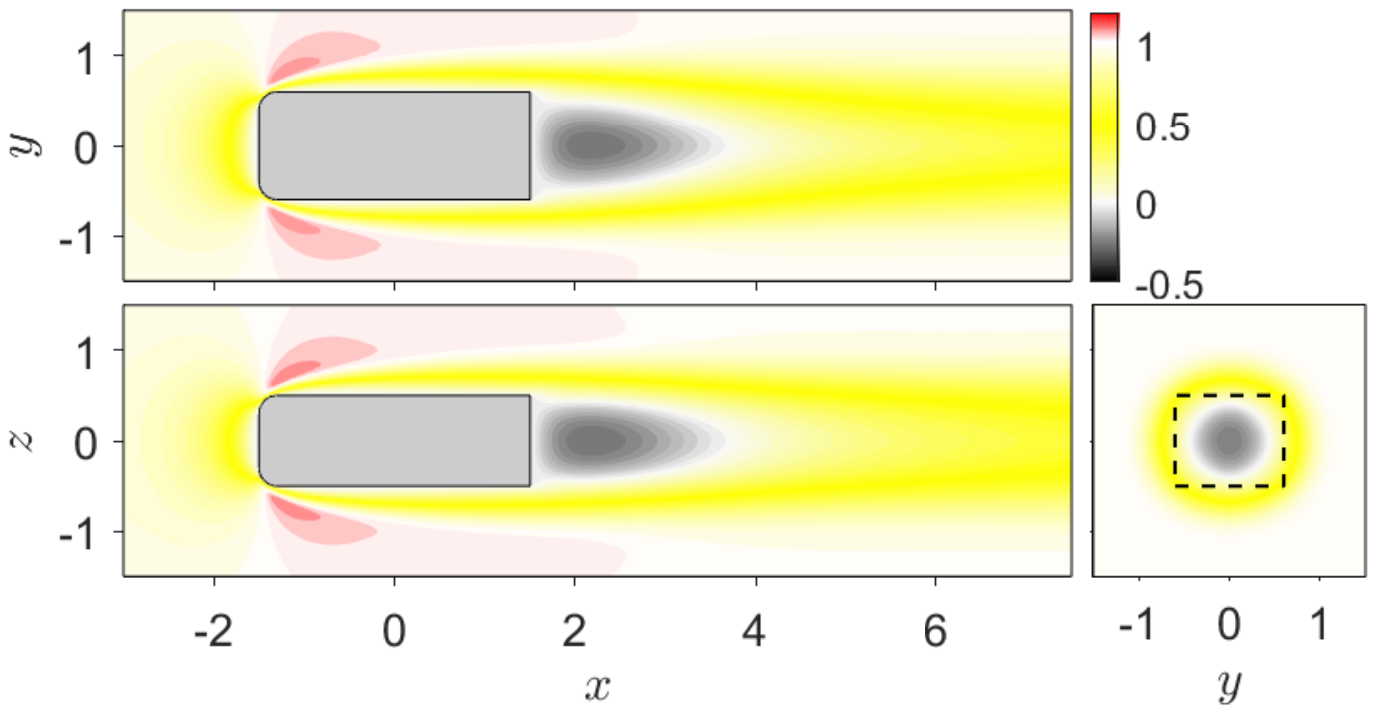}
      \put(-3,48){$(d)$}   
 	\end{overpic}
}
\centerline{   
    \begin{overpic}[width=8cm, trim=32mm 105mm 39mm 118mm, clip=true]{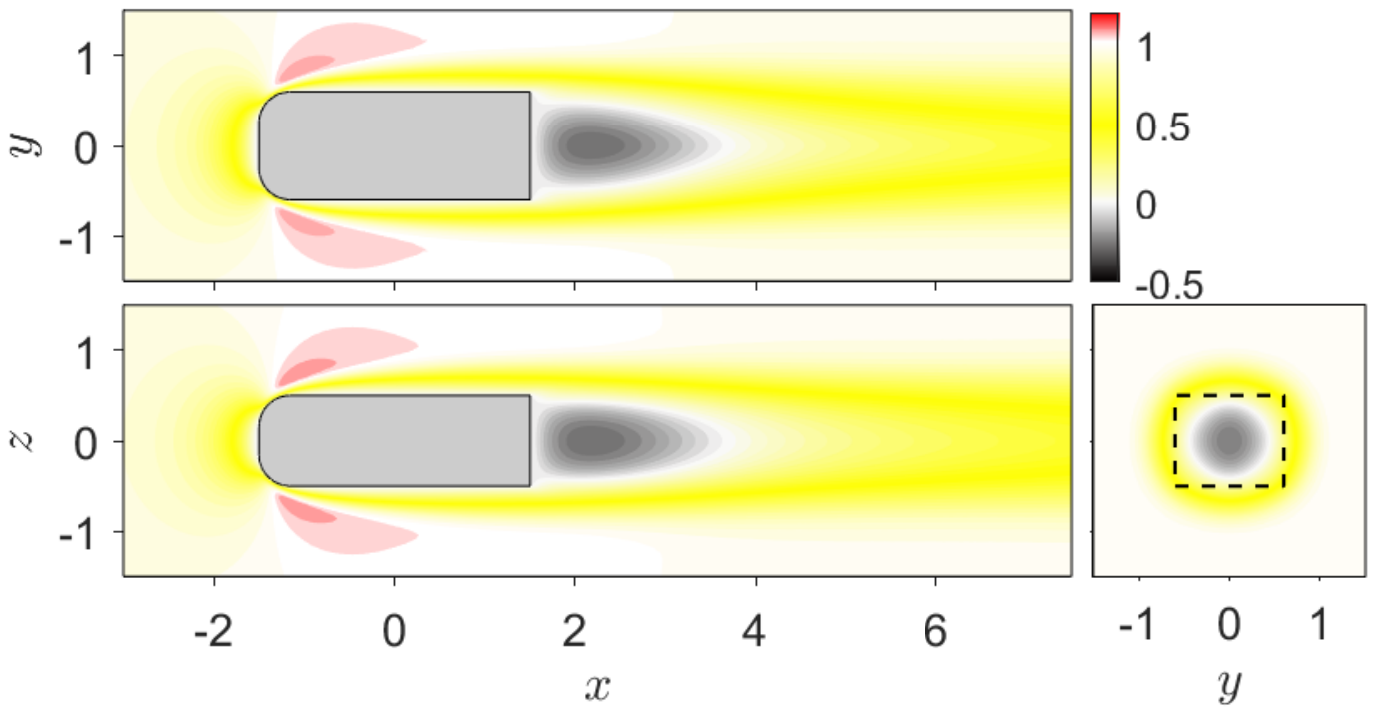}
      \put(-3,48){$(e)$}   
 	\end{overpic}
    \begin{overpic}[width=8cm, trim=32mm 105mm 39mm 118mm, clip=true]{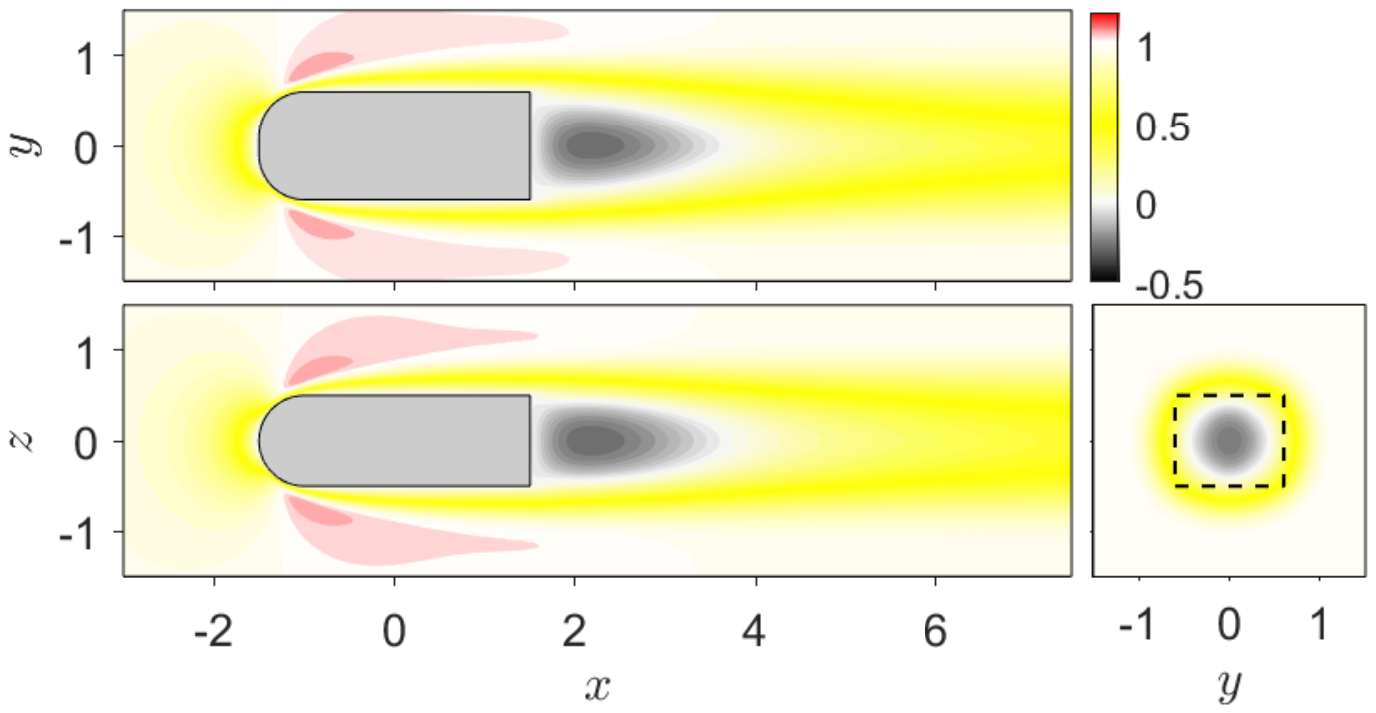}
      \put(-3,48){$(f)$}   
 	\end{overpic}
}
\caption{
Streamwise velocity $u_0$ of the base flow past rectangular prisms,  $W=1.2$, $L=3$, at $Re=250$, in the planes $z=0$ (top view), $y=0$ (side view) and $x=2.5$ (rear view):
$(a)$~$R=0$,
$(b)$~$R=0.05$,
$(c)$~$R=0.1$,
$(d)$~$R=0.2$,
$(e)$~$R=0.3472$,
$(f)$~$R=0.5$.
}
\label{fig:BF_R_effect}
\end{figure}

\begin{figure}
\centerline{  
    \begin{overpic}[width=8cm, trim=30mm 90mm 40mm 90mm, clip=true]{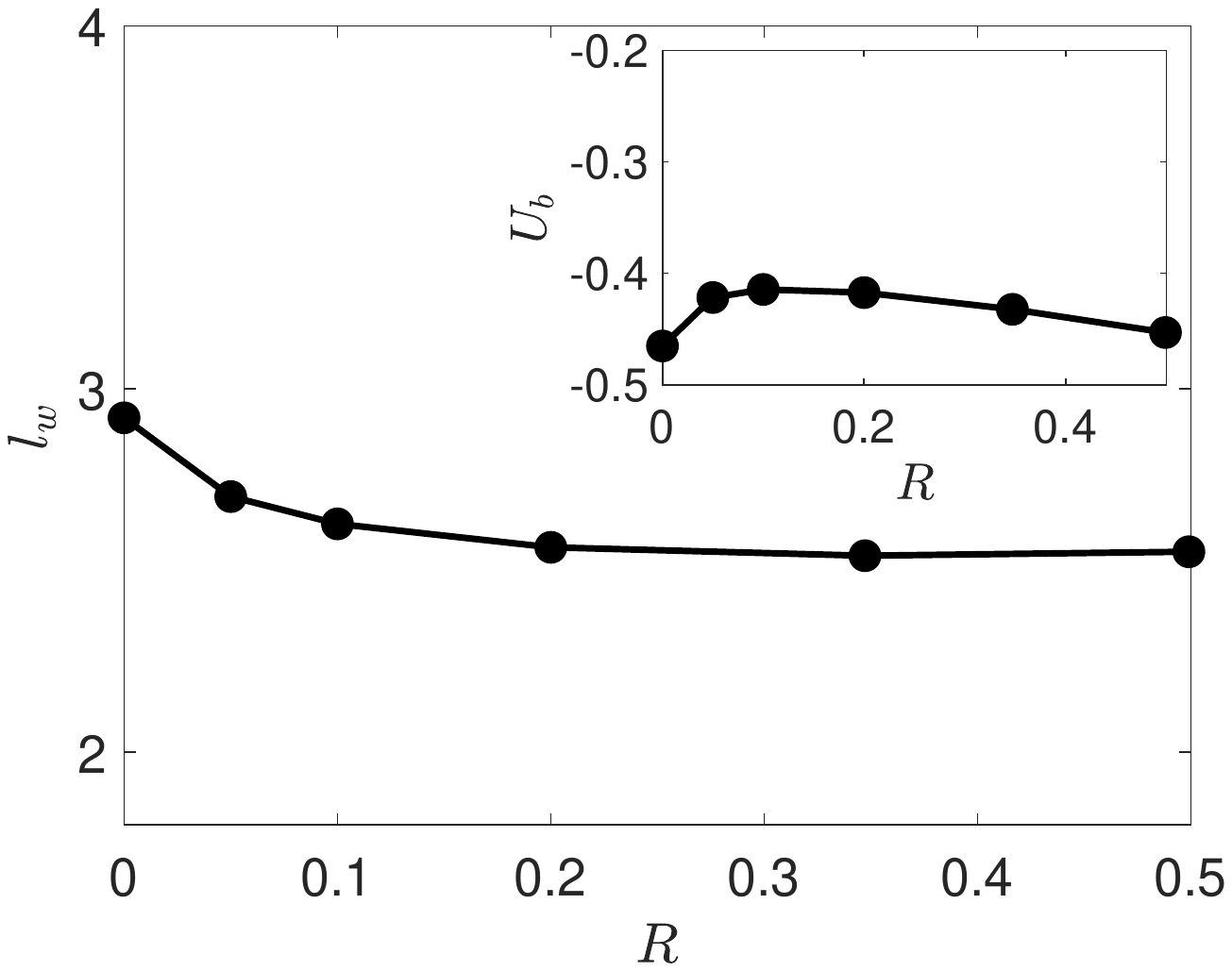}
      \put(0,74){$(a)$}
 	\end{overpic}
    \begin{overpic}[width=8cm, trim=30mm 90mm 40mm 90mm, clip=true]{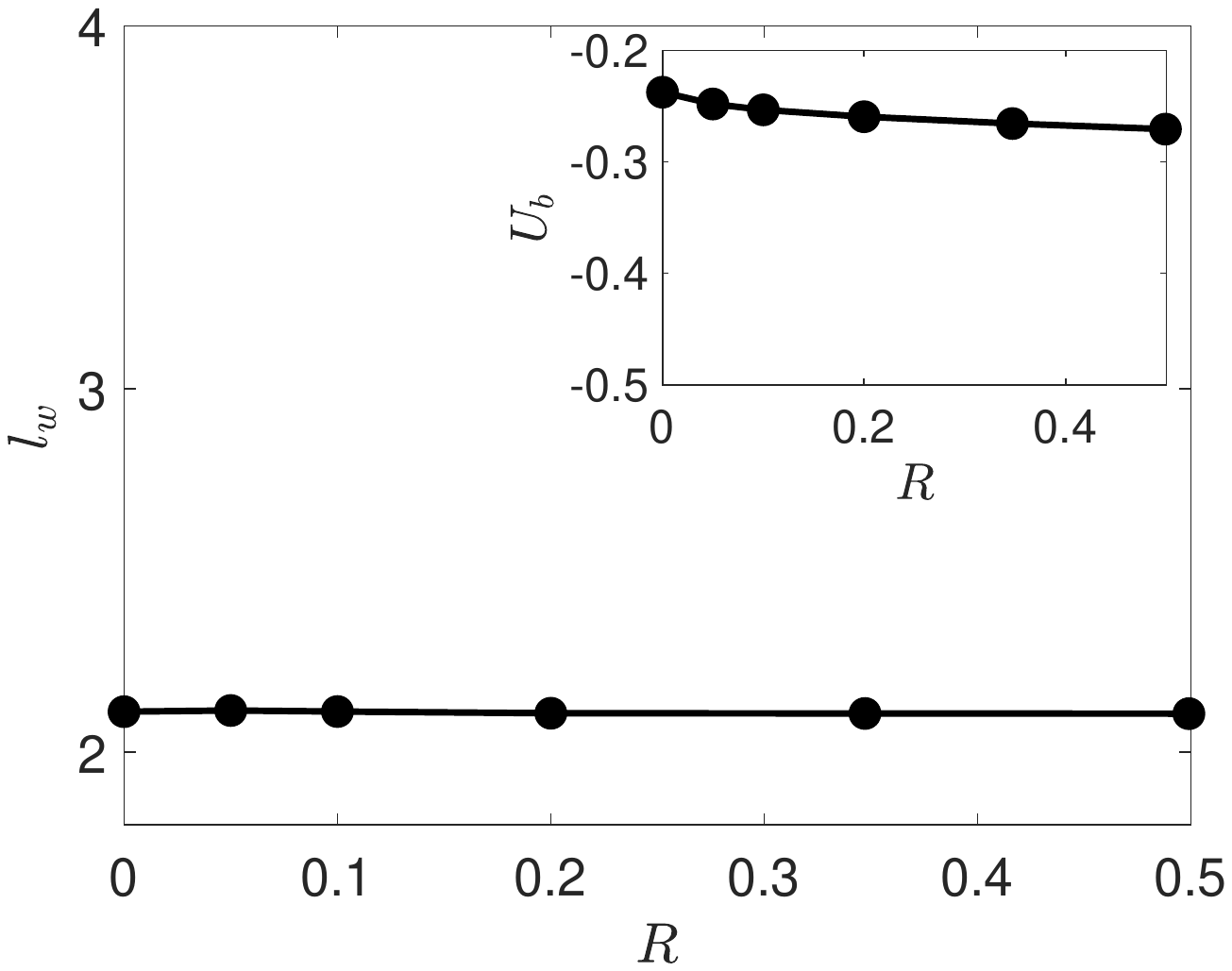}
      \put(0,74){$(b)$}
 	\end{overpic}
}
\caption{
Length of the wake recirculation in the flow past rectangular prisms, $W=1.2$, $Re=250$.
$(a)$~$L=1$, $(b)$~$L=3$.
}
\label{fig:recirc_length_L1_L3_Re250}
\end{figure}

We now investigate the effect of the fillet radius $R$ on the steady base flow past bodies of length $L=1$ and $L=3$. 
Figure~\ref{fig:BF_R_effect} shows the streamwise velocity field for $L=3$ and several  fillet radius values. 
Rounding the front edges clearly suppresses the side recirculations. It turns out that a very small fillet, $R \gtrsim 0.05$, is sufficient to keep the flow attached to the body. 
For the shorter body, $L=1$, this also makes the wake recirculation narrower and slightly shorter. For the longer body, $L=3$, this has no visible effect on the wake recirculation.
Figure~\ref{fig:recirc_length_L1_L3_Re250} confirms that the wake recirculation length is slightly reduced for $L=1$ and barely affected for $L=3$.
The backflow (inset) follows different trends for $L=1$ and $L=3$.

\begin{figure}
\centerline{    
    \begin{overpic}[width=8cm, trim=30mm 90mm 40mm 90mm, clip=true]{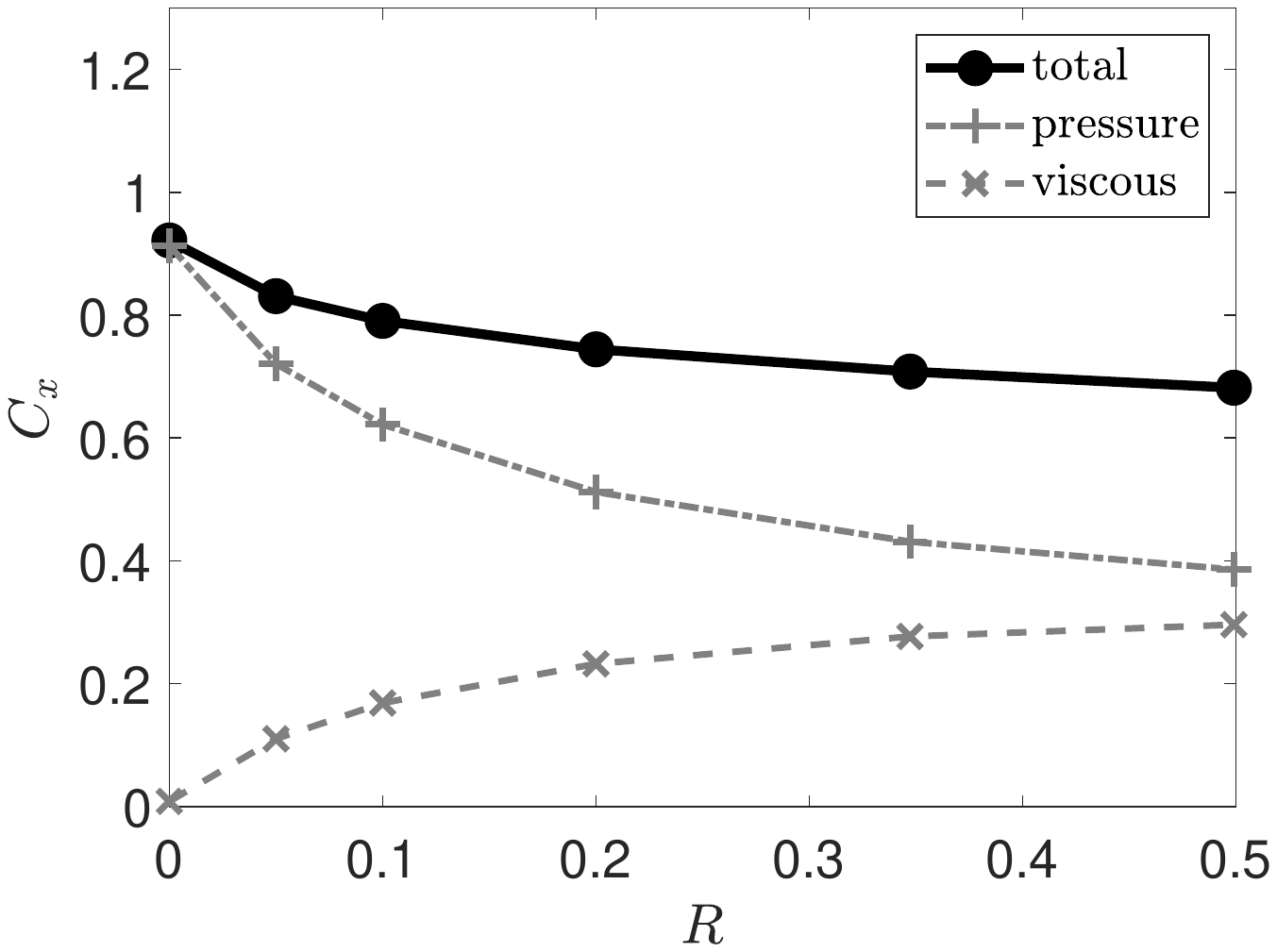}
      \put(0,74){$(a)$}
 	\end{overpic}    
    \begin{overpic}[width=8cm, trim=30mm 90mm 40mm 90mm, clip=true]{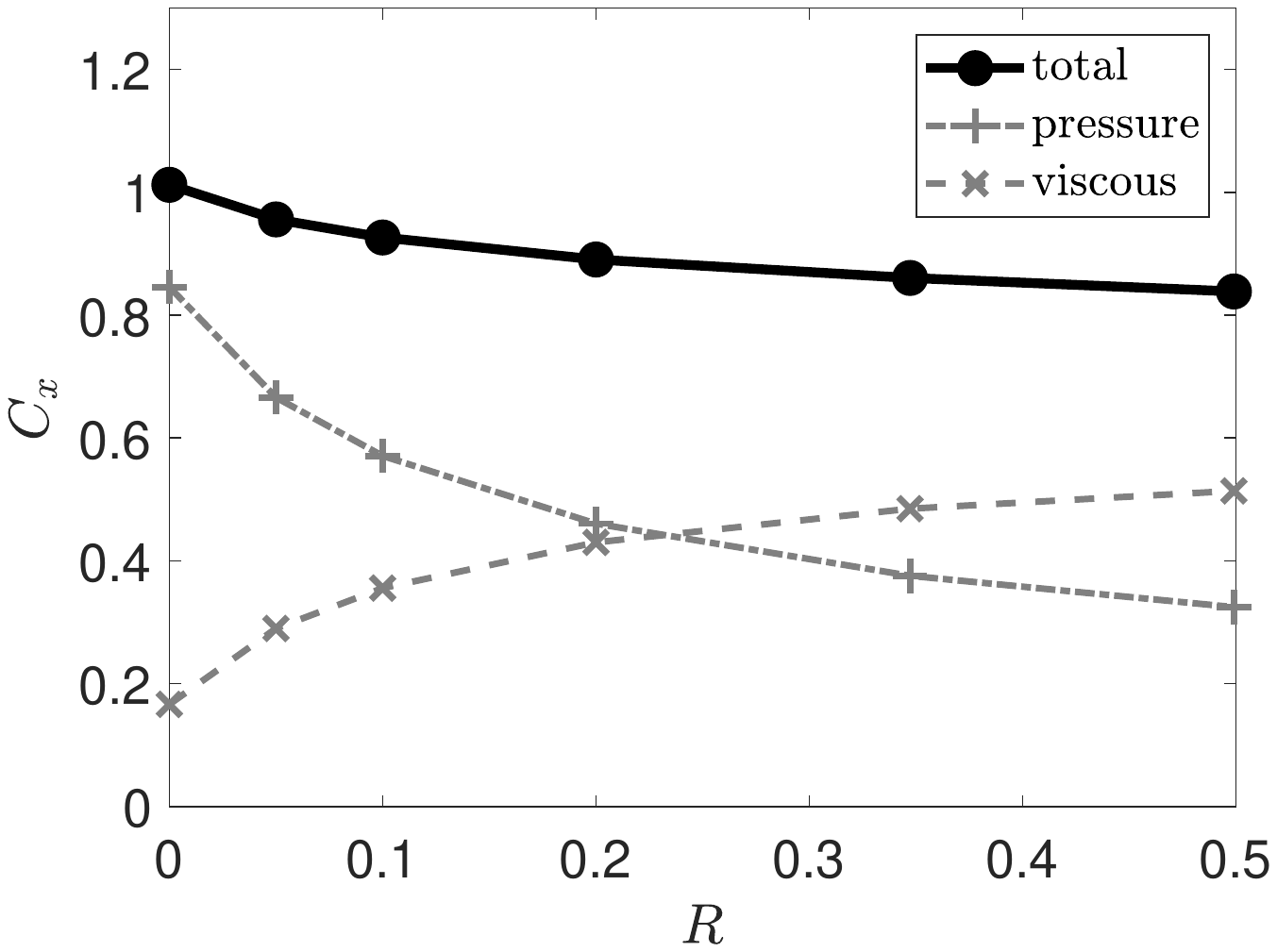}
      \put(0,74){$(b)$}
 	\end{overpic}
}
\caption{
Drag coefficient of  rectangular prisms, $W=1.2$, $Re=250$.
$(a)$~$L=1$, $(b)$~$L=3$.
}
\label{fig:drag_L1_L3_Re250}
\end{figure}

Finally, figure~\ref{fig:drag_L1_L3_Re250} shows that $R$ has a strong effect on drag. 
For both $L=1$ and $L=3$, pressure drag decreases with $R$, as a result of the reduced front area.
By contrast, viscous drag increases because boundary layers stay attached to the body. Overall,   pressure effects are stronger and $C_x$ decreases.

%-----------------------------------------------------
%-----------------------------------------------------
%-----------------------------------------------------
\section{Linear stability analysis \label{sec:LSA}}

%-----------------------------------------------------
%\subsection{Presentation}

We now investigate the linear stability of the steady base flows $\bm q_0(t)$ computed in section~\ref{sec:BF}.
We therefore consider small-amplitude perturbations,  
\begin{align} 
\bm q(\bm x,t) = \bm q_0(\bm x) + \epsilon \bm q_1(\bm x,t),
\label{eq:q0_eps_q1}
\end{align}
where $0 < \epsilon \ll 1$.
Injecting (\ref{eq:q0_eps_q1}) in the NS equations (\ref{eq:NS}) yields at order $\epsilon$ the linear dynamics of the perturbations $\bm q_1 (\bm x,t) = (\bm u_1, p_1)^T$:
\begin{align}
\bnabla \cdot \bm u_1 = 0,
\quad
\partial_t \bm u_1 +
(\bm{u}_0 \cdot \bnabla) \bm{u}_1 +
(\bm{u}_1 \cdot \bnabla) \bm{u}_0 = -\bnabla p_1 + \frac{1}{Re} \bnabla^2 \bm{u}_1.
\label{eq:LNS}
\end{align}
We rewrite (\ref{eq:LNS}) in compact form,
\begin{align}
\mathcal B \partial_t  {\bm q}_1 + \mathcal A {\bm q}_1 = \bm{0},
\label{eq:LNS_compact}
\end{align}
with the linear operators
\begin{align}
\mathcal A = \left( 
\begin{array}{cc}
\mathcal C(\bm{u}_0, \bcdot)  
 - Re^{-1} \bnabla^2 (\bcdot)
& \bnabla (\bcdot) 
\\ 
\bnabla \cdot (\bcdot) & 0
\end{array}
\right),
\quad
\mathcal B = \left( 
\begin{array}{cc}
\mathcal I & 0 \\ 
0 & 0
\end{array}
\right),
\label{eq:LNS_A_B}
\end{align}
%are the NS operator linearised about the base flow $\bm q_0$,
%the mass operator,
with
$\mathcal C(\bm a, \bm b) = (\bm a \cdot \bnabla) \bm b + (\bm b \cdot \bnabla) \bm a$
the (symmetric) convection operator, and $\mathcal I$ the identity operator.

Using the normal mode ansatz $\bm q_1(\bm x,t) = \hat {\bm q}_1(\bm x) e^{\lambda t} + c.c.$ (where $c.c.$ stands for complex conjugate) yields an eigenvalue problem for the complex eigenvalues $\lambda = \sigma + i \omega$ and complex-valued eigenmodes $\hat{\bm q}_1(\bm x)$:
\begin{align}
\bnabla \cdot \hat{\bm u}_1 = 0,
\quad
\lambda \hat{\bm u}_1 +
(\bm{u}_0      \cdot \bnabla) \hat{\bm u}_1 +
(\hat{\bm u}_1 \cdot \bnabla) \bm{u}_0 
= -\bnabla \hat{p}_1 + \frac{1}{Re} \bnabla^2 \hat{\bm u}_1,
\label{eq:EVP}
\end{align}
or in compact form
\begin{align}
\lambda \mathcal B \hat {\bm q}_1 + \mathcal A \hat {\bm q}_1 = \bm{0}.
\label{eq:EVP_compact}
\end{align}

Unlike the base flow, which is symmetric in both $y$ and $z$ directions, there are four families of eigenmodes corresponding to all possible combinations of symmetry ($S$) and anti-symmetry ($A$) in the $y$ and $z$ directions: 
$S_y S_z$, $S_y A_z$, $A_y S_z$, $A_y A_z$.
For each base flow, i.e. each set of parameters $(W,L,R,Re)$, we therefore compute four eigenspectra.
On the symmetry planes, boundary conditions for these four families are 
as follows:
\begin{align}
S_y S_z: \quad
& \partial_y u_1 = v_1 = \partial_y w_1 = 0 \quad \mbox{on the } y=0 \mbox{ plane},
\nonumber 
\\
& \partial_z u_1 = \partial_z v_1 = w_1 = 0 \quad \mbox{on the } z=0 \mbox{ plane};
\label{eq:BC_EVP_SS}
\\
\nonumber 
\\
S_y A_z: \quad
& \partial_y u_1 = v_1 = \partial_y w_1 = 0 \quad \mbox{on the } y=0 \mbox{ plane},
\nonumber 
\\
& u_1 = v_1 = \partial_z w_1 = p_1 = 0 \quad \mbox{on the } z=0 \mbox{ plane};
\label{eq:BC_EVP_SA}
\\
\nonumber 
\\
A_y S_z: \quad
& u_1 = \partial_y v_1 = w_1 = p_1 = 0 \quad \mbox{on the } y=0 \mbox{ plane},
\nonumber 
\\
& \partial_z u_1 = \partial_z v_1 = w_1 = 0 \quad \mbox{on the } z=0 \mbox{ plane};
\label{eq:BC_EVP_AS}
\\
\nonumber 
\\
A_y A_z: \quad
& u_1 = \partial_y v_1 = w_1 = p_1 = 0 \quad \mbox{on the } y=0 \mbox{ plane},
\nonumber 
\\
& u_1 = v_1 = \partial_z w_1 = p_1 = 0 \quad \mbox{on the } z=0 \mbox{ plane}.
\label{eq:BC_EVP_AA}
\end{align}
Elsewhere, boundary conditions are directly derived from and consistent with the base flow boundary conditions:
homogeneous Dirichlet boundary condition $\bm{u}_1 = \bm 0$ on the inlet plane,
no-slip condition $\bm u_1 = \bm 0$ on the body surface,
stress-free condition $-p_1 \bm n + Re^{-1} \bnabla \bm u_1 \cdot \bm n = \bm 0$ on the outlet plane,
and symmetry conditions similar to (\ref{eq:BC_BF}) on the remaining (lateral and upper) boundaries.

Eigenmodes are defined up to a complex-valued factor.
For the sake of comparison, in the following we choose  to normalise the modes as $\langle \mathcal{B}\hat{\bm q}_1 ,\, \hat{\bm q}_1 \rangle 
=\langle \hat{\bm u}_1 ,\, \hat{\bm u}_1 \rangle
= 1$,
where the inner product is defined by $\langle \bm{a} ,\, \bm{b} \rangle = \int_{\Omega'} \bm{a}^* \cdot \bm{b} \, \mathrm{d} \bm{x}$, the superscript $*$ stands for complex conjugate,
and whether $\bm{a}$ and $\bm{b}$  contain both velocity and pressure fields or velocity fields only is clear from the context.

We also compute adjoint modes ${\bm q}_1^{\dag}$ solution of the eigenvalue problem
\begin{align}
\lambda \mathcal B \hat{\bm q}_1^{\dag} + \mathcal A^\dag \hat{\bm q}_1^{\dag} = \bm{0},
\label{eq:ADJ_compact}
\end{align}
where the adjoint linearised NS operator is defined by 
$ \langle \mathcal{A} \bm a ,\, \bm b \rangle = \langle  \bm a ,\, \mathcal{A}^\dag \bm b \rangle $ for any $\bm a$, $
\bm b$.
Integration by parts yields
\begin{align}
\mathcal{A}^\dag = \left( 
\begin{array}{cc}
\mathcal{C}^\dag(\bcdot, \bm{u}_0) - Re^{-1} \bnabla^2 (\bcdot)
& -\bnabla (\bcdot) 
\\ 
\bnabla \cdot (\bcdot) & 0
\end{array}
\right),
\label{eq:ADJ_A}
\end{align}
where 
$\mathcal{C}^\dag(\bm{a},\bm{b})
=
\bnabla \bm{b}^T \cdot \bm{a} -(\bm{b} \cdot \bnabla) \bm{a} 
$
is the (non-symmetric) adjoint advection operator, responsible for convective non-normality in open flows \citep{Chomaz05}accounte.
Again, adjoint modes are defined up to a complex-valued factor, and we choose the normalisation $\langle \mathcal{B}\hat{\bm q}_1^\dag ,\, \hat{\bm q}_1 \rangle = \langle \hat{\bm u}_1^\dag ,\, \hat{\bm u}_1 \rangle = 1.$

%-----------------------------------------------------
\subsection{Results}

\begin{figure}
  \centerline{   
    \begin{overpic}[width=8cm, trim=32mm 92mm 42mm 95mm, clip=true]{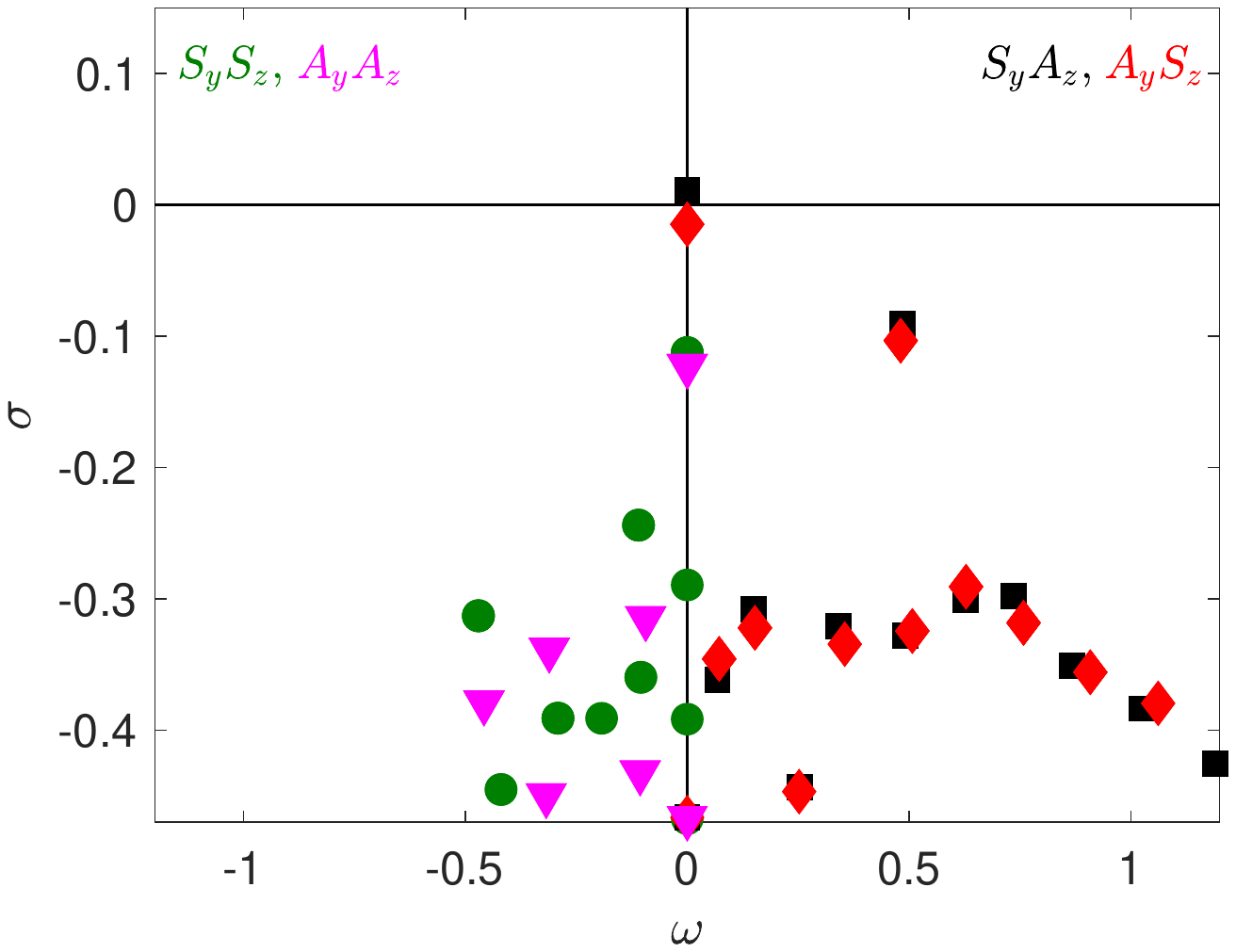}
      \put(-1,74){$(a)$}
 	\end{overpic}
 	\hspace{0.2cm}  
    \begin{overpic}[width=8cm, trim=32mm 92mm 42mm 95mm, clip=true]{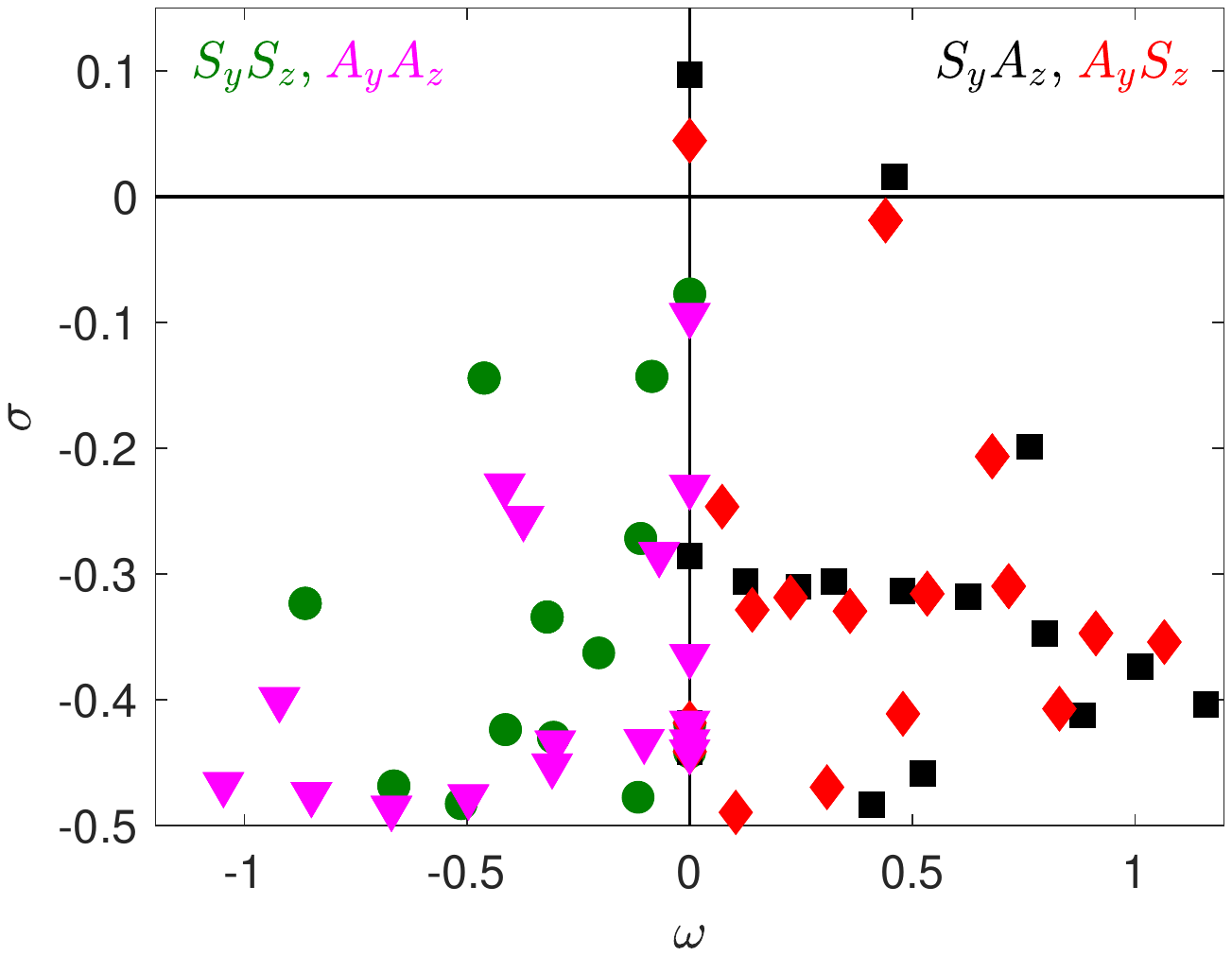}
      \put(-1,74){$(b)$}
 	\end{overpic}
  }
  \vspace{0.5cm}
  \centerline{   
   \begin{overpic}[width=16.1cm, trim=10mm 5mm 20mm 30mm, clip=true]{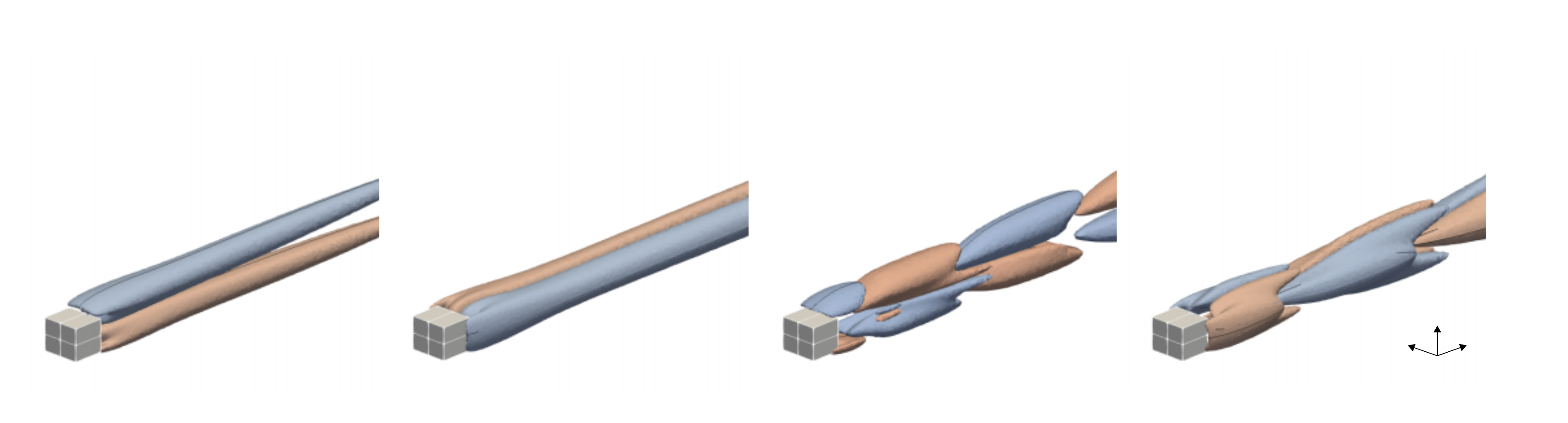}
      \put(   0,15){$(c)$}
      \put(51.5,15){$(d)$}
      \put( 8,  15){\footnotesize     $S_y A_z$}
      \put(33.7,15){\footnotesize\tcr{$A_y S_z$}}
      \put(58.4,15){\footnotesize     $S_y A_z$}
      \put(85.1,15){\footnotesize\tcr{$A_y S_z$}}
      \put(99.5,3.5){\footnotesize $x$}
      \put(93.5,3.5){\footnotesize $y$}
      \put(96.5,5.7){\footnotesize $z$}
 	\end{overpic}
  }
\caption{
Eigenvalue spectra for the rectangular prism $W=1.2$, $L=1$, $R=0$, at 
$(a)$~$Re=185$, between the  first and second pitchfork (stationary) bifurcations, and
$(b)$~$Re=220$, between the  first and second Hopf (oscillating) bifurcations.
The first four bifurcating eigenmodes belong to  the families $S_y A_z$ and $A_y S_z$, i.e.  break either the top/down  symmetry or the left/right symmetry, respectively: 
$(c)$~stationary modes and $(d)$~oscillatory modes (isosurfaces of streamwise velocity).
}
\label{fig:spectrum-example}
\end{figure}

In the following, we fix $W=1.2$ and compute eigenspectra for different body length $L$, fillet radius $R$ and Reynolds number $Re$. 
Figure~\ref{fig:spectrum-example} shows typical spectra for  $L=1$, $R=0$, at $Re=185$ in panel $(a)$ and $Re=220$ in panel $(b)$.
In each panel, the left half-plane shows  $S_y S_z$ and $A_y A_z$ eigenmodes (circles and  triangles, respectively), and the right half-plane shows $S_y A_z$ and $A_y S_z$ eigenmodes (squares and diamonds, respectively). 
Full spectra can be reconstructed from half-plane spectra because eigenvalues are frequency-symmetric: they come either as purely real values (stationary modes, $\lambda=\sigma$, $\omega=0$) or as complex conjugate pairs (oscillatory modes, $\lambda = \sigma \pm i \omega$ with $\omega \neq 0$). 
In figure~\ref{fig:spectrum-example}$(a)$, it appears that two  modes are marginally stable (small growth rate $\sigma$): one $S_y A_z$ mode 
and one $A_y S_z$ mode. 
These two modes are both stationary, i.e. they become unstable via pitchfork bifurcations.
At larger $Re$, in figure~\ref{fig:spectrum-example}$(b)$, two other modes are marginally stable, and belong again to the $S_y A_z$ and $A_y S_z$ families.
These two modes are both oscillatory, i.e. they become unstable via Hopf bifurcations.
Other modes, in particular doubly-symmetric $S_y S_z$ modes and doubly-antisymmetric $A_y A_z$ modes, are all strongly stable.

%-----------------------------------------------------
\subsubsection{Effect of $L$ \label{sec:LSA-L}}

\begin{figure}
%\centerline{   
%    \begin{overpic}[width=9cm, trim=30mm 106mm 25mm 95mm, clip=true]{C:/Users/boujo/Documents/_Projects_MSc/2021_Adrien_Gimonnet/results/Rec_vs_L_fixed_R_0.pdf}
%      \put(-1,56){$(a)$}
%      \put(25,45){\footnotesize oscillatory}
%      \put(65,35){\footnotesize stationary}
% 	\end{overpic}
%}
%\centerline{   
%    \begin{overpic}[width=9cm, trim=30mm 135mm 25mm 97mm, clip=true]{C:/Users/boujo/Documents/_Projects_MSc/2021_Adrien_Gimonnet/results/omc_vs_L_fixed_R_0.pdf}
%      \put(-1,38){$(b)$}
% 	\end{overpic}
%}
\centerline{   
    \begin{overpic}[width=9cm, trim=30mm 106mm 25mm 95mm, clip=true]{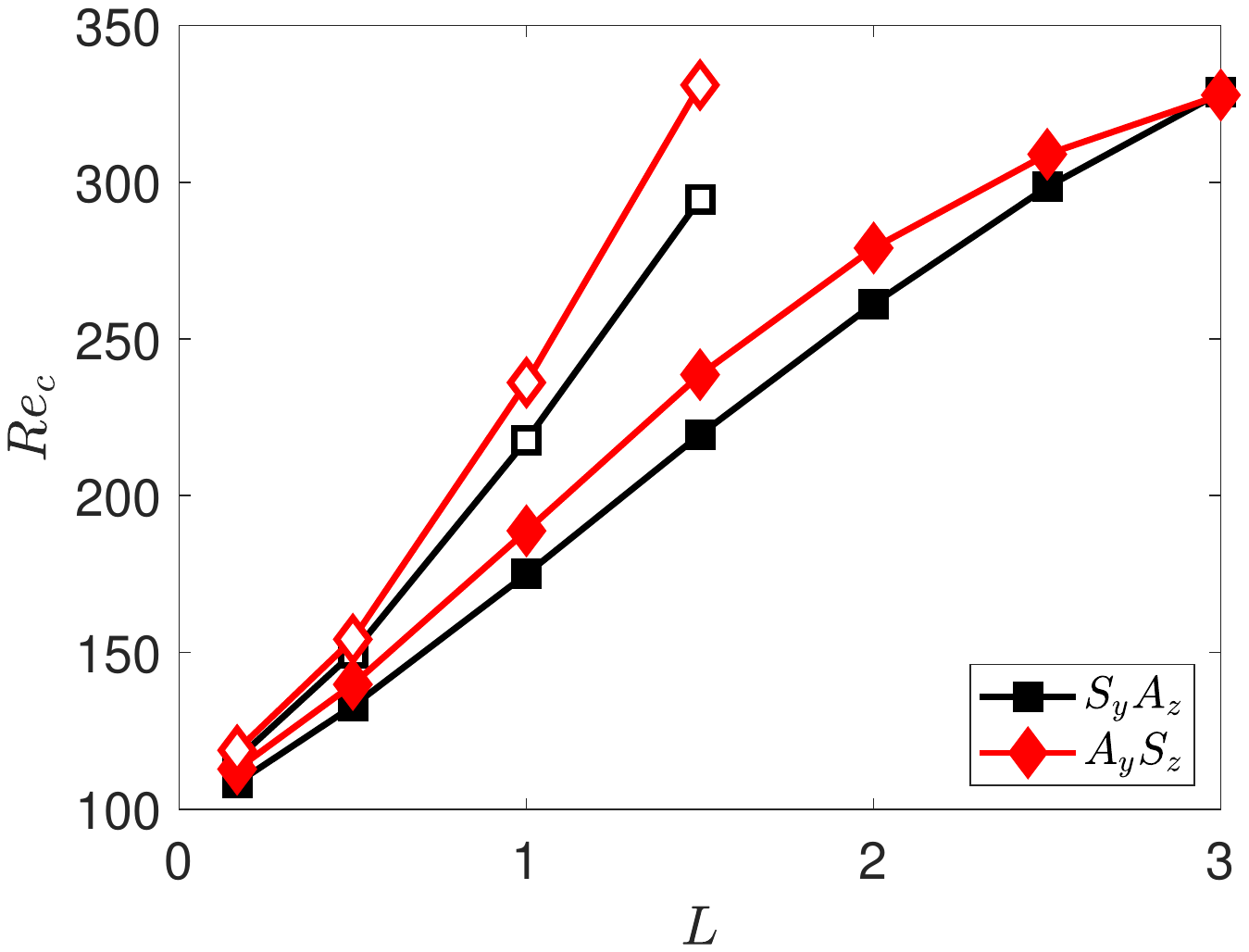}
      \put(-1,56){$(a)$}
      \put(25,45){\footnotesize oscillatory}
      \put(65,35){\footnotesize stationary}
%      \put(25,54){\footnotesize oscillatory}
%      \put(63,43){\footnotesize stationary}
 	\end{overpic}
}
\centerline{   
    \begin{overpic}[width=9cm, trim=30mm 135mm 25mm 97mm, clip=true]{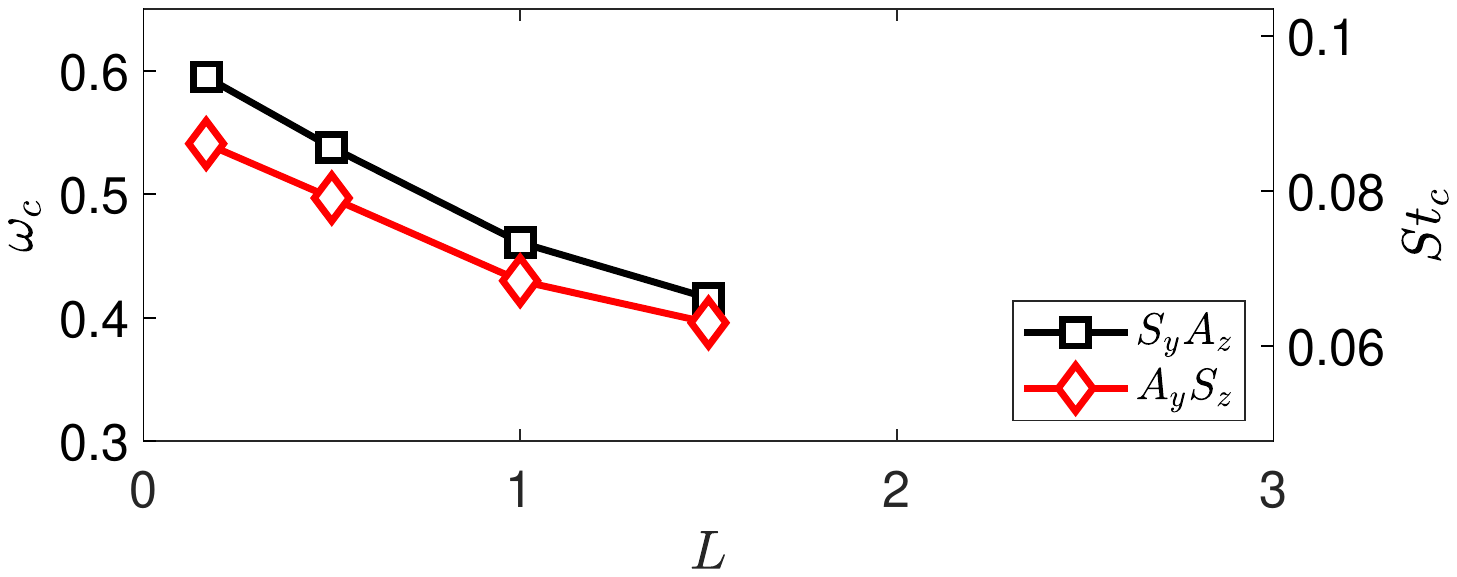}
      \put(-1,36){$(b)$}
 	\end{overpic}
}
\caption{
Properties of the first bifurcations in the wake flow past rectangular prisms of various length $L$ and with sharp edges ($R=0$).
$(a)$~Critical Reynolds number $Re_c$.
Filled symbols: stationary (pitchfork) bifurcations; open symbols: oscillatory (Hopf) bifurcations.
$(b)$~Critical angular frequency $\omega_c=\omega(Re_c)$ (left axis) and Strouhal number $St_c=\omega_c/(2\pi)$ (right axis) of the oscillatory modes.
}
\label{fig:Rec_R_0}
\end{figure}

In this section we investigate the effect of the length $L$ on the linear stability of wake flows past bodies with sharp edges ($R=0$). 
The corresponding steady base flows were described in section~\ref{sec:BF-L}.
Figure~\ref{fig:Rec_R_0} shows the critical Reynolds number of the first bifurcations observed.
Throughout this study, critical Reynolds numbers are computed by cubic interpolation of $Re(\sigma)$ from at least three values of $Re$  that bracket $\sigma=0$, with a maximum step $\Delta Re=5$ between two successive Reynolds number.
For all the values of $L$ investigated in this study, two stationary $S_y A_z$ and $A_y S_z$ modes become unstable first (pitchfork bifurcations), followed by two pairs of oscillatory $S_y A_z$ and $A_y S_z$  modes (Hopf bifurcations). 
These four bifurcating modes all involve one symmetry breaking, while $S_y S_z$ and $A_y A_z$ modes remain stable until much larger Reynolds numbers.
The critical Reynolds numbers of the first four bifurcations shown in figure~\ref{fig:Rec_R_0} all increase with $L$. 
Approximations of these increasing trends are $Re_c \simeq 100 ( 1 + 0.8L)$ for the two pitchfork bifurcations, and $Re_c \simeq 100(1 + 0.8L + 0.4 L^2)$ for the two Hopf bifurcations. 
For the flat plate, $L=1/6$, the stationary and oscillatory modes become unstable almost simultaneously, $110 < Re_c < 120$. As $L$ increases, the oscillatory modes remain stable much longer than the steady ones.

For all values of $L$, the two stationary modes become unstable in close succession, which is certainly related to the body width-to-height ratio being close to 1. 
For thin plates, $L=1/6$, Marquet observed that the gap between the critical Reynolds numbers of these two modes is exactly zero for $W=H$, as expected by symmetry, and increases with $W/H$.

The increase of $Re_c$ with $L$ is fully consistent with results reported for other three-dimensional bluff bodies.
For example, for axisymmetric bodies such as a thin disk, a sphere and bullet-shaped bodies, the first bifurcating mode is always stationary, followed by an oscillatory mode.
As shown in table~\ref{tab:Rec_axisym}, the associated critical Reynolds numbers increase with the length-to-diameter ratio, $L/D$, and the overall trend is in qualitative agreement with that of the rectangular prisms of the present study (choosing $D$ approximately between the height $H$ and the width $W=1.2H$).
Note that the first two bifurcating modes reported in table~\ref{tab:Rec_axisym} are of azimuthal wavenumber $m=1$ and break the axisymmetry of the wake with an arbitrary azimuthal orientation. 
By contrast, rectangular prisms have two planar symmetries, which selects the orientation  of the modes: vertical wake deflection for modes $S_y A_z$ and lateral deflection for modes $A_y S_z$.

%\begin{itemize}
%\item
%Disque (Meliga, Chomaz and Sipp 2009;  Natarajan and Acrivos 2006): 116-117 pour mode steady ($m=1$) et 125 mode unsteady ($m=1$ aussi, $\omega=0.76-0.79$)
%
%\item
%Sphere: 210-213 steady, 273-278 unsteady ($\omega=0.71$) (Natarajan and Acrivos 2006, Przadka et al 2008; regarder aussi Johnson and Patel 1999; Ghidersa and Dusek 2000;
%Tomboulides and Orszag 2000; Magnaudet and Mougin 2007)
%
%\item
%Missile, $L=9.8D$, compressible $M=0.5$: 485 steady, 999 unsteady ($\omega=0.40$) 
%
%\item
%Missile (Bohorquez et al 2011), elliptical nose, for different $L=1$ to 6.
%For example $L=2$: 327 and 518.
%
%
%\item
%Missile (Jimenez-Gonzalez et al), $L=2$: 325
%\end{itemize}
%
%Regarder aussi Pier (2008) and Fabre (2008)

\begin{table}
  \begin{center}
\def~{\hphantom{0}}
  \begin{tabular}{lcccc}
Body   & $L/D$   & $Re_c^s$ & $Re_c^o$ \\[3pt]
Disk   & $\ll 1$ & 116-117  & 125      \\[3pt]
Sphere & 1       & 210-213  & 273-278  \\[3pt]
Bullet & 1       & 216      & 285  \\
       & 1.5     & 286      & 415  \\
       & 2       & 326      & 517  \\
       & 2.5     & 352      & 596  \\
       & 3       & 372      & 652  \\
       & 3.5     & 388      & 696 
  \end{tabular}
  \caption{Critical Reynolds numbers of the first two bifurcations in wake flows past axisymmetric bodies.
$Re_c^s$: stationary (pitchfork) bifurcation;
$Re_c^o$: oscillatory (Hopf) bifurcation.
All bifurcating modes have an azimuthal wavenumber $m=1$.
Ranges reflect variations found in the literature \citep{natarajan_acrivos_1993, Gumowski08, Fabre08, meliga_global_2009, bohorquez2011}.
}
  \label{tab:Rec_axisym}
  \end{center}
\end{table}

%-----------------------------------------------------
\subsubsection{Effect of $R$ \label{sec:LSA-R}}

\begin{figure}
\centerline{   
   \begin{overpic}[width=8cm, trim=32mm 92mm 40mm 95mm, clip=true]{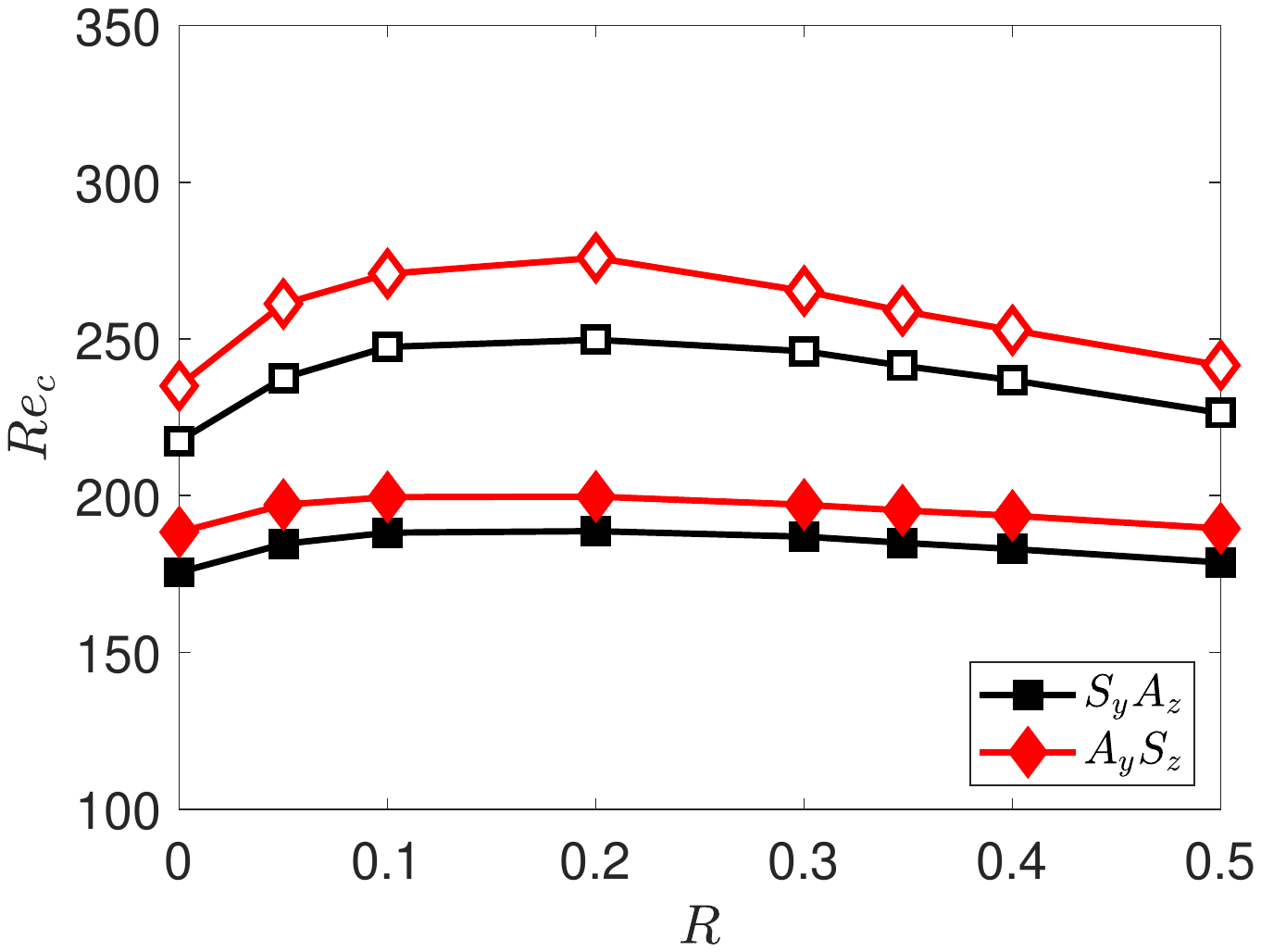}
      \put(-2,74){$(a)$}
  \end{overpic}
  \begin{overpic}[width=8cm, trim=32mm 92mm 40mm 95mm, clip=true]{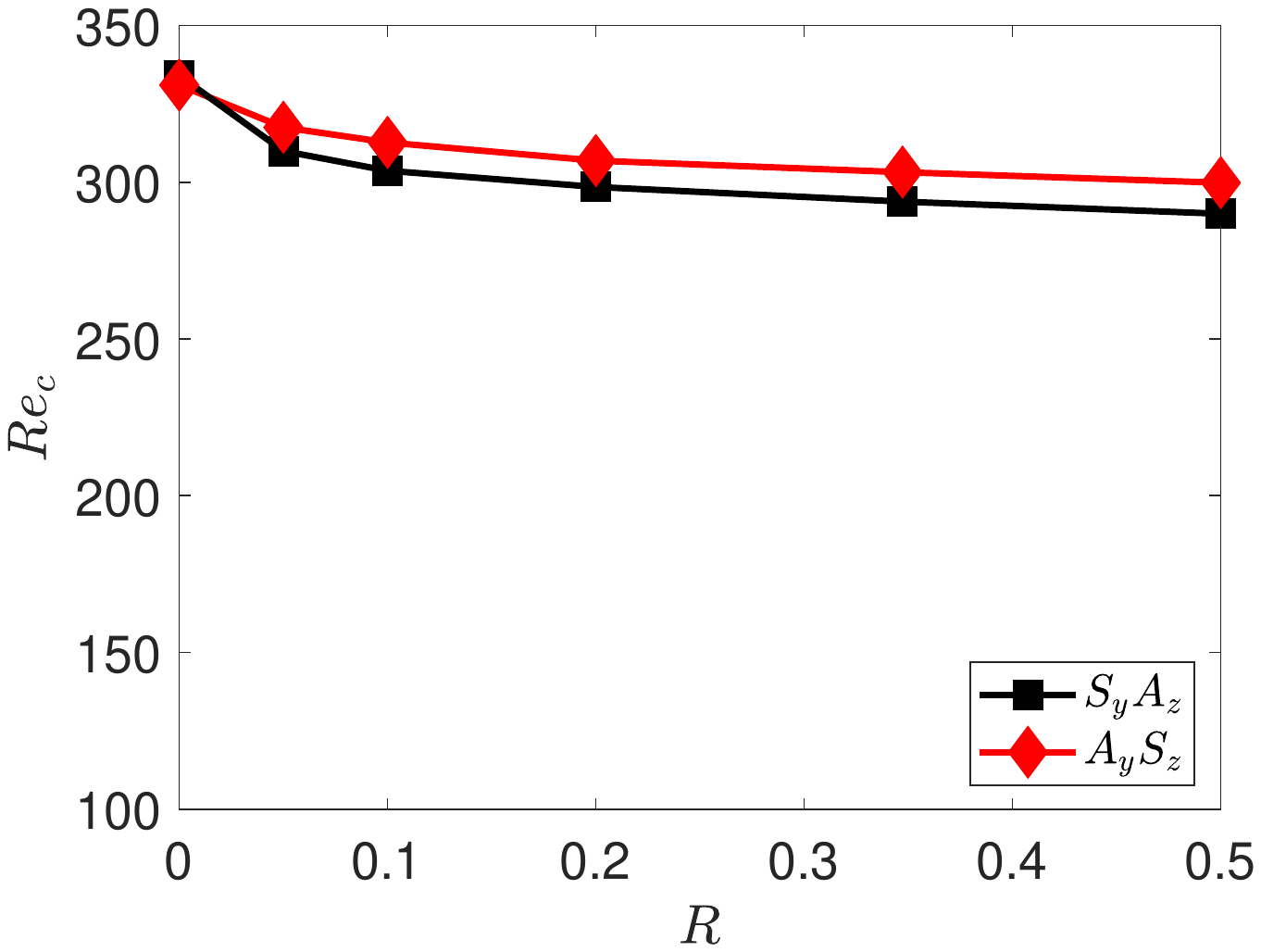}
      \put(-2,74){$(b)$}
  \end{overpic}
}
\caption{
Critical Reynolds number $Re_c$ of the first stationary bifurcations in the wake flow past rectangular prisms of length $(a)$~$L=1$ and $(b)$~$L=3$, and various fillet radius $R$.
Filled symbols: stationary (pitchfork) bifurcations; open symbols: oscillatory (Hopf) bifurcations.
%\tcr{To be updated.}
}
\label{fig:Rec_L_1_3}
\end{figure}

We now investigate the effect of the fillet radius $R$ on the linear stability of wake flows past bodies of length $L=3$, similar to our reference Ahmed body ($L=3$, $R=0.3472$). 
The corresponding steady base flows were described in section~\ref{sec:BF-R}.
Figure~\ref{fig:Rec_L_1_3}$(a)$ shows that for the shorter body, $L=1$, rounding the front edges  makes the first four bifurcating modes more stable until $R \simeq 0.1-0.2$, before making them more unstable. 
Figure~\ref{fig:Rec_L_1_3}$(b)$ shows that for $L=3$, however, rounding makes the stationary modes more unstable. 
These trends of $Re_c$  are correlated with those of the backflow (Fig.~\ref{fig:recirc_length_L1_L3_Re250}), but
not to those of the drag coefficient, which always decreases with $R$ (Fig.~\ref{fig:drag_L1_L3_Re250}).
They are also consistent with those reported in \citet{chiarini_linear_2021} for the first oscillatory mode past 2D rectangular cylinders.

%-----------------------------------------------------
%-----------------------------------------------------
%-----------------------------------------------------
\section{Weakly non-linear stability analysis \label{sec:WNL}}

For  the geometries investigated in section~\ref{sec:LSA}, two stationary modes, with $S_y A_z$ and $A_y S_z$ symmetries, are the first to become unstable, and their critical Reynolds numbers are close to one another. 
In the following, we refer to these modes as modes $A$ and mode $B$, respectively.
%(labelled $\hat{\bm u}_1^A$, or mode $A$) 
Linear stability analysis cannot predict the state of the flow once the two modes are unstable. 
Therefore, in this section we perform a weakly non-linear stability analysis in order to clarify the non-linear interactions between modes $A$ and $B$, and the non-linear states expected to be observed in an experiment or a DNS close to the bifurcation thresholds. 
Rigorous WNL analyses based on global eigenmodes have been applied to the first (Hopf) bifurcation of the 2D circular cylinder wake by \citet{sipp_global_2007} and to the first two bifurcations (pitchfork and Hopf) of the axisymmetric disk wake by \cite{meliga_global_2009}.

Motivated by the strong interest in wake deflection in simplified car models, as already mentioned in section~\ref{sec:intro},
we focus on geometries representative of Ahmed bodies. 
We first consider the reference Ahmed body $W=1.2$, $L=3$,  $R=0.3472$, and present detailed WNL results in section~\ref{sec:WNL-bifurc_diag}, complemented with DNS results in section~\ref{sec:WNL-DNS}.
We will then show in Sec.~\ref{sec:WNL-effect_W_L} that the WNL bifurcation sequence obtained for this reference geometry is robust to variations in $L$ and $W$ for usual Ahmed body geometries.

%-----------------------------------------------------
\subsection{Derivation of the amplitude equations \label{sec:WNL-derivation}}

\begin{figure}
  \centerline{  
    \begin{overpic}[width=8cm, trim=32mm 92mm 42mm 95mm, clip=true]{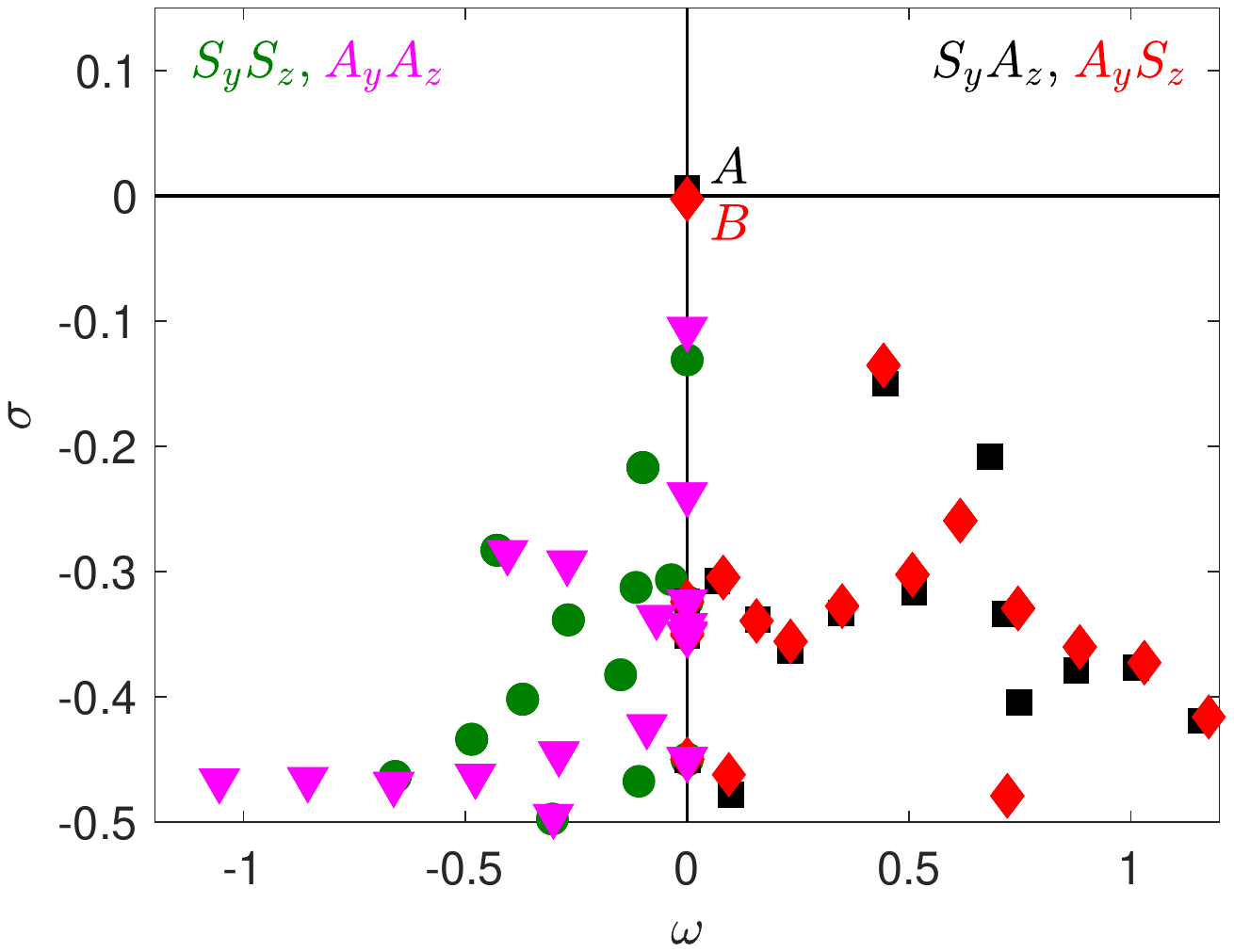}
      \put(-1,74){$(a)$}
 	\end{overpic}
 	%%%\hspace{0.2cm}
    \begin{overpic}[height=6.4cm, trim=25mm 88mm 51mm 97mm, clip=true]{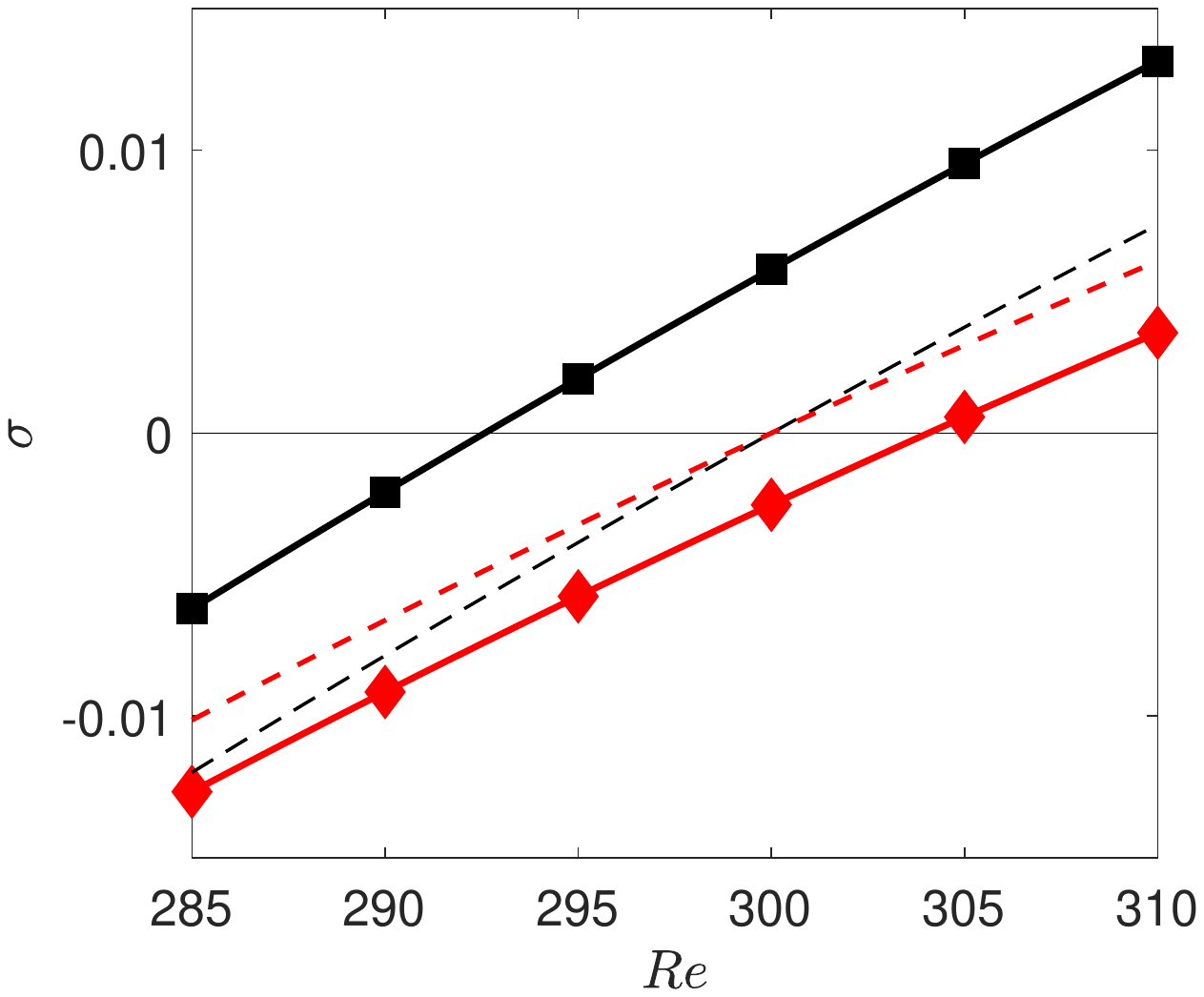}
      \put(1,77){$(b)$}
      %
%      \put(42,49){$A$}
%      \put(72,37){$\tcr{B}$}
%
      \put(53,58){$A$}
      \put(57,32){$\tcr{B}$}
 	\end{overpic}
  }
\caption{
$(a)$~Eigenvalue spectrum for the reference Ahmed body ($W=1.2$, $L=3$, $R=0.3472$) at $Re=300$, just between the  first and second pitchfork bifurcations ($Re_c^A = 293$, $Re_c^B = 304$).
$(b)$~Symbols: growth rates of modes $A$ and $B$, the two leading stationary modes.
Dashed lines: shifted growth rates $\sigma_A-\sigma_A(Re_c)$ and $\sigma_B-\sigma_B(Re_c)$ for $Re_c=300$ (see the WNL analysis in section~\ref{sec:WNL}).
}
\label{fig:growthrate}
\end{figure}

\begin{figure}
\centerline{   
    \begin{overpic}[width=8cm, trim=32mm 105mm 39mm 118mm, clip=true]{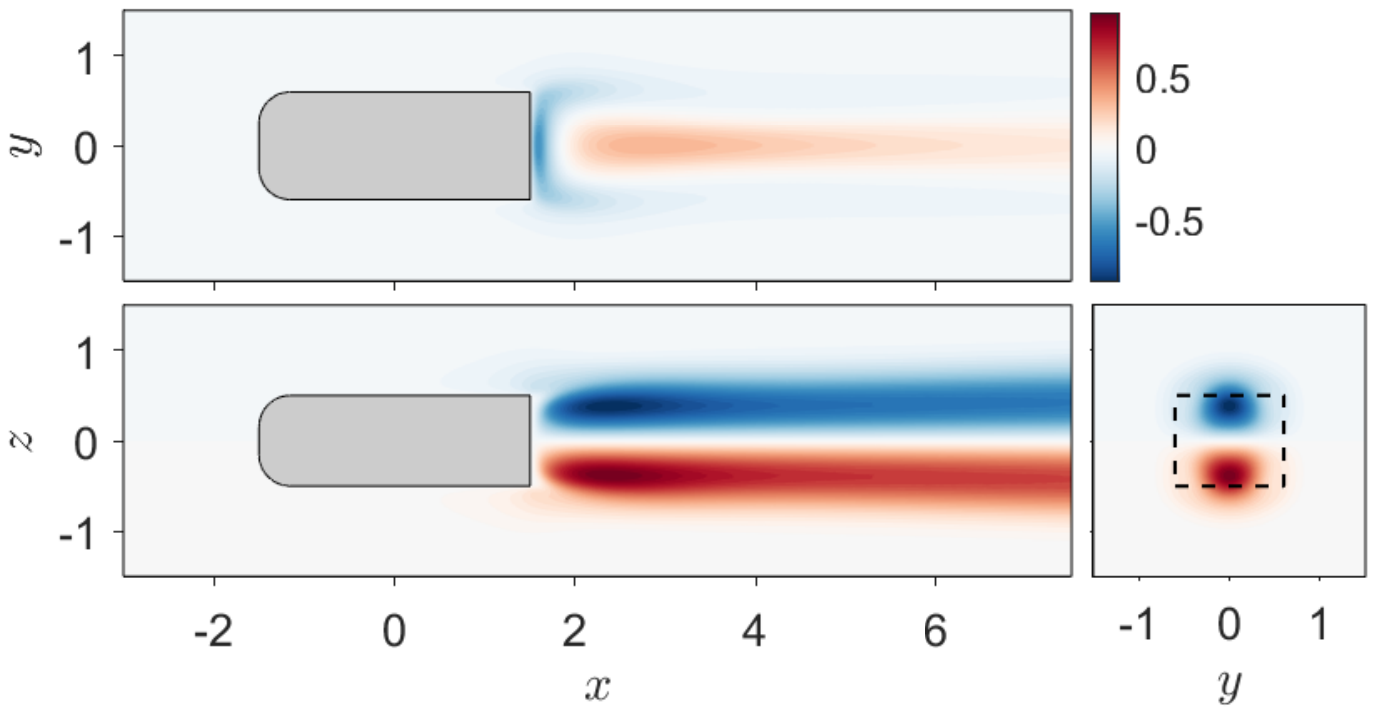}
      \put(-3,48){$(a)$}   
      \put(10,46)  {\footnotesize $\hat{{w}}_1^A$}
      \put(10,24.5){\footnotesize $\hat{{u}}_1^A$} 
      \put(81,24.5){\footnotesize $\hat{{u}}_1^A$}        
 	\end{overpic}
    \begin{overpic}[width=8cm, trim=32mm 105mm 39mm 118mm, clip=true]{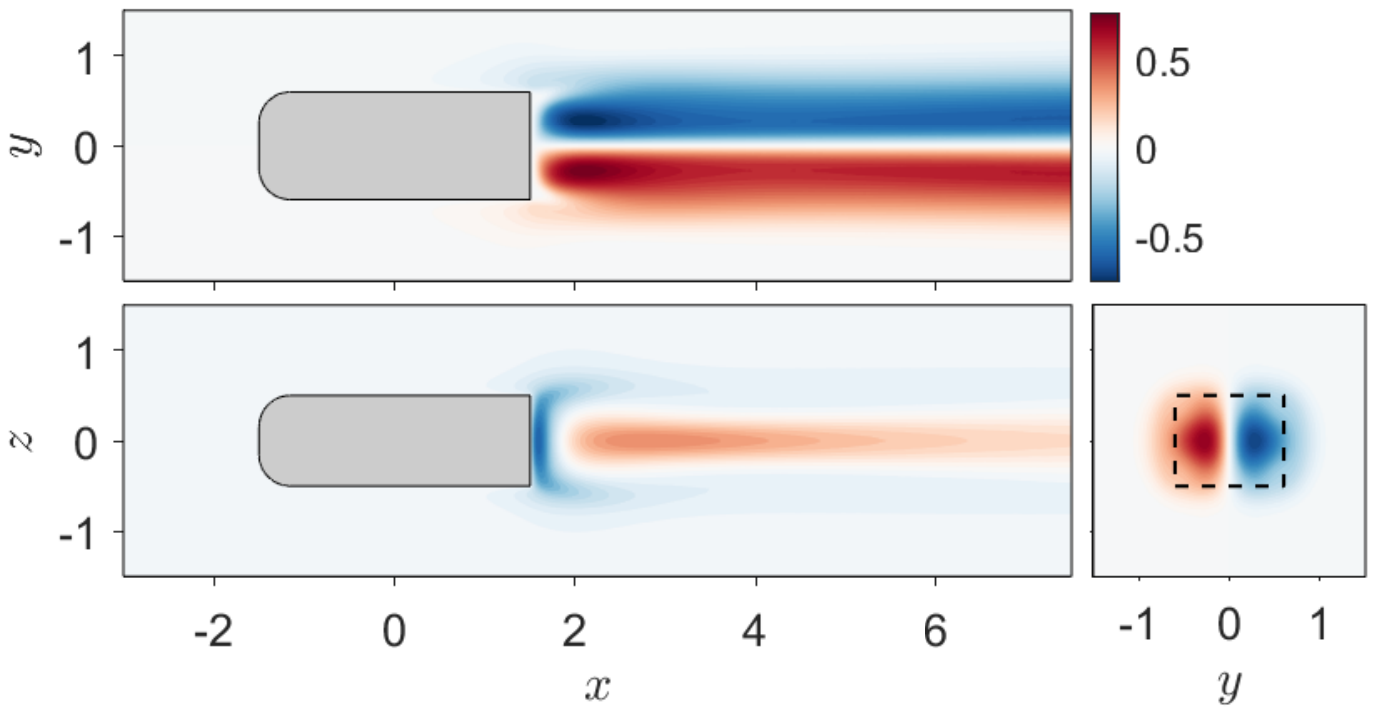}
      \put(-3,48){$(b)$}    
      \put(10,46)  {\footnotesize $\hat{{u}}_1^B$}
      \put(10,24.5){\footnotesize $\hat{{v}}_1^B$} 
      \put(81,24.5){\footnotesize $\hat{{u}}_1^B$}         
 	\end{overpic}
}
\centerline{   
    \begin{overpic}[width=8cm, trim=32mm 105mm 39mm 118mm, clip=true]{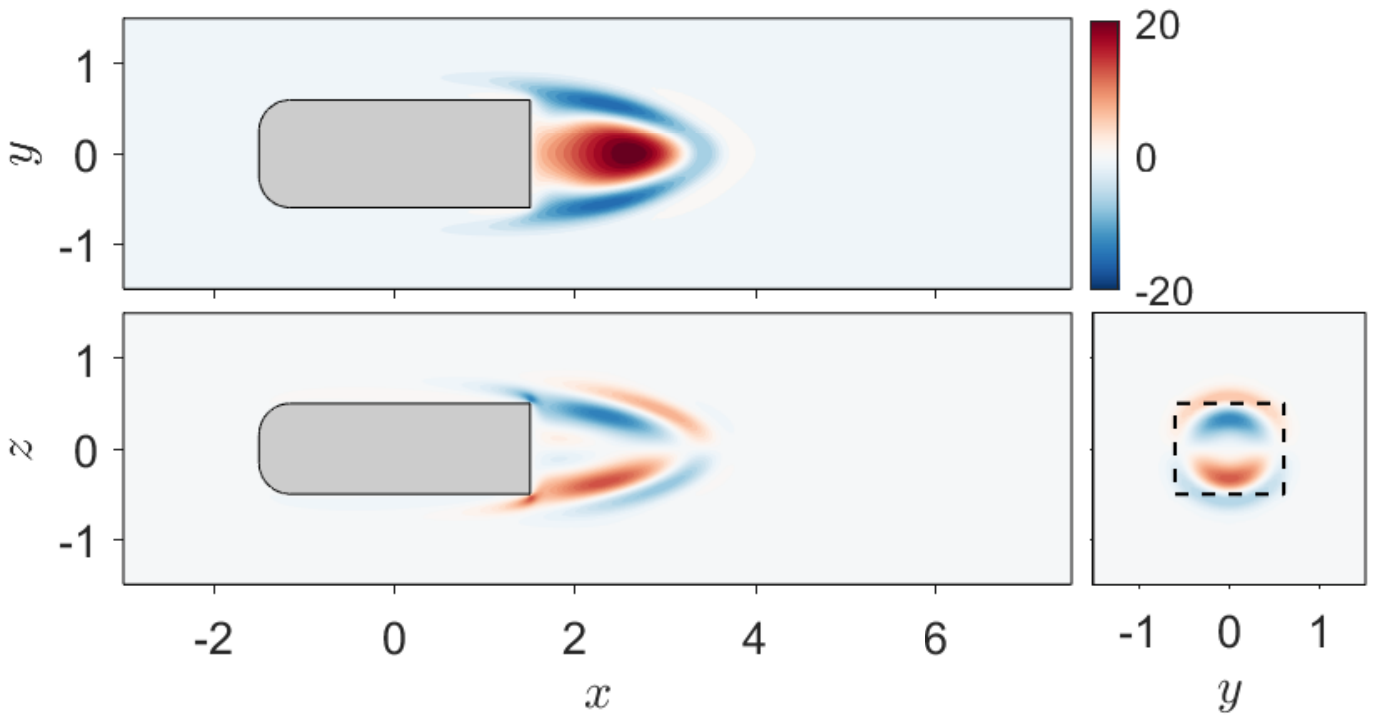}
      \put(-3,48){$(c)$}  
      \put(10,46)  {\footnotesize $\hat{{w}}_1^{A\dag}$}
      \put(10,24.5){\footnotesize $\hat{{u}}_1^{A\dag}$} 
      \put(81,24.5){\footnotesize $\hat{{u}}_1^{A\dag}$}     
 	\end{overpic}
    \begin{overpic}[width=8cm, trim=32mm 105mm 39mm 118mm, clip=true]{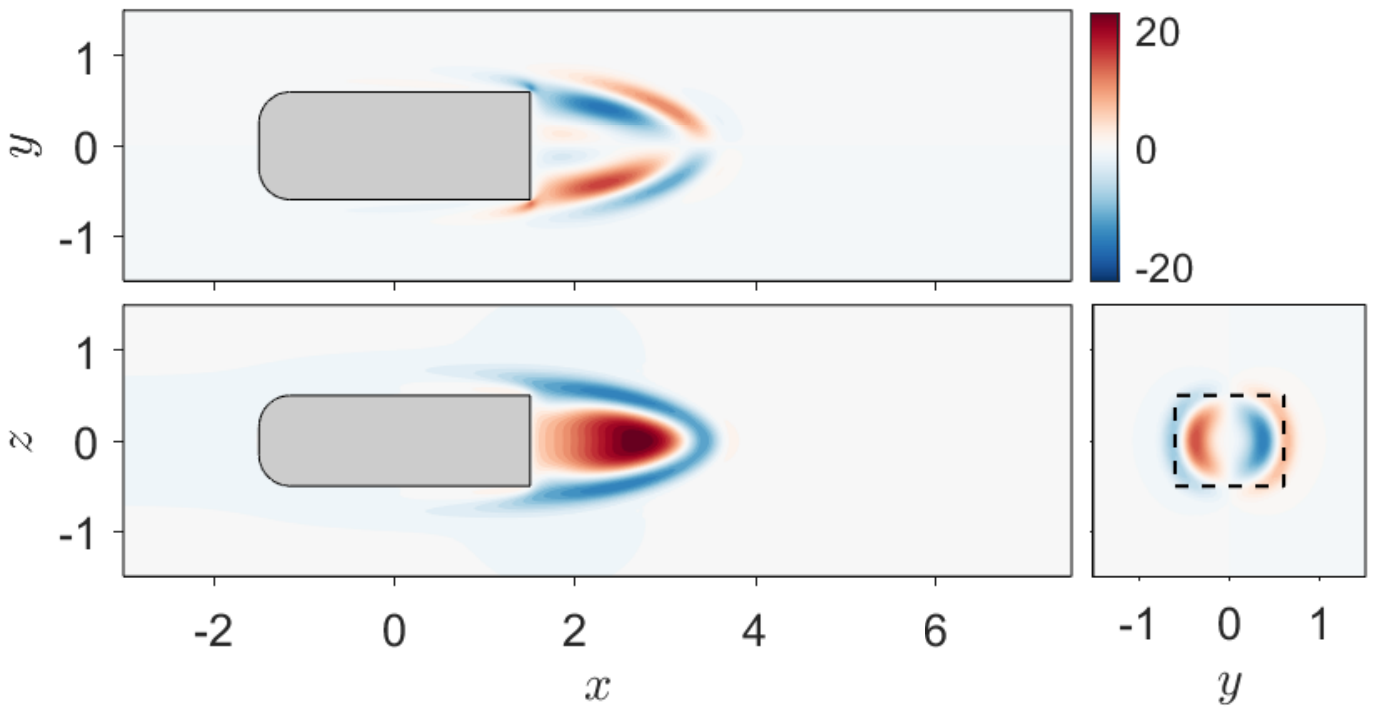}
      \put(-3,48){$(d)$}   
      \put(10,46)  {\footnotesize $\hat{{u}}_1^{B\dag}$}
      \put(10,24.5){\footnotesize $\hat{{v}}_1^{B\dag}$} 
      \put(81,24.5){\footnotesize $\hat{{u}}_1^{B\dag}$}          
 	\end{overpic}
}
\caption{
Eigenmodes and adjoint modes for the reference Ahmed body ($W=1.2$, $L=3$, $R=0.3472$) at $Re=300$, in the planes $z=0$ (top view), $y=0$ (side view) and $x=2.5$ (rear view):
$(a)$~$\hat{\bm{u}}_1^A$,
$(b)$~$\hat{\bm{u}}_1^B$,
$(c)$~$\hat{\bm{u}}_1^{A\dag}$,
$(d)$~$\hat{\bm{u}}_1^{B\dag}$.
}
\label{fig:WNL_1st_order_fields}
\end{figure}

For the reference Ahmed body, figure~\ref{fig:growthrate} shows the eigenspectrum at $Re=300$, as well as the evolution of the growth rates of modes $A$ and $B$ for Reynolds numbers in the vicinity of $Re=300$. 
Mode $A$ becomes unstable slightly before mode $B$: $Re_c^A = 293$ and $Re_c^B = 304$.
These two modes are depicted in figure~\ref{fig:WNL_1st_order_fields}$(a)-(b)$, together with the associated adjoint modes in figure~\ref{fig:WNL_1st_order_fields}$(c)-(d)$.
As already observed, mode $A$ breaks the top-down symmetry while mode $B$ breaks the left-right symmetry. 
Velocity perturbations extend far downstream and, as expected for stationary modes, do not show the wavepacket structures typical of oscillatory modes.
Adjoint perturbations have the same symmetries and are localised in the recirculation region.
We note that although eigenmodes are generally complex-valued,
stationary modes like modes $A$ and $B$ are associated with real eigenvalues ($\lambda=\sigma$) and therefore can always be defined as purely real-valued.

We assume that the Reynolds number is close to the bifurcation thresholds of modes $A$ and $B$. 
We introduce a reference critical Reynolds $Re_c$ that we choose typically in $[Re_c^A, Re_c^B]$, for example $Re_c=300$ for the reference Ahmed body.
%we define for simplicity as
%\begin{align}
%Re_c = \frac{Re_{c}^A+Re_{c}^B}{2}.
%\end{align}
In practice, the specific choice of $Re_c$ has little effect, as shown in Appendix B.
We quantify the departure from criticality with
\begin{align}
\frac{1}{Re_c} - \frac{1}{Re} = \alpha = \epsilon^2 \tilde\alpha,
%\qquad \mbox{with } \epsilon\ll 1 \mbox{ and } \tilde\alpha=O(1).
\end{align}
where $0 <\epsilon\ll 1 $, and where $\tilde\alpha$ is a parameter of order one (negative for $Re<Re_c$ and  positive for $Re>Re_c$).
%By definition, $\alpha(Re)$ is an increasing function, with $\alpha<0$ for $Re<Re_c$ and  $\alpha>0$ for $Re>Re_c$. 
At $Re_c$, the growth rates $\sigma_A$ and $\sigma_B$ are small but non-zero: mode $A$ is slightly unstable and mode $B$ slightly stable (figure~\ref{fig:growthrate}). 
Following \citet{meliga_global_2009}, we introduce rescaled order-one growth rates,
\begin{align}
\tilde\sigma_A = \frac{\sigma_A}{\epsilon^2},
\quad 
\tilde\sigma_B = \frac{\sigma_B}{\epsilon^2},
\end{align}
together with a shift operator $\mathcal S$ such that
% for $Re = Re_c$,
\begin{align}
\mathcal S \hat{\bm q}_1^A = \tilde \sigma_A(Re_c) \mathcal B \hat{\bm q}_1^A,
\quad
\mathcal S \hat{\bm q}_1^B = \tilde \sigma_B(Re_c) \mathcal B \hat{\bm q}_1^B,
\end{align}
and $\mathcal S \hat{\bm q}_1 = \bm 0$ for all other modes.
The interest of the shift operator is that although the linearised NS operator is not singular at $Re=Re_c$ (since $ \mathcal A \hat{\bm q}_1^A = -\sigma_A \mathcal B \hat{\bm q}_1^A \neq \bm 0$ and  $ \mathcal A \hat{\bm q}_1^B = -\sigma_B \mathcal B \hat{\bm q}_1^B \neq \bm 0$),
the shifted linearised NS operator defined by
\begin{align}
\tilde{\mathcal A} = \mathcal{A}  +  \epsilon^2 \mathcal S
\label{eq:WNL_shifted_LNS}
\end{align}
is exactly singular at $Re=Re_c$ for both modes $A$ and $B$, i.e.
\begin{align}
 \tilde{\mathcal A} \hat{\bm q}_1^A =\tilde{\mathcal A} \hat{\bm q}_1^B = \bm 0.
\label{eq:Atilde_singular}
\end{align}
In other words, modes $A$ and $B$ are exactly neutral for $\tilde{\mathcal A}$ at $Re=Re_c$, as shown by the shifted growth rates $\sigma_A-\sigma_A(Re_c)$ and $\sigma_B-\sigma_B(Re_c)$ in figure~\ref{fig:growthrate}.

%
%$$$$
%%
%$$ \Rightarrow \quad 
%\tilde{\bm{A}} \bm{u}_1^A =
%\bm A(Re_c) \bm{u}_1^A \tcr{\, + \,} \epsilon^2 \bm S \bm{u}_1^A
%= 
%\tcr{\, - \,} \sigma_A(Re_c) \bm{u}_1^A \tcr{\, + \,} \epsilon^2 \tilde \sigma_A(Re_c) \bm{u}_1^A = \bm 0, 
%$$
%%
%$$ \Rightarrow \quad 
%\tilde{\bm{A}} \bm{u}_1^B =
%\bm A(Re_c) \bm{u}_1^B \tcr{\, + \,} \epsilon^2 \bm S \bm{u}_1^B
%= 
%\tcr{\, - \,} \sigma_B(Re_c) \bm{u}_1^B \tcr{\, + \,} \epsilon^2 \tilde \sigma_B(Re_c) \bm{u}_1^B = \bm 0, $$
%i.e. $\bm{A}(Re_c)$ is not neutral for modes $A$, $B$, but $\tilde{\bm{A}}$ is.
%

We perform our WNL analysis with the method of multiple scales, therefore we  introduce the slow time scale $T = \epsilon^2 t$.
The flow field expansion
%\begin{align} 
%\bm q(\bm x,t) = \bm q_0(\bm x) + \epsilon \bm q_1(\bm x,t) + \epsilon^2 \bm q_2(\bm x,t) + \epsilon^3 \bm q_3(\bm x,t) + \ldots
%\label{eq:WNL_expansion}
%\end{align}
\begin{align} 
\bm q(\bm x,t,T) = \bm q_0 + \epsilon \bm q_1+ \epsilon^2 \bm q_2 + \epsilon^3 \bm q_3 + \ldots
\label{eq:WNL_expansion}
\end{align}
is injected in the NS equations (\ref{eq:LNS}) at $Re=Re_c$,
where $\partial_t$ is now transformed into $\partial_t + \epsilon^2 \partial_T$.
Collecting like-order terms in $\epsilon$ and making use of (\ref{eq:WNL_shifted_LNS}) yields a series of problems detailed hereafter.

%-----------------------------------------------------
\subsubsection{Orders $\epsilon^0$ and $\epsilon^1$}
% steady base flow}

The equations at order $\epsilon^0$ are the non-linear NS equations~(\ref{eq:NS}), and the zeroth-order field $\bm q_0(\bm x)$ is the steady base flow computed in section~\ref{sec:BF} at the reference Reynolds number $Re_c$.

The equations at order $\epsilon^1$ are the linearised NS equations~(\ref{eq:LNS_compact}) at $Re=Re_c$, now with the shifted linearised NS operator:
\begin{align}
\mathcal B \partial_t {\bm q}_1 + \tilde{\mathcal A} {\bm q}_1 = \bm{0}.
\end{align}
We assume that the first-order field $\bm q_1$ is a superposition of the two stationary modes $A$ and $B$ computed in section~\ref{sec:LSA} and shown in figure~\ref{fig:WNL_1st_order_fields},
\begin{align}
{\bm q}_1(\bm x,T) = A(T) \hat{\bm q}_1^A + B(T) \hat{\bm q}_1^B,
\label{eq:q1}
\end{align}
where $A$ and $B$ are two real-valued slowly-varying amplitudes, yet to be determined.

%-----------------------------------------------------
\subsubsection{Order $\epsilon^2$}

\begin{figure}
\centerline{   
    \begin{overpic}[width=8cm, trim=32mm 105mm 39mm 118mm, clip=true]{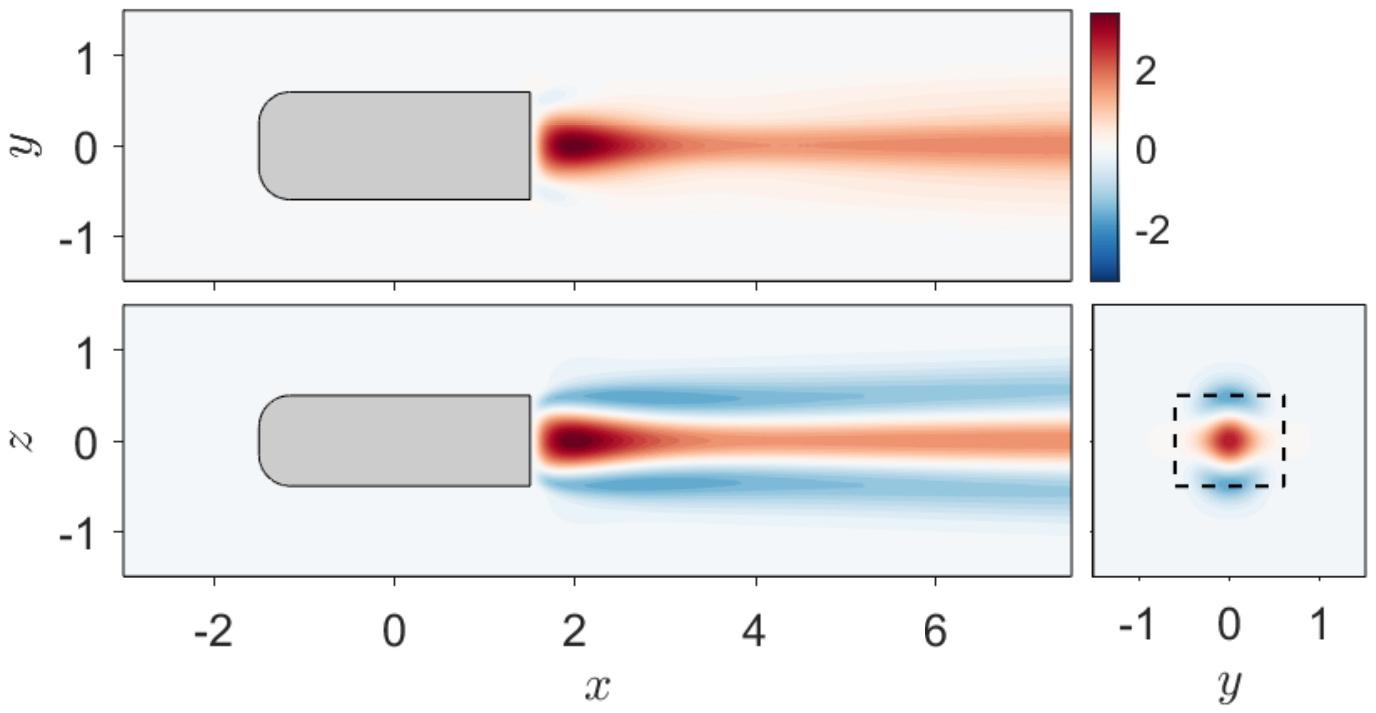}
      \put(-3,48){$(a)$}   
      \put(10,46)  {\footnotesize $u_2^{A^2}$}
      \put(10,24.5){\footnotesize $u_2^{A^2}$} 
      \put(80.5,24.5){\footnotesize $u_2^{A^2}$}             
 	\end{overpic}
    \begin{overpic}[width=8cm, trim=32mm 105mm 39mm 118mm, clip=true]{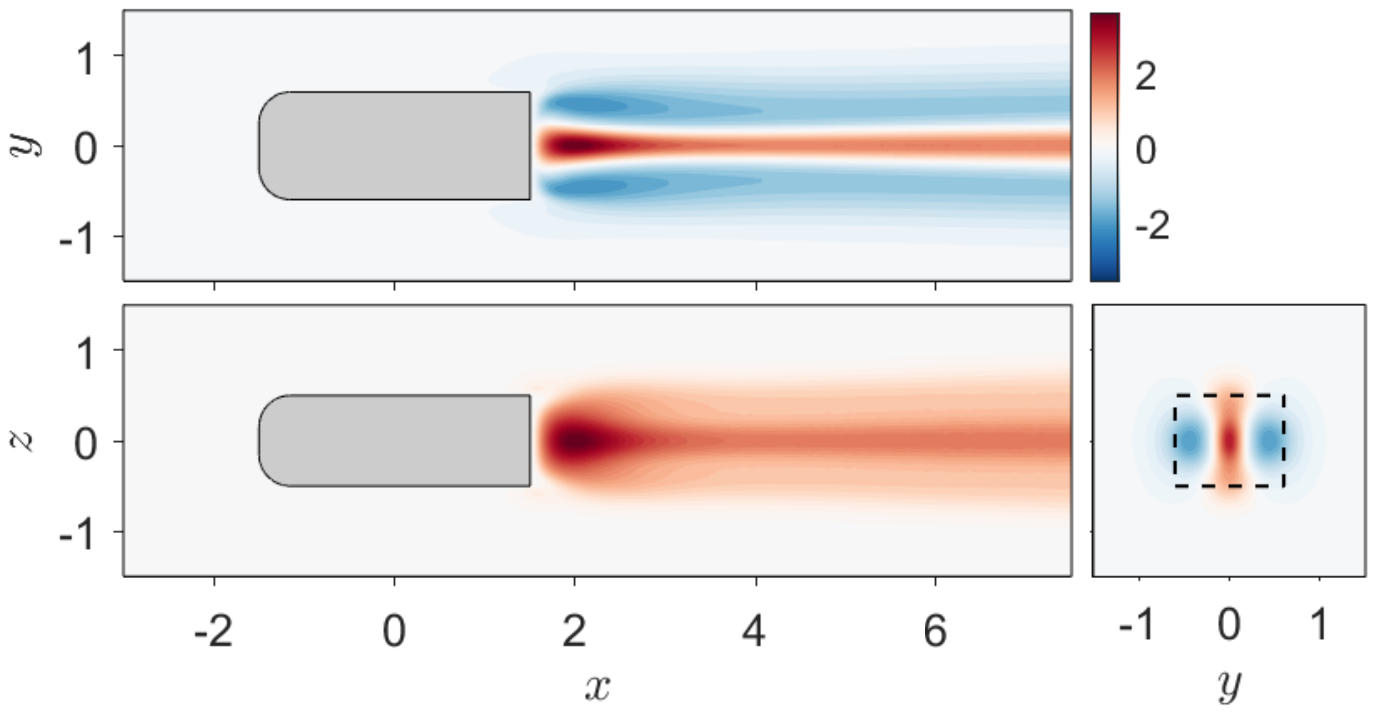}
      \put(-3,48){$(b)$}    
      \put(10,46)  {\footnotesize $u_2^{B^2}$}
      \put(10,24.5){\footnotesize $u_2^{B^2}$} 
      \put(80.5,24.5){\footnotesize $u_2^{B^2}$}             
 	\end{overpic}
}
\centerline{   
    \begin{overpic}[width=8cm, trim=32mm 105mm 39mm 118mm, clip=true]{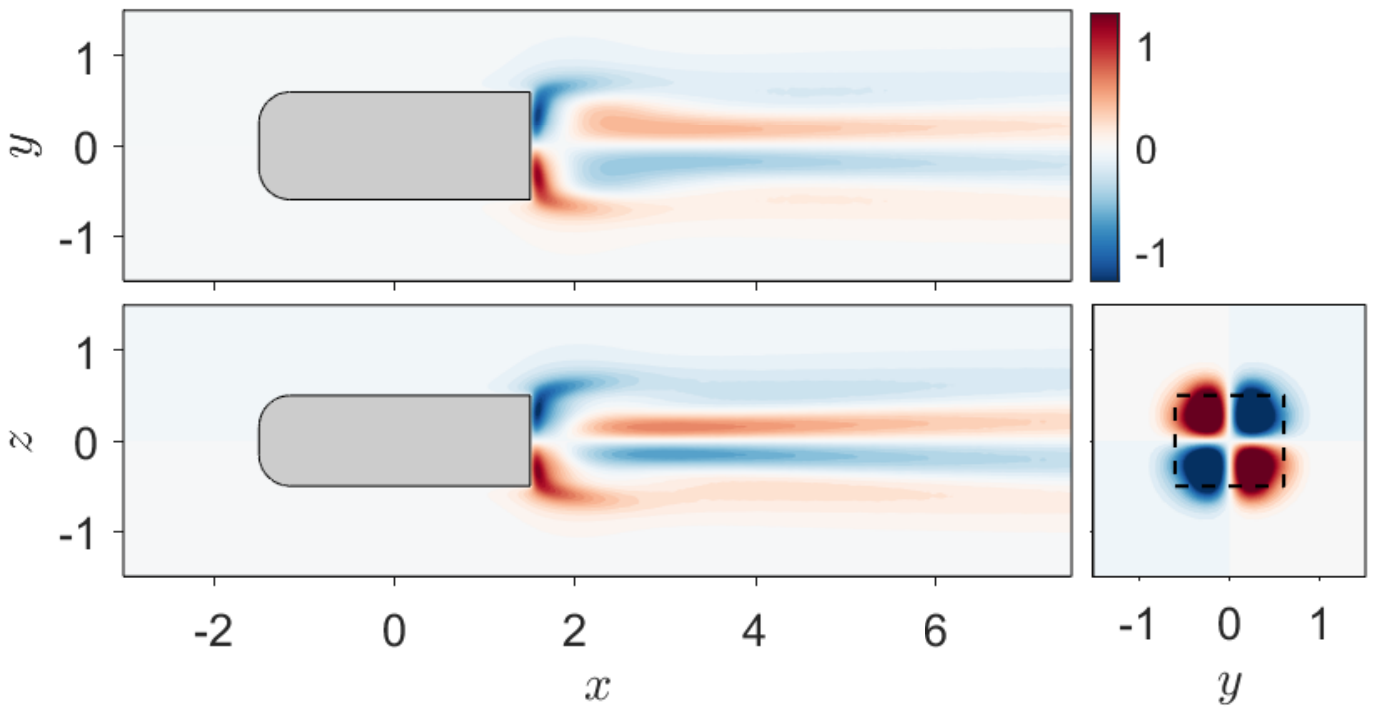}
      \put(-3,48){$(c)$}    
      \put(10,46)  {\footnotesize $w_2^{AB}$}
      \put(10,24.5){\footnotesize $v_2^{AB}$} 
      \put(80.5,24.5){\footnotesize $u_2^{AB}$}          
 	\end{overpic}
    \begin{overpic}[width=8cm, trim=32mm 105mm 39mm 118mm, clip=true]{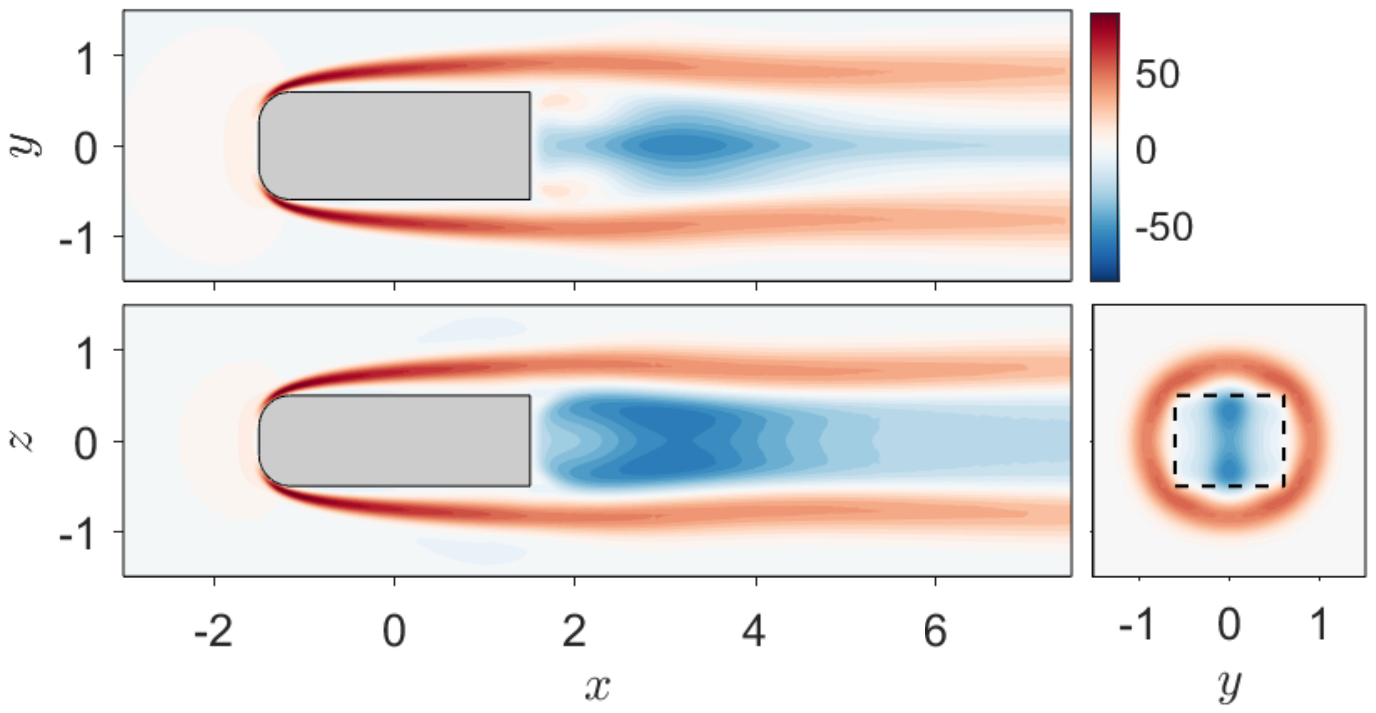}
      \put(-3,48){$(d)$}    
      \put(10,47)  {\footnotesize $u_2^\alpha$}
      \put(10,25.5){\footnotesize $u_2^\alpha$} 
      \put(80.5,25.5){\footnotesize $u_2^\alpha$}          
 	\end{overpic}
}
\caption{
Second-order fields, shown in the planes $z=0$ (top view), $y=0$ (side view) and $x=2.5$ (rear view):
$(a)$~$\bm u_2^{A^2}$,
$(b)$~$\bm u_2^{B^2}$,
$(c)$~$\bm u_2^{AB}$,
$(d)$~$\bm u_2^\alpha$.
}
\label{fig:WNL_2nd_order_fields}
\end{figure}

At order $\epsilon^2$, the second-order field $\bm q_2$ is a solution of the linearised NS equations,
\begin{align}
\mathcal B \partial_t {\bm q}_2 + \tilde{\mathcal A} {\bm q}_2 = (\bm F_2,0)^T
\label{eq:epsilon2}
\end{align}
forced by a term that depends on lower-order fields only,
\begin{align}
\bm F_2 
= - \tilde\alpha \nabla^2 \bm{u}_0 
-(\bm{u}_1 \cdot \nabla) \bm{u}_1.
\label{eq:F2_1}
\end{align}
The first term in (\ref{eq:F2_1}) is due to Reynolds number variations (proportional to $\tilde\alpha$) in the base flow $\bm{u}_0$,
and the second term is due to the transport of $\bm{u}_1$ by itself.
With the expression (\ref{eq:q1}) of the first-order field, this forcing reads
%
%\begin{align}
%\bm F_2 = 
%- \tilde\alpha \nabla^2 \bm{u}_0
%-A^2 (\hat{\bm u}_1^A\cdot \nabla)\hat{\bm u}_1^A 
%   -B^2 (\hat{\bm u}_1^B \cdot \nabla) \hat{\bm u}_1^B 
%   -AB \left[ (\hat{\bm u}_1^A \cdot \nabla) \hat{\bm u}_1^B 
%           + (\hat{\bm u}_1^B \cdot \nabla) \hat{\bm u}_1^A \right].
%\label{eq:F2_2}
%\end{align}
%
\begin{align}
\bm F_2 = 
- \tilde\alpha \nabla^2 \bm{u}_0
-A^2 (\hat{\bm u}_1^A\cdot \nabla)\hat{\bm u}_1^A 
   -B^2 (\hat{\bm u}_1^B \cdot \nabla) \hat{\bm u}_1^B 
   -AB \, \mathcal{C} (\hat{\bm u}_1^A , \hat{\bm u}_1^B).
\label{eq:F2_2}
\end{align}
All terms in (\ref{eq:F2_2}) are potentially resonant since they are forcing at zero frequency an operator that is singular precisely at zero frequency, as expressed by~(\ref{eq:Atilde_singular}).
Considering the spatial symmetries of these terms, however, shows that none of them is resonant, in a manner reminiscent of the cross-junction studied by \citet{Bongarzone2021}. 
Indeed, $\nabla^2 \bm{u}_0$, $(\hat{\bm u}_1^A\cdot \nabla)\hat{\bm u}_1^A$ and $(\hat{\bm u}_1^B\cdot \nabla)\hat{\bm u}_1^B$ are $S_y S_z$-symmetric,
and $(\hat{\bm u}_1^A \cdot \nabla) \hat{\bm u}_1^B + (\hat{\bm u}_1^B \cdot \nabla) \hat{\bm u}_1^A$ is $A_y A_z$-symmetric,
whereas $\tilde{\mathcal A}$ is singular to the $S_y A_z$-symmetric mode $A$ and the $A_y S_z$-symmetric mode $B$.
It follows that the forced equation~(\ref{eq:epsilon2}) can be inverted.
We look for a second-order field of the form
\begin{align}
\bm{q}_2 (\bm x, T) = \tilde\alpha \hat{\bm{q}}_2^{\alpha} 
+ A^2 \hat{\bm{q}}_2^{A^2} 
+ B^2 \hat{\bm{q}}_2^{B^2} 
+ AB  \hat{\bm{q}}_2^{AB},
\label{eq:q2}
\end{align}
where each term is the response to the individual forcing terms in~(\ref{eq:F2_2}):
\begin{align}
\mathcal B \partial_t {\bm q}_2^\alpha + \tilde{\mathcal A} {\bm q}_2^\alpha 
&= - \tilde\alpha   (\nabla^2 \bm{u}_0, 0)^T,
\label{eq:epsilon2-detail_alpha}
\\
\mathcal B \partial_t {\bm q}_2^{A^2} + \tilde{\mathcal A} {\bm q}_2^{A^2} 
&= -A^2  ( (\hat{\bm u}_1^A\cdot \nabla)\hat{\bm u}_1^A, 0)^T,
\label{eq:epsilon2-detail_A2}
\\
\mathcal B \partial_t {\bm q}_2^{B^2} + \tilde{\mathcal A} {\bm q}_2^{B^2} 
&= -B^2  ( (\hat{\bm u}_1^B \cdot \nabla) \hat{\bm u}_1^B, 0)^T,
\label{eq:epsilon2-detail_B2}
\\
\mathcal B \partial_t {\bm q}_2^{AB} + \tilde{\mathcal A} {\bm q}_2^{AB} 
&= -AB  ( \, \mathcal{C} (\hat{\bm u}_1^A , \hat{\bm u}_1^B), 0)^T.
\label{eq:epsilon2-detail_AB}
\end{align}
For each of the above problems, boundary conditions are similar to those used for the eigenmodes. 
In particular, the symmetry  of the forcing terms imposes the symmetry  of the second-order fields, therefore we use on the symmetry planes the corresponding  boundary conditions (\ref{eq:BC_EVP_SS}) or (\ref{eq:BC_EVP_AA}).
Figure~\ref{fig:WNL_2nd_order_fields} shows the second-order fields.
As expected from the symmetries of the forcing terms, 
$\bm u_2^{A^2}$, $\bm u_2^{B^2}$ and $\bm u_2^\alpha$ are all doubly symmetric ($S_y S_z$), while $\bm u_2^{AB}$ is doubly antisymmetric ($A_y A_z$).

%-----------------------------------------------------
\subsubsection{Order $\epsilon^3$}

At order $\epsilon^3$, the third-order field $\bm q_3$ is a solution of the linearised NS equations,
\begin{align}
\mathcal B \partial_t {\bm q}_3 + \tilde{\mathcal A} {\bm q}_3 = (\bm F_3,0)^T
\label{eq:epsilon3}
\end{align}
forced by a term that depends on lower-order fields only,
%
%\begin{align}
%\bm F_3 
%= \pm\partial_T \bm{u}_1
%- \tilde\alpha \nabla^2 \bm{u}_1 
%-\left[ (\bm{u}_1 \cdot\nabla) \bm{u}_2 
%       +(\bm{u}_2 \cdot\nabla) \bm{u}_1
%\right]
%\pm \mathcal S \bm{u}_1.
%\label{eq:F3_1}
%\end{align}
%%
\begin{align}
\bm F_3 
= -\partial_T \bm{u}_1
- \tilde\alpha \nabla^2 \bm{u}_1 
-\mathcal{C}(\bm{u}_1 , \bm{u}_2)
+ \mathcal S \bm{u}_1.
\label{eq:F3_1}
\end{align}
The first term in (\ref{eq:F3_1}) is due to the slow variation of the first-order field $\bm{u}_1$, i.e. the yet unknown amplitudes $A(T)$ and $B(T)$.
The second term is due to Reynolds number variations in the first-order field $\bm{u}_1$.
The third term is due to the mutual transport of the first- and second-order fields $\bm{u}_1$ and $\bm{u}_2$.
The last term is due to the non-zero values of the growth rates $\sigma_A$ and $\sigma_B$ at $Re=Re_c$, accounted for by the action of the shift operator.
With the expressions (\ref{eq:q1}) and (\ref{eq:q2}) of the first- and second-order fields, the forcing becomes
%
%\begin{align}
%\bm F_3 
%=&  -\partial_T  A \, \hat{\bm u}_1^A -\partial_T B \,  \hat{\bm u}_1^B 
%%--------------------------
%- \tilde\alpha A \nabla^2 \hat{\bm u}_1^A 
%- \tilde\alpha B \nabla^2 \hat{\bm u}_1^B  
%\nonumber
%\\
%%--------------------------
%& -\mathcal{C}( A \hat{\bm u}_1^A + B \hat{\bm u}_1^B  , \tilde\alpha \hat{\bm{u}}_2^{\alpha} 
%+ A^2 \hat{\bm{u}}_2^{A^2} 
%+ B^2 \hat{\bm{u}}_2^{B^2} 
%+ AB  \hat{\bm{u}}_2^{AB} )
%%--------------------------
%+ A \sigma_A \hat{\bm u}_1^A 
%+ B \sigma_B \hat{\bm u}_1^B.
%\label{eq:F3_2}
%\end{align}
%
%\begin{align}
%\bm F_3 
%=&  -\partial_T  A \, \hat{\bm u}_1^A 
%- \tilde\alpha A \nabla^2 \hat{\bm u}_1^A 
%+ A \sigma_A \hat{\bm u}_1^A 
%-\partial_T B \,  \hat{\bm u}_1^B 
%- \tilde\alpha B \nabla^2 \hat{\bm u}_1^B  
%+ B \sigma_B \hat{\bm u}_1^B
%\nonumber
%\\
%%--------------------------
%& -\mathcal{C}( A \hat{\bm u}_1^A + B \hat{\bm u}_1^B  , \tilde\alpha \hat{\bm{u}}_2^{\alpha} 
%+ A^2 \hat{\bm{u}}_2^{A^2} 
%+ B^2 \hat{\bm{u}}_2^{B^2} 
%+ AB  \hat{\bm{u}}_2^{AB} )
%%--------------------------
%\label{eq:F3_2}
%\end{align}
%
\begin{align}
\bm F_3 
=&  \left( -\partial_T  A 
- A \tilde\alpha \nabla^2 
+ A \sigma_A  \right) \hat{\bm u}_1^A
+
 \left(-\partial_T B 
- B \tilde\alpha \nabla^2   
+ B \sigma_B \right) \hat{\bm u}_1^B 
\nonumber
\\
%--------------------------
& -\mathcal{C}( A \hat{\bm u}_1^A + B \hat{\bm u}_1^B  , \,
\tilde\alpha \hat{\bm{u}}_2^{\alpha} 
+ A^2 \hat{\bm{u}}_2^{A^2} 
+ B^2 \hat{\bm{u}}_2^{B^2} 
+ AB  \hat{\bm{u}}_2^{AB} ).
%--------------------------
\label{eq:F3_2}
\end{align}
It turns out that all terms in (\ref{eq:F3_2}) are either $S_y A_z$-symmetric or $A_y S_z$-symmetric, and are therefore resonant with mode $A$ or mode $B$. To avoid secular terms, i.e. to ensure that (\ref{eq:epsilon3}) can be inverted, a compatibility condition must be enforced. The Fredholm alternative states that the forcing $\bm{F}_3$ must be orthogonal to the kernel of the adjoint linearised NS operator. This leads to the following  equations for the amplitudes $A$ and $B$:
\begin{align}
\partial_t A &=  \lambda_A A 
- \chi_A A^3
- \eta_A B^2 A,
\label{eq:A}
\\
\partial_t B &=  \lambda_B B 
- \chi_B B^3
- \eta_B A^2 B,
\label{eq:B}
\end{align}
where the fast time $t=T/\epsilon^2$ has been reintroduced.
All the WNL coefficients in (\ref{eq:A})-(\ref{eq:B}), whose values reported in Appendix A, are computed as scalar products between the adjoint modes, $\hat{\bm u}_1^{A\dag}$, $\hat{\bm u}_1^{B\dag}$, and resonant forcing terms:
%
%\begin{align}
%\lambda_A \langle \bm{u}^{A\dag},\bm{u}_1^A \rangle 
%&
%= -\alpha \langle \bm{u}^{A\dag}, 
% \bm{u}_2^\alpha \cdot\nabla \bm{u}_1^A
%+ \bm{u}_1^A \cdot\nabla \bm{u}_2^\alpha 
%+ \nabla^2 \bm{u}_1^A \rangle
%\tcr{\, + \,} \sigma_A, 
%\label{eq:lambdaA}
%\\
%\lambda_B \langle \bm{u}^{B\dag},\bm{u}_1^B \rangle 
%&
%= -\alpha \langle \bm{u}^{B\dag}, 
% \bm{u}_2^\alpha \cdot\nabla \bm{u}_1^B
%+ \bm{u}_1^B \cdot\nabla \bm{u}_2^\alpha 
%+ \nabla^2 \bm{u}_1^B \rangle
%\tcr{\, + \,} \sigma_B, 
%\label{eq:lambdaB}
%\\
%\eta_A  \langle \bm{u}^{A\dag},\bm{u}_1^A \rangle 
%&
%= \epsilon^2  \langle \bm{u}^{A\dag}, 
%  \bm{u}_2^{B^2} \cdot\nabla \bm{u}_1^A
%+ \bm{u}_1^A \cdot\nabla \bm{u}_2^{B^2}
%+ \bm{u}_2^{AB} \cdot\nabla \bm{u}_1^B
%+ \bm{u}_1^B \cdot\nabla \bm{u}_2^{AB}
%\rangle, 
%\label{eq:etaA}
%\\
%\eta_B \langle \bm{u}^{B\dag},\bm{u}_1^B \rangle 
%&
%= \epsilon^2 \langle \bm{u}^{B\dag}, 
%  \bm{u}_2^{A^2} \cdot\nabla \bm{u}_1^B
%+ \bm{u}_1^B \cdot\nabla \bm{u}_2^{A^2}
%+ \bm{u}_2^{AB} \cdot\nabla \bm{u}_1^A
%+ \bm{u}_1^A \cdot\nabla \bm{u}_2^{AB}
%\rangle, 
%\label{eq:etaB}
%\\
%\chi_A  \langle \bm{u}^{A\dag},\bm{u}_1^A \rangle 
%&
%= \epsilon^2 \langle \bm{u}^{A\dag}, 
%  \bm{u}_2^{A^2} \cdot\nabla \bm{u}_1^A
%+ \bm{u}_1^A \cdot\nabla \bm{u}_2^{A^2}
%\rangle,
%\label{eq:chiA}
%\\
%\chi_B \langle \bm{u}^{B\dag},\bm{u}_1^B \rangle 
%&
%= \epsilon^2 \langle \bm{u}^{B\dag}, 
%  \bm{u}_2^{B^2} \cdot\nabla \bm{u}_1^B
%+ \bm{u}_1^B \cdot\nabla \bm{u}_2^{B^2}
%\rangle.
%\label{eq:chiB}
%\end{align}
%
%
\begin{align}
\lambda_A 
&
= \epsilon^2 \tilde\lambda_A
=  \sigma_A -\alpha  \langle \hat{\bm{u}}_1^{A\dag},\, 
\mathcal{C}( \hat{\bm{u}}_2^\alpha , \hat{\bm{u}}_1^A)
+ \nabla^2 \hat{\bm{u}}_1^A \rangle, 
\label{eq:lambdaA}
\\
\lambda_B 
&
= \epsilon^2 \tilde\lambda_B
=  \sigma_B -\alpha  \langle \hat{\bm{u}}_1^{B\dag},\, 
\mathcal{C}( \hat{\bm{u}}_2^\alpha , \hat{\bm{u}}_1^B)
+ \nabla^2 \hat{\bm{u}}_1^B \rangle, 
\label{eq:lambdaB}
\\
\eta_A   
&
= \epsilon^2 \tilde\eta_A
= \epsilon^2  \langle \hat{\bm{u}}_1^{A\dag},\, 
\mathcal{C}(  \hat{\bm{u}}_2^{B^2}, \hat{\bm{u}}_1^A)
+ \mathcal{C}( \hat{\bm{u}}_2^{AB} , \hat{\bm{u}}_1^B)
\rangle, 
\label{eq:etaA}
\\
\eta_B 
&
= \epsilon^2 \tilde\eta_B
= \epsilon^2  \langle \hat{\bm{u}}_1^{B\dag},\, 
\mathcal{C}(  \hat{\bm{u}}_2^{A^2} , \hat{\bm{u}}_1^B)
+ \mathcal{C}( \hat{\bm{u}}_2^{AB} , \hat{\bm{u}}_1^A)
\rangle, 
\label{eq:etaB}
\\
\chi_A   
&
= \epsilon^2 \tilde\chi_A
= \epsilon^2  \langle \hat{\bm{u}}_1^{A\dag},\, 
 \mathcal{C}( \hat{\bm{u}}_2^{A^2} , \hat{\bm{u}}_1^A
)
\rangle,
\label{eq:chiA}
\\
\chi_B  
&
= \epsilon^2 \tilde\chi_B
= \epsilon^2  \langle \hat{\bm{u}}_1^{B\dag},\, 
\mathcal{C}(  \hat{\bm{u}}_2^{B^2} , \hat{\bm{u}}_1^B)
\rangle,
\label{eq:chiB}
\end{align}
where we recall that the adjoint and direct modes are normalised such that $\langle \hat{\bm{u}}_1^{A\dag} ,\, \hat{\bm{u}}_1^A \rangle = \langle \hat{\bm{u}}_1^{B\dag} ,\, \hat{\bm{u}}_1^B \rangle = 1$.

For any given solution of the equation amplitudes  (\ref{eq:A})-(\ref{eq:B}) (see section \ref{sec:WNL-bifurc_diag}),
the total flow field up to second order can be reconstructed as
\begin{align}
\bm{q} = \bm q_0 + \epsilon \left(
 A \hat{\bm q}_1^A + B \hat{\bm q}_1^B
\right)
+ \epsilon^2 \left(
\tilde\alpha \hat{\bm{q}}_2^{\alpha} 
+ A^2 \hat{\bm{q}}_2^{A^2} 
+ B^2 \hat{\bm{q}}_2^{B^2} 
+ AB  \hat{\bm{q}}_2^{AB}
\right)
+ O(\epsilon^3).
\label{eq:WNL_reconstruct}
\end{align}
We note that the  values of the WNL coefficients, $\lambda_A, \ldots \chi_B$, and of the amplitudes, $A$, $B$,  depend on the  choices of $\epsilon$ and of the direct/adjoint modes normalisation,
but the final reconstructed field (\ref{eq:WNL_reconstruct}) itself is  independent of these specific choices.

%-----------------------------------------------------
\subsection{Bifurcation diagram \label{sec:WNL-bifurc_diag}}

The amplitude equations (\ref{eq:A})-(\ref{eq:B}) are coupled ordinary differential equations. 
They are partly similar to the classic Stuart-Landau equation describing the pitchfork/Hopf bifurcation of a single stationary/oscillating mode, but differ by the two non-linear coupling terms $\eta_A A B^2$ and $\eta_B B A^2 $.
Their structure is well known in the literature (see for instance \citet{Kuznetsov}).
There are in general four equilibrium solutions ($\partial_t A = \partial_t B = 0$):
\begin{itemize}
\item 
Symmetric state: $ A=B=0$;

\item 
Pure state A (top/down symmetry breaking): 
\begin{align} 
A^2 = \frac{\lambda_A}{\chi_A}, \quad B=0;
\end{align}

\item 
Pure state B (left/right symmetry breaking):
\begin{align} 
A=0, \quad B^2 = \frac{\lambda_B}{\chi_B};
\end{align} 

\item 
Mixed state (A,B) (double symmetry breaking):
\begin{align} 
A^2 =  \frac{\chi_B \lambda_A - \eta_A \lambda_B}{\chi_A \chi_B -\eta_A\eta_B},  \qquad 
B^2 = \frac{\chi_A \lambda_B - \eta_B \lambda_A}{\chi_A \chi_B -\eta_A\eta_B}.
\end{align} 
%
%
%i.e. when $A\neq0$, $B\neq 0$:
%\begin{align}
%\left[ 
%\begin{array}{cc}
%\chi_A  & \eta_A \\ 
%\eta_B & \chi_B
%\end{array} 
%\right]
%\left(
%\begin{array}{c}
%A^2 \\ B^2
%\end{array} 
%\right)
%=
%\left(
%\begin{array}{c}
%\lambda_A \\ \lambda_B
%\end{array} 
%\right).
%\end{align}
\end{itemize}
We assess the linear stability of each state in two ways: 
(i)~by computing the eigenvalues of the Jacobian $\mathcal{J}$ of the system linearised about the state of interest,
\begin{align}
\mathcal{J} = \left[ \begin{array}{cc}
\lambda_A - 3 \chi_A A^2 - \eta_A B^2  & 
- 2 \eta_A B A  \\ 
- 2 \eta_B A B
& \lambda_B - 3 \chi_B B^2 - \eta_B A^2 
\end{array} \right],
\label{eq:jacobian}
\end{align}
that state being stable if both  eigenvalues have a negative  real part;
(ii)~by solving in time the amplitude equations (\ref{eq:A})-(\ref{eq:B}) from an initial condition close to the state of interest, that state being stable if it is an attractor of the  final solution.

\begin{figure}
  \centerline{
    \begin{overpic}[width=8cm, trim=33mm 91mm 35mm 95mm, clip=true]{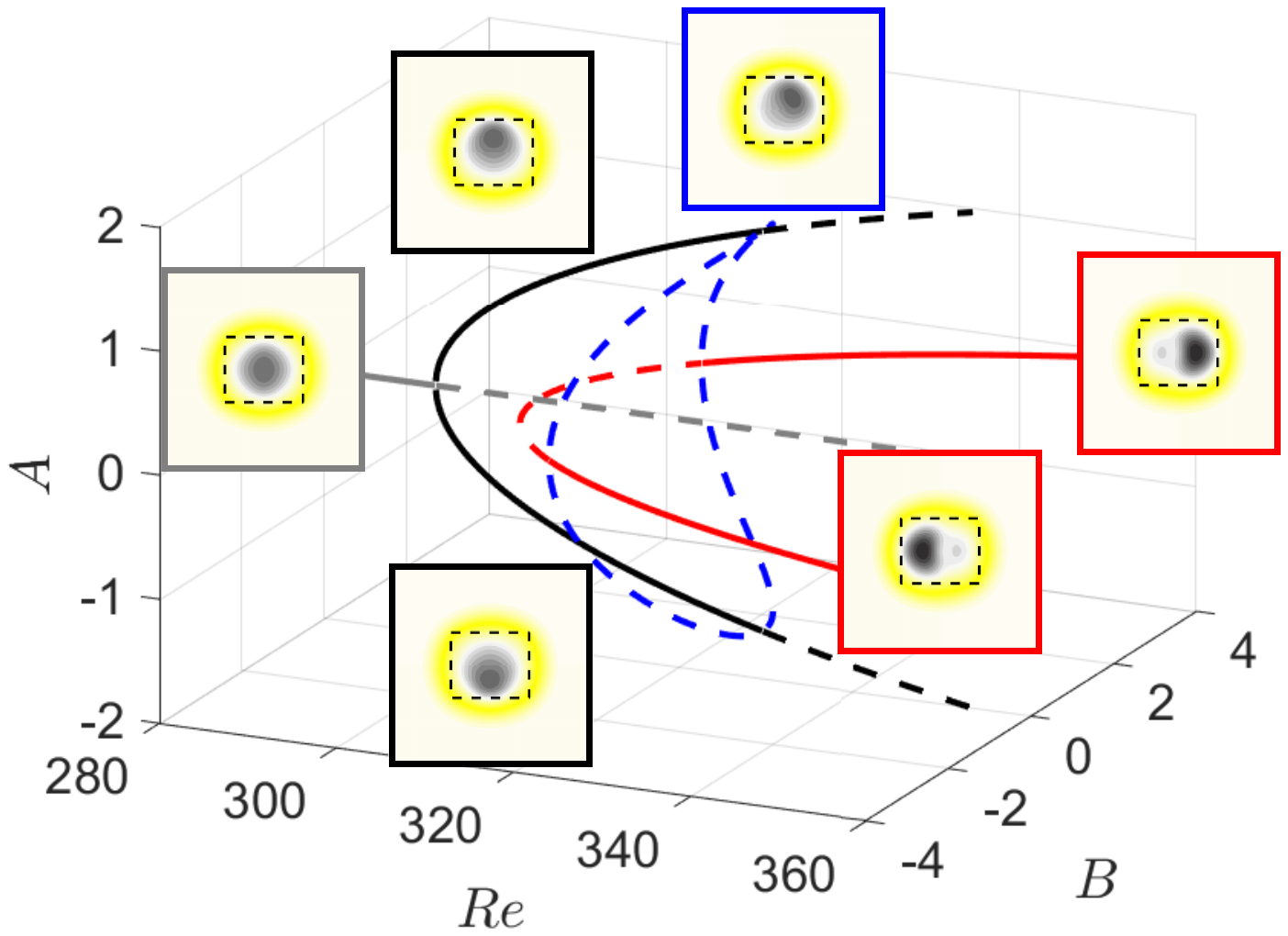}
      \put(0,74){$(a)$}
 	\end{overpic}    
 	\begin{overpic}[width=8cm, trim=35mm 92mm 40mm 90mm, clip=true]{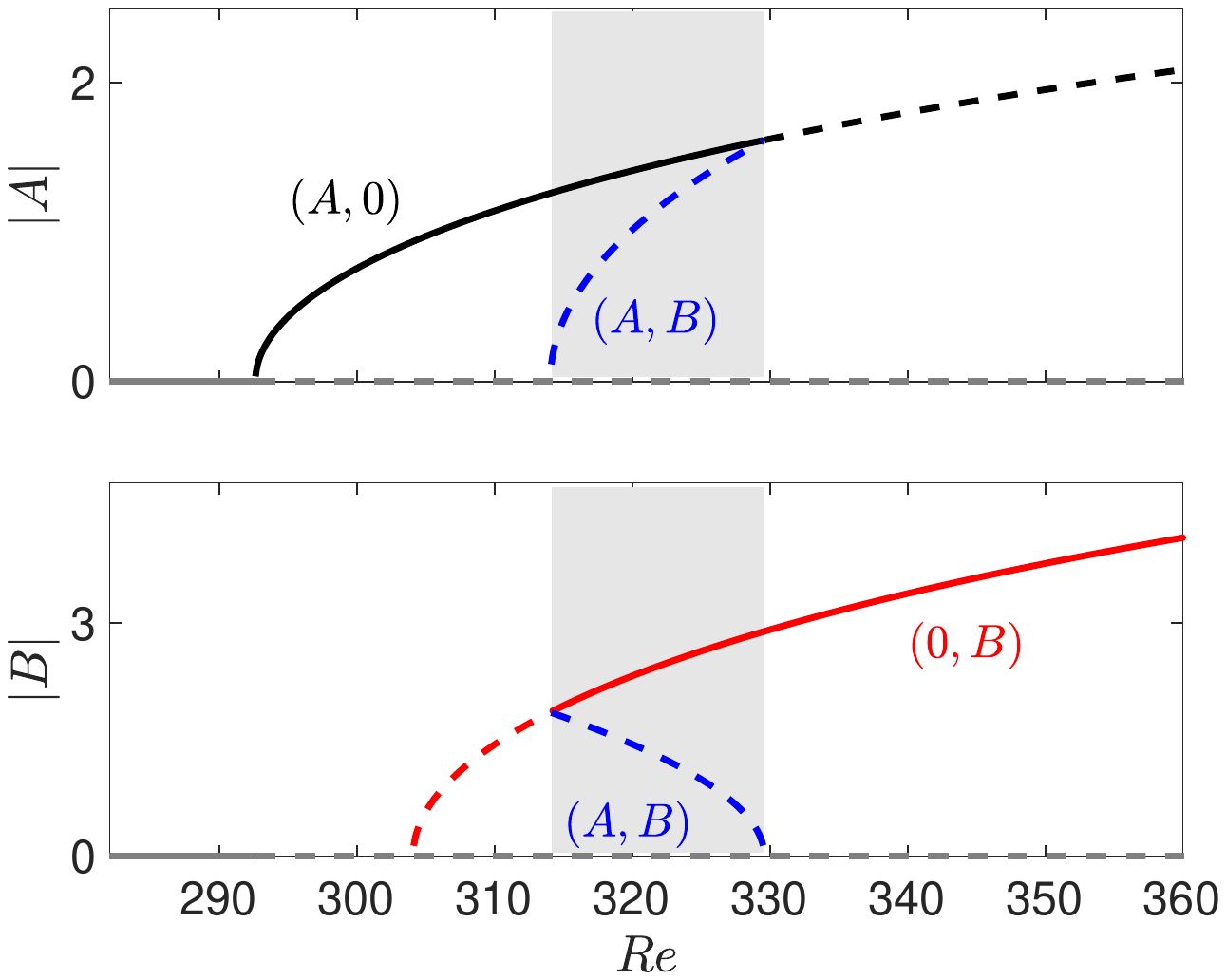}
      \put(0,74){$(b)$}
      \put(15,42.2){\scriptsize $\circled{1}$}
      \put(26,42.2){\scriptsize $\circled{2}$}
      \put(38,42.2){\scriptsize $\circled{3}$}
      \put(52,42.2){\scriptsize $\circled{4}$}
      \put(77,42.2){\scriptsize $\circled{5}$}
 	\end{overpic}
  }
\caption{
Bifurcation diagram associated with the amplitude equations (\ref{eq:A})-(\ref{eq:B}) obtained from a weakly non-linear analysis, for the reference Ahmed body ($W=1.2$, $L=3$, $R=0.3472$), using $Re_c=300$ and $\epsilon=0.1$. 
Stable and unstable states are shown with solid and dashed lines, respectively.
$(a)$~3D view in the $(A,B,Re)$ space. Insets show a rear view of the reconstructed field (see figure~\ref{fig:WNL_reconstruct}).
$(b)$~2D views in the  $(|A|,Re)$  and $(|B|,Re)$ spaces.
The shaded region corresponds to bistability and hysteresis.
}
\label{fig:bifurc_diagram}
\end{figure}

\begin{figure}
\centerline{ 
    \begin{overpic}[width=8cm, trim=32mm 105mm 39mm 118mm, clip=true]{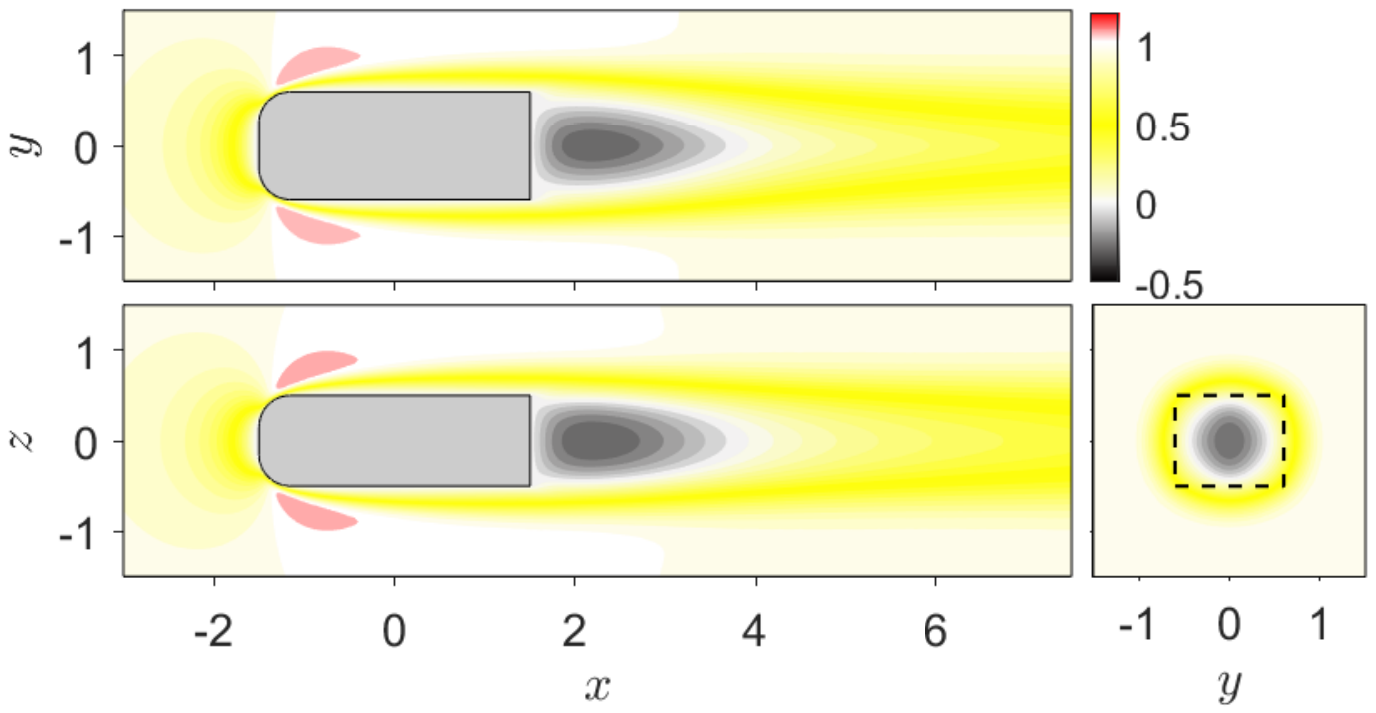}
      \put(-3,48){$(a)$}              
 	\end{overpic}
    \begin{overpic}[width=8cm, trim=32mm 105mm 39mm 118mm, clip=true]{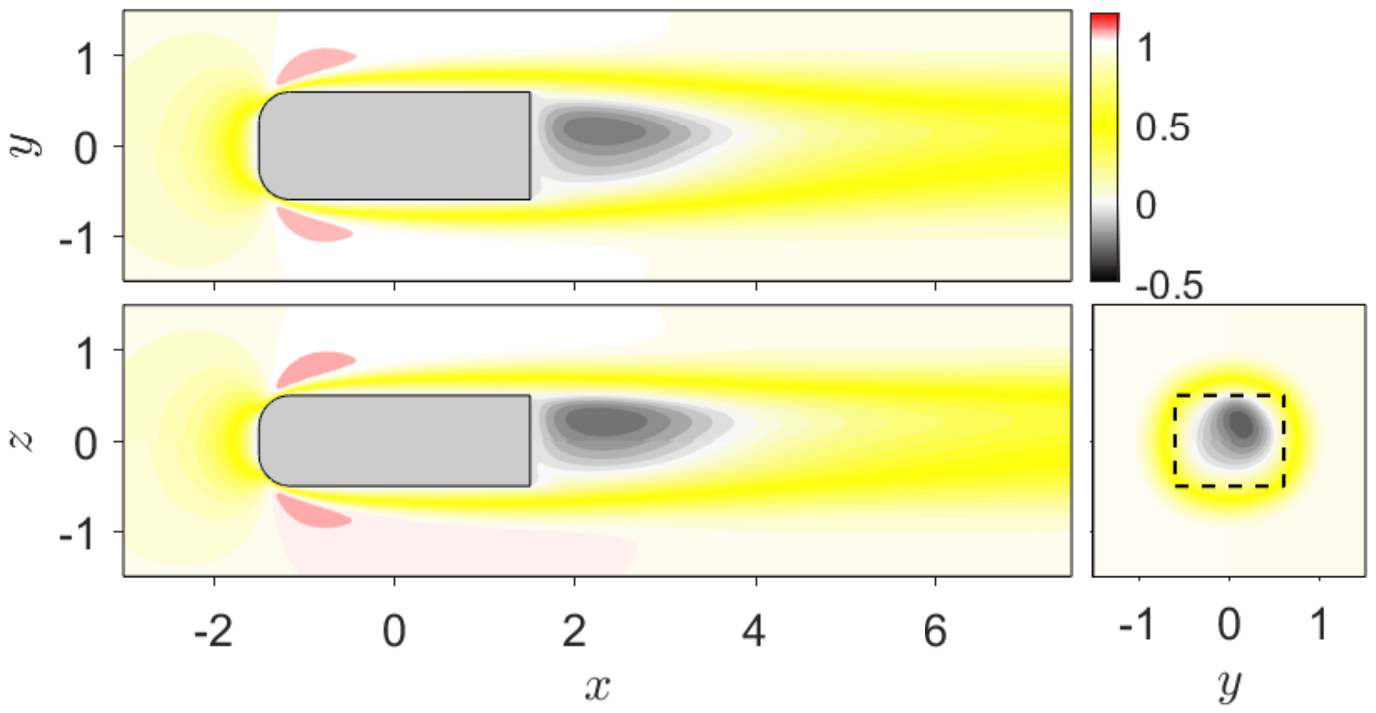}
      \put(-3,48){$(b)$}              
 	\end{overpic}
}
\centerline{   
    \begin{overpic}[width=8cm, trim=32mm 105mm 39mm 118mm, clip=true]{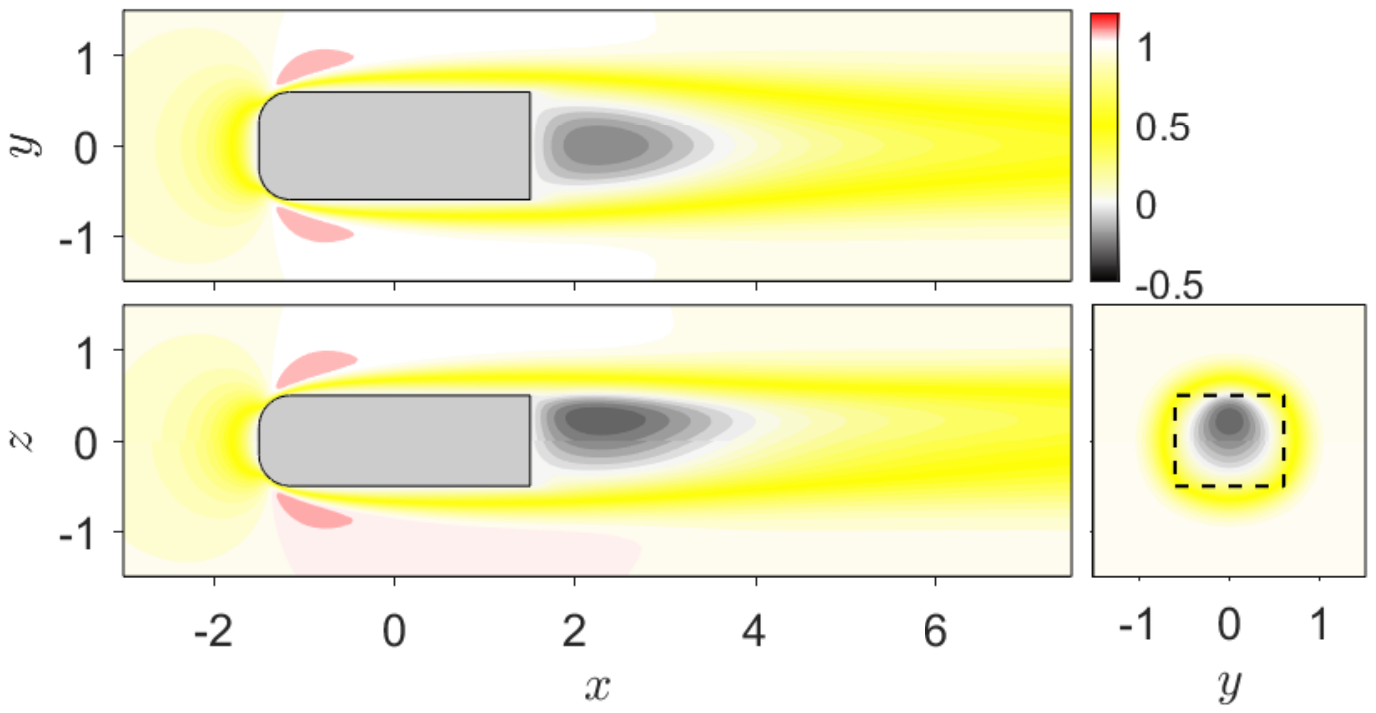}
      \put(-3,48){$(c)$}               
 	\end{overpic}
    \begin{overpic}[width=8cm, trim=32mm 105mm 39mm 118mm, clip=true]{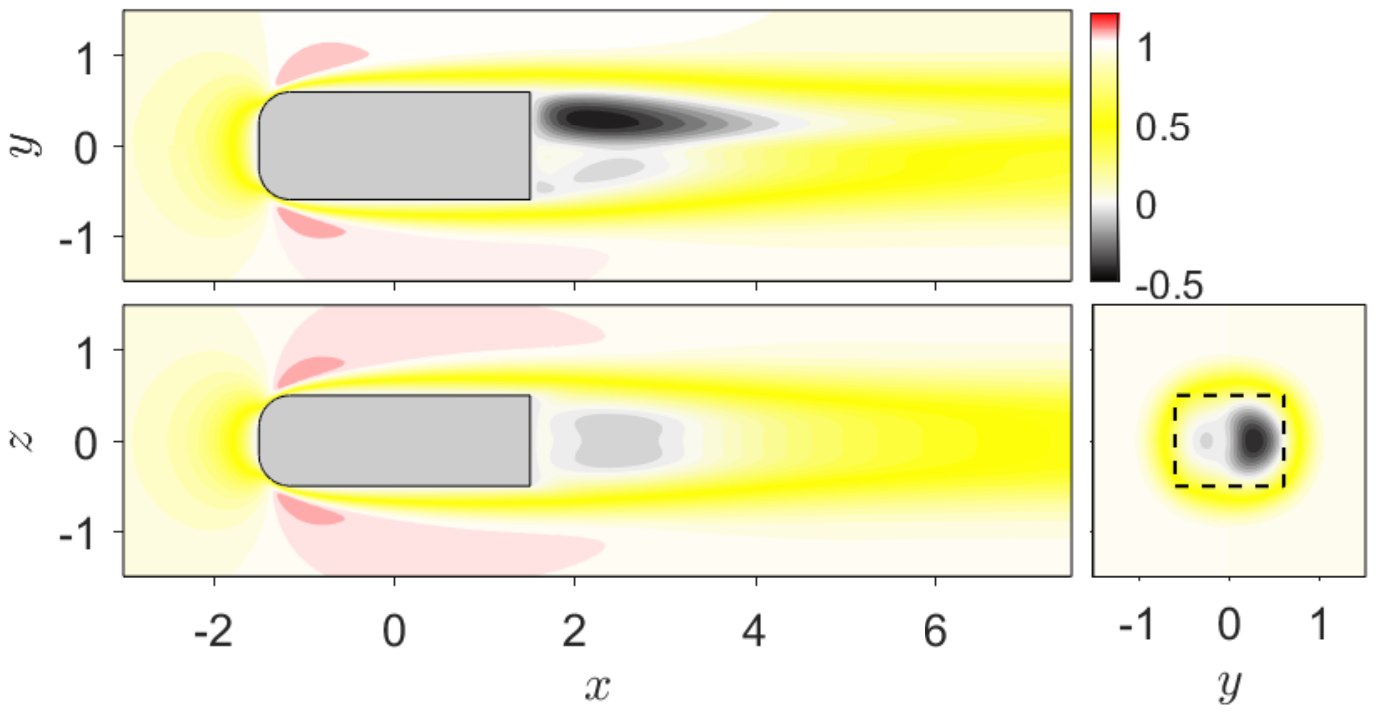}
      \put(-3,48){$(d)$}                
 	\end{overpic}
} 
\caption{
Streamwise velocity $u$ of reconstructed flows (\ref{eq:WNL_reconstruct}) at $Re=325$, from the weakly non-linear analysis up to second-order, shown in the planes $z=0$ (top view), $y=0$ (side view) and $x=2.5$ (rear view): states
$(a)$~$(0,0)$ (unstable),
$(b)$~$(A,B)$ (unstable),
$(c)$~$(A,0)$ (stable),
$(d)$~$(0,B)$ (stable).
}
\label{fig:WNL_reconstruct}
\end{figure}

We obtain the bifurcation diagram shown in figure~\ref{fig:bifurc_diagram}
for the reference Ahmed body ($W=1.2$, $L=3$, $R=0.3472$),  using $Re_c=300$ and $\epsilon=0.1$. 
We use a standard representation, where solid lines indicate linearly stable states and dashes lines unstable state.
Pitchfork bifurcations are invariant under reflections $A \rightarrow -A$ and $B \rightarrow -B$, so it is sufficient to report results for $|A|$ and $|B|$ (i.e. for  half of the symmetric diagram).
The bifurcation sequence is as follows:
\begin{enumerate}
\item
For $Re_c < 293$, only the  symmetric base flow is stable;

\item
At $Re = Re_c^A=293$ mode $A$ bifurcates,  leading to the stable pure state $(A,0)$ and breaking the top/down symmetry of the base flow;

\item
At $Re = Re_c^B=304$ mode $B$  bifurcates,
but the pure state $(0,B)$ is unstable at first, and $(A,0)$ remains the only stable state;

\item
At $Re=314$ the pure state $(0,B)$ becomes stable, thus leading to bistability: either single  symmetry-breaking state (top/down or left/right) can be expected to be observed;

\item
At $Re=330$  $(A,0)$ becomes unstable, leaving $(0,B)$ as the only stable state at larger Reynolds numbers.
\end{enumerate}
We note that the mixed state $(A,B)$ exists in the bistable region but is never stable, meaning that this double symmetry-breaking state is not expected to be observed in a stable manner.
Figure~\ref{fig:WNL_reconstruct} shows the four states  at $Re=325$, reconstructed up to second order according to (\ref{eq:WNL_reconstruct}). 
The wake is clearly deflected in the vertical and horizontal directions in the (stable) pure states $(A,0)$ and  $(0,B)$, respectively, and in both directions in the (unstable) mixed state $(A,B)$.

Interestingly, although mode $A$ bifurcates before mode $B$,  state $(A,0)$ quickly becomes unstable and eventually $(0,B)$ remains the only stable state. 
In other words,  the top/down symmetry-breaking state  cannot be observed except in the range $293 \leq Re \leq 330$, while the left/right symmetry-breaking state can be observed for $Re \geq 316$. 
This is precisely the same symmetry breaking as that observed in the turbulent regime, where Ahmed body wakes deflect along the longer dimension of the body. 
Therefore, unlike wakes past disks, spheres and bullet-shaped axisymmetric bluff bodies,  the preferred turbulent symmetry breaking in the Ahmed body wake is not reminiscent of the  first laminar bifurcation, but of the second one. 
% occurring at a slightly larger Reynolds number. 
Our WNL analysis rationalises this observation by unveiling the bifurcation sequence.
In section~\ref{sec:WNL-effect_W_L}, we will demonstrate the robustness of this  sequence for other geometries typical of Ahmed bodies.

%-----------------------------------------------------
\subsection{Comparison with DNS \label{sec:WNL-DNS}}

In this section, we perform fully non-linear direct numerical simulations of the flow past our reference Ahmed body ($W=1.2$, $L=3$, $R=0.3472$) at various Reynolds numbers, with the aim of confirming the bifurcation sequence obtained in section~{\ref{sec:WNL-bifurc_diag}}.
In particular, we investigate whether the single symmetry-breaking states $(A,0)$ and $(0,B)$ are indeed observed, and if they appear and disappear in the same order. 
We also look for evidence of bistability and hysteresis,  increasing and decreasing the Reynolds number in two independent sequences: $Re=290 \rightarrow 310 \rightarrow 330$, etc.,  and $Re=410 \rightarrow 390 \rightarrow 370$, etc. 
In each case, simulations are run for a  time sufficiently long to allow the flow to reach a truly stationary state.
We determine the flow state by monitoring the horizontal velocity $v(x,0,0)$ and the vertical velocity $w(x,0,0)$  on the symmetry axis. Specifically, we should obtain:
\begin{itemize}
\item 
$v=w=0$ for the symmetric base flow,

\item 
$v=0$, $w\neq 0$ for the top/down symmetry-breaking state $(A,0)$,

\item 
$v\neq 0$, $w=0$ for the left/right symmetry-breaking state $(0,B)$,

\item 
$v\neq 0$, $w\neq 0$ for the double-symmetry breaking state $(A,B)$, although we do not expect to observe this state in the stationary regime.
\end{itemize}

\begin{figure}
  \centerline{
    \begin{overpic}[width=8cm, trim=25mm 90mm 40mm 95mm, clip=true]{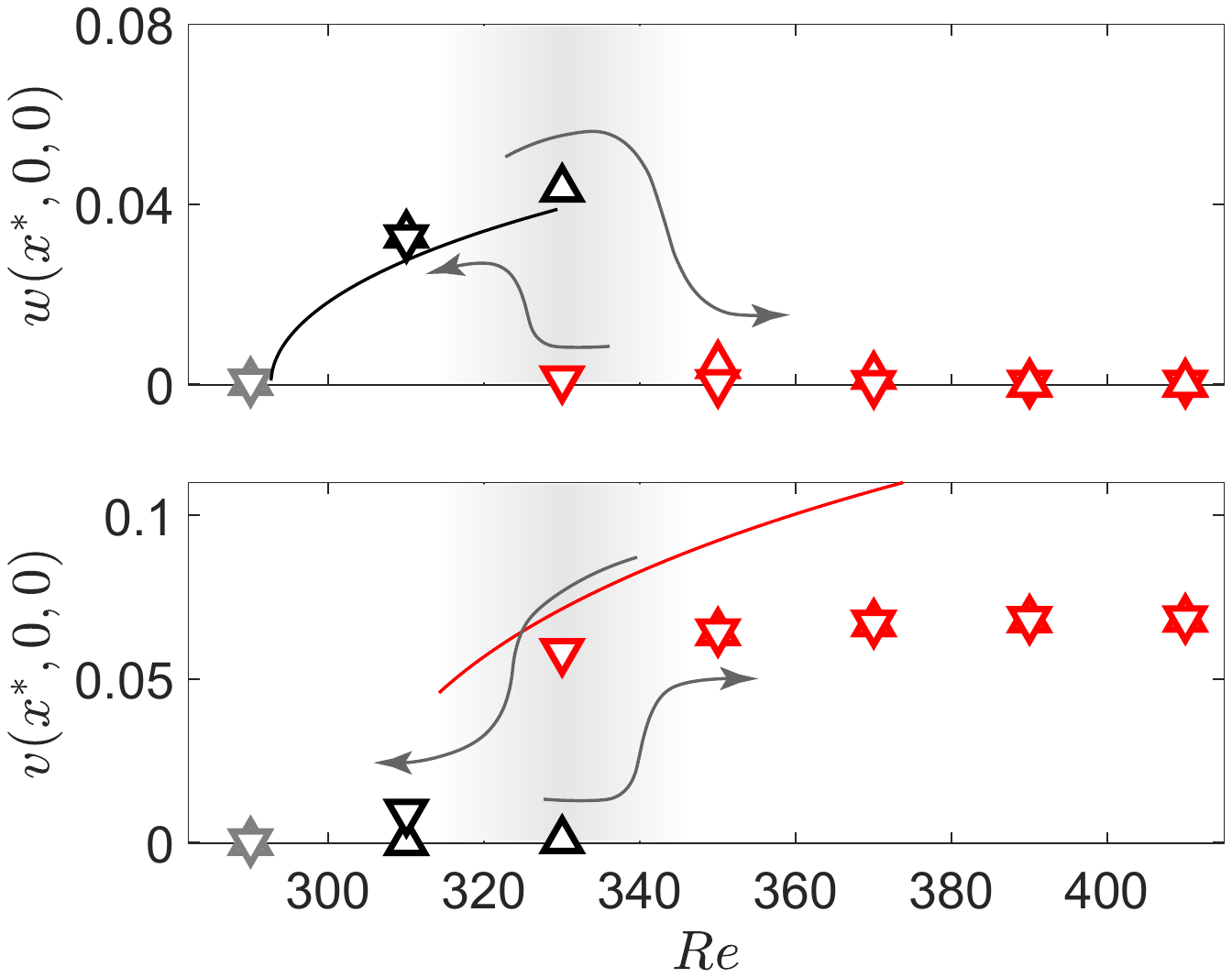}
 	   \put(25,62){\footnotesize $(A,0)$}
 	  %\put(25,15){\footnotesize $(A,0)$}
 	   \put(70,32){\footnotesize $\tcr{(0,B)}$}
 	  %\put(80,60){\footnotesize $\tcr{(0,B)}$}
    \end{overpic}
  }
\caption{
Vertical and horizontal velocities in the wake in 
%$\bm{x}=(x^*,0,0)^T$, $x^*=2.5$
$(x^*,0,0) = (2.5,0,0)$,
 obtained from DNS of the reference Ahmed body ($W=1.2$, $L=3$, $R=0.3472$).
Upward-pointing symbols ($\vartriangle$) correspond to increasing Reynolds numbers, and downward-pointing symbols ($\triangledown$) to decreasing Reynolds numbers.
Shaded region: bistability and hysteresis.
Solid lines: stable reconstructed WNL states.
}
\label{fig:DNS}
\end{figure}

Figure~\ref{fig:DNS} shows the velocities obtained in the stationary regime at the location $(x^*,0,0) = (2.5,0,0)$. 
In all simulations, the final flow is steady, without any oscillations.
When increasing $Re$, the vertical velocity $w$ first becomes non-zero when $Re \approx 300$, while $v=0$. In other words, state $(A,0)$ is observed.
This state loses stability when $Re \approx 340$ and the flow shifts to state $(0,B)$, with a non-zero horizontal velocity $v$, while $w=0$, for this and larger Reynolds numbers.
When decreasing $Re$, state $(0,B)$ persists until $Re \approx 320$, at which point the flow shifts back to state $(A,0)$. 
There is therefore a clear hysteresis in the approximate range of Reynolds number $]320 \pm 10, 340 \pm 10[$.

It is worth mentioning that the flow transitions between  $(A,0)$ and $(0,B)$ via a state reminiscent of the double-symmetry breaking state $(A,B)$ (rather than via the symmetric base flow), although this state is only visited transiently.
We also note that transitions are rather slow (of the order of $10^3$ convective time units), consistent with the small growth rates of the Jacobian (\ref{eq:jacobian}) calculated near the corresponding transitions, i.e. near the boundaries between stages 3 and 4 and between stages 4 and 5 in the bifurcation diagram of figure~\ref{fig:bifurc_diagram}.

The agreement between DNS and WNL is excellent for the transition Reynolds numbers between stages 1, 2 and 3, and for the bifurcated velocity.
At larger $Re$, the agreement deteriorates for the transition Reynolds numbers  between stages 3, 4 and 5, and for the bifurcated velocities, which may be ascribed to the increasing departure from criticality.
Nonetheless, the whole bifurcation sequence is perfectly predicted by the WNL analysis.

As a concluding remark, we note that since the two bifurcations of interest are stationary, it would also  be possible to compute the fully non-linear states with a continuation method or, alternatively, with the self-consistent model proposed by \citet{camarri2019}.

%-----------------------------------------------------
\subsection{Effect of $W$ and $L$ \label{sec:WNL-effect_W_L}}

\begin{table}
  \begin{center}
\def~{\hphantom{0}}
  \begin{tabular}{lcc}
%      $a/d$  & $M=4$   &   $M=8$ & Callan \etal \\[3pt]
Study                        & $W/H$ & $L/H$ \\ [3pt]
\citet{bonnavion_cadot_2018,EVRARD2016} & 1.174 & 3.336 \\
\citet{barros2017}                  & 1.179 & 3.007 \\
%Aider 2018                   & 1.172 & 3.103 \\
\citet{Varon2017}                   & 1.347 & 3.619 \\
\citet{brackston2016}               & 1.350 & 3.752 \\
\citet{Grandemange12PRE, Cadot2015} & 1.350 & 3.625 \\
\citet{Evstafyeva17}                & 1.370 & 3.360 \\
  \end{tabular}
\caption{Typical Ahmed body widths and lengths (normalised by the body height) found in the literature.
}
  \label{tab:Ahmed_W_L}
  \end{center}
\end{table}

We have presented the bifurcation diagram of our reference Ahmed body ($W=1.2$, $L=3$, $R=0.3472$) in section~\ref{sec:WNL-bifurc_diag}. 
One may wonder whether this sequence  is robust to geometric modifications. 
Indeed, Ahmed bodies studied in the literature show some variations in dimension, especially in width and length (table~\ref{tab:Ahmed_W_L}).
Therefore we investigate the effect of $W$ and $L$,  keeping the fillet radius constant ($R=0.3472$).
Specifically, we consider narrower and wider bodies ($W \in [1, 1.35]$, $L=3$), as well as 
shorter and longer bodies ($W=1.2$, $L \in [2.6, 3.8]$).
We compute the base flow and eigenmodes for these bodies and determine the critical Reynolds numbers $Re_c^A$ and $Re_c^B$;
we then set $Re_c$ by rounding $(Re_c^A+Re_c^B)/2$ to the nearest multiple of 5, and finally we compute the coefficients of the amplitude equations (\ref{eq:A})-(\ref{eq:B}).

\begin{figure}
  \centerline{
    \begin{overpic}[height=6.5cm, trim=47mm 95mm 55mm 95mm, clip=true]{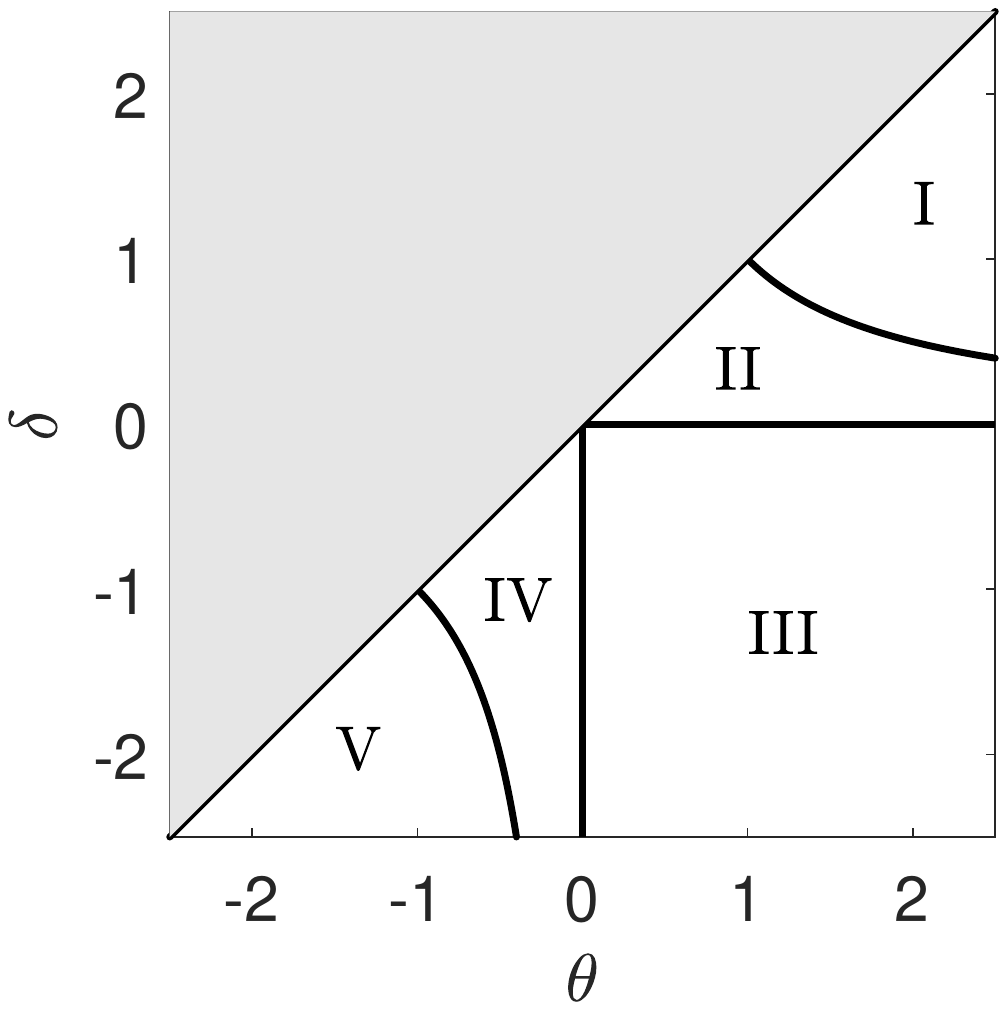}
      \put(0,90){$(a)$}
 	\end{overpic}
 	\hspace{0.5cm}
    \begin{overpic}[height=6.5cm, trim=30mm 95mm 40mm 95mm, clip=true]{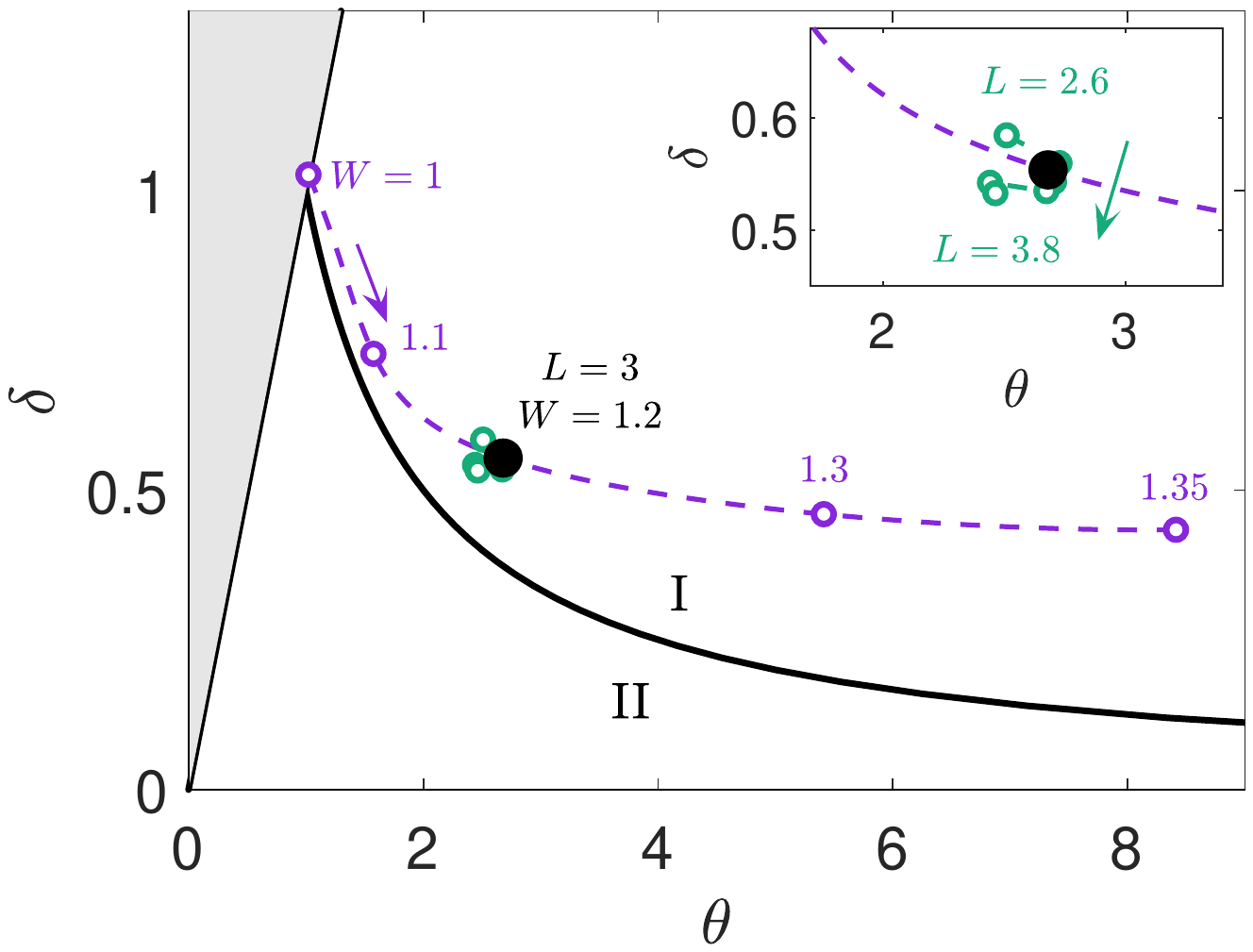}
      \put(0,70){$(b)$}
 	\end{overpic}
  }
\caption{$(a)$~Five regions of the $\theta$--$\delta$ plane, associated with the five possible  bifurcation diagrams for the amplitude equations (\ref{eq:A})-(\ref{eq:B}).
$(b)$~Variation of $\theta$ and $\delta$ with $W$ and  $L$ for rectangular prisms of fixed fillet radius $R=0.3472$.
Dashed lines are a guide to the eye.
In the considered range of widths, $1 \leq W \leq 1.35$, and lengths, $2.6 \leq L \leq 3.8$, representative of most Ahmed bodies, $\theta$ and $\delta$ stay in region I,  i.e. the bifurcation diagram remains as in figures \ref{fig:bifurc_diagram} and 
\ref{fig:sketch_BD_regions_I_II}$(a)$.
}
\label{fig:sketch_Kuznetsov}
\end{figure}

\begin{figure}
\centerline{
    \begin{overpic}[width=7.8cm, trim=41mm 97mm 44mm 97mm, clip=true]{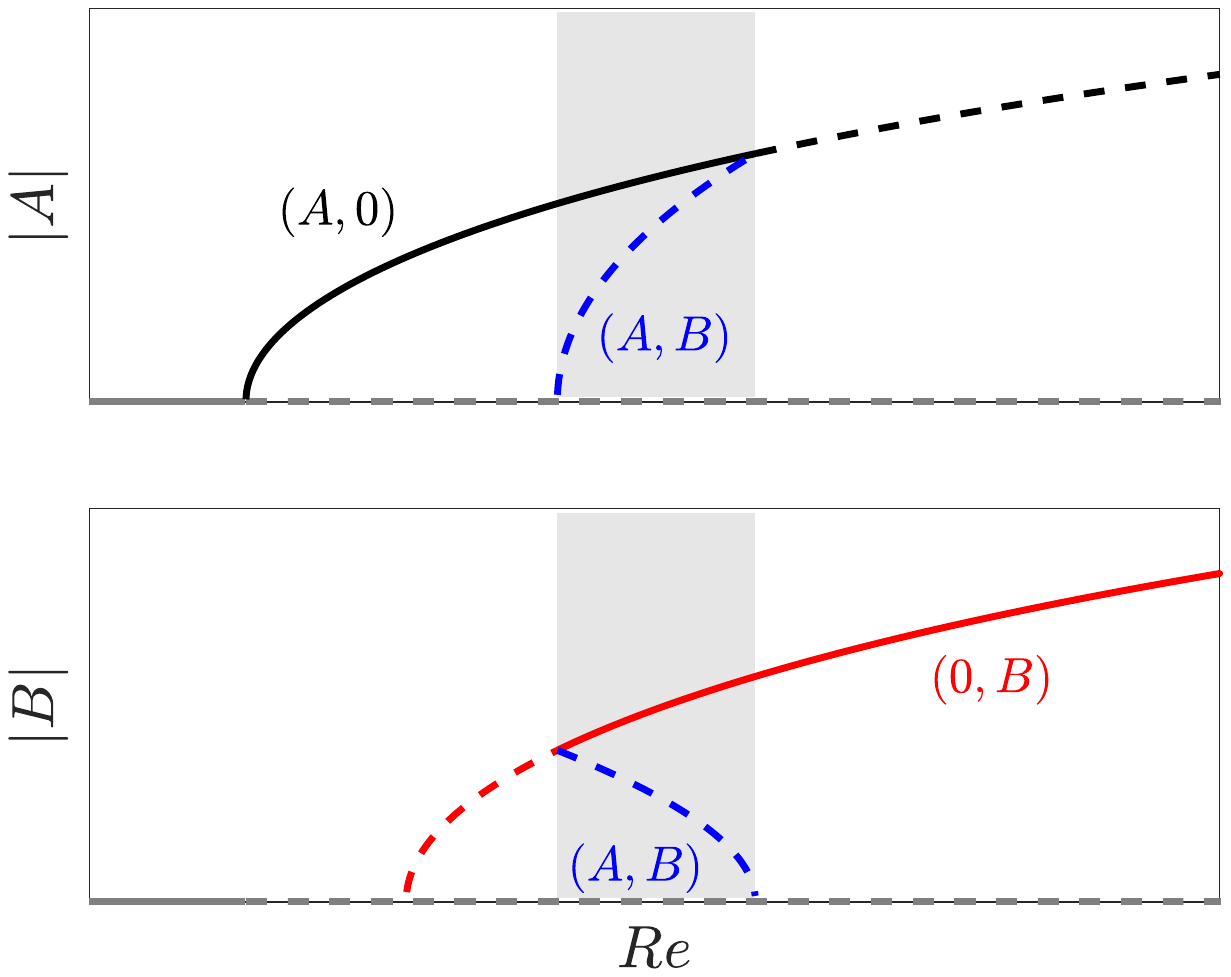}
      \put(-1,82){$(a)$}
      \put(8,82){Region I}
      \put(11,41.7){\scriptsize $\circled{1}$}
      \put(24,41.7){\scriptsize $\circled{2}$}
      \put(35,41.7){\scriptsize $\circled{3}$}
      \put(50,41.7){\scriptsize $\circled{4}$}
      \put(77,41.7){\scriptsize $\circled{5}$}
 	\end{overpic}
    \begin{overpic}[width=7.8cm, trim=41mm 97mm 44mm 97mm, clip=true]{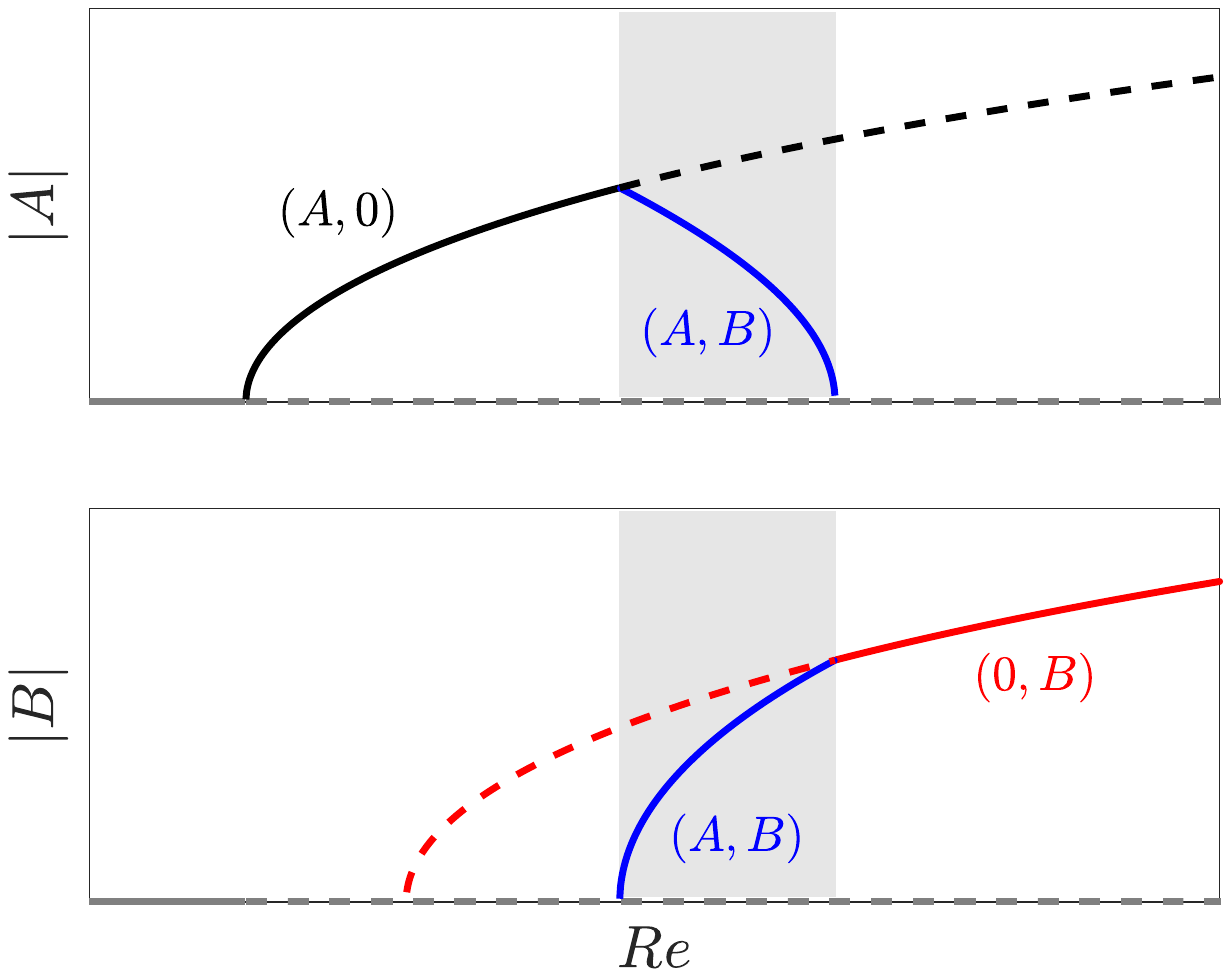}
      \put(-1,82){$(b)$}
      \put(8,82){Region II}
      \put(11,41.7){\scriptsize $\circled{1}$}
      \put(24,41.7){\scriptsize $\circled{2}$}
      \put(38,41.7){\scriptsize $\circled{3}$}
      \put(56,41.7){\scriptsize $\circled{4}$}
      \put(80,41.7){\scriptsize $\circled{5}$}
 	\end{overpic}
}
\vspace{0.5cm}
\centerline{
  \hspace{2cm}
    \begin{overpic}[width=14cm, trim=0mm 0mm 0mm 50mm, clip=true]{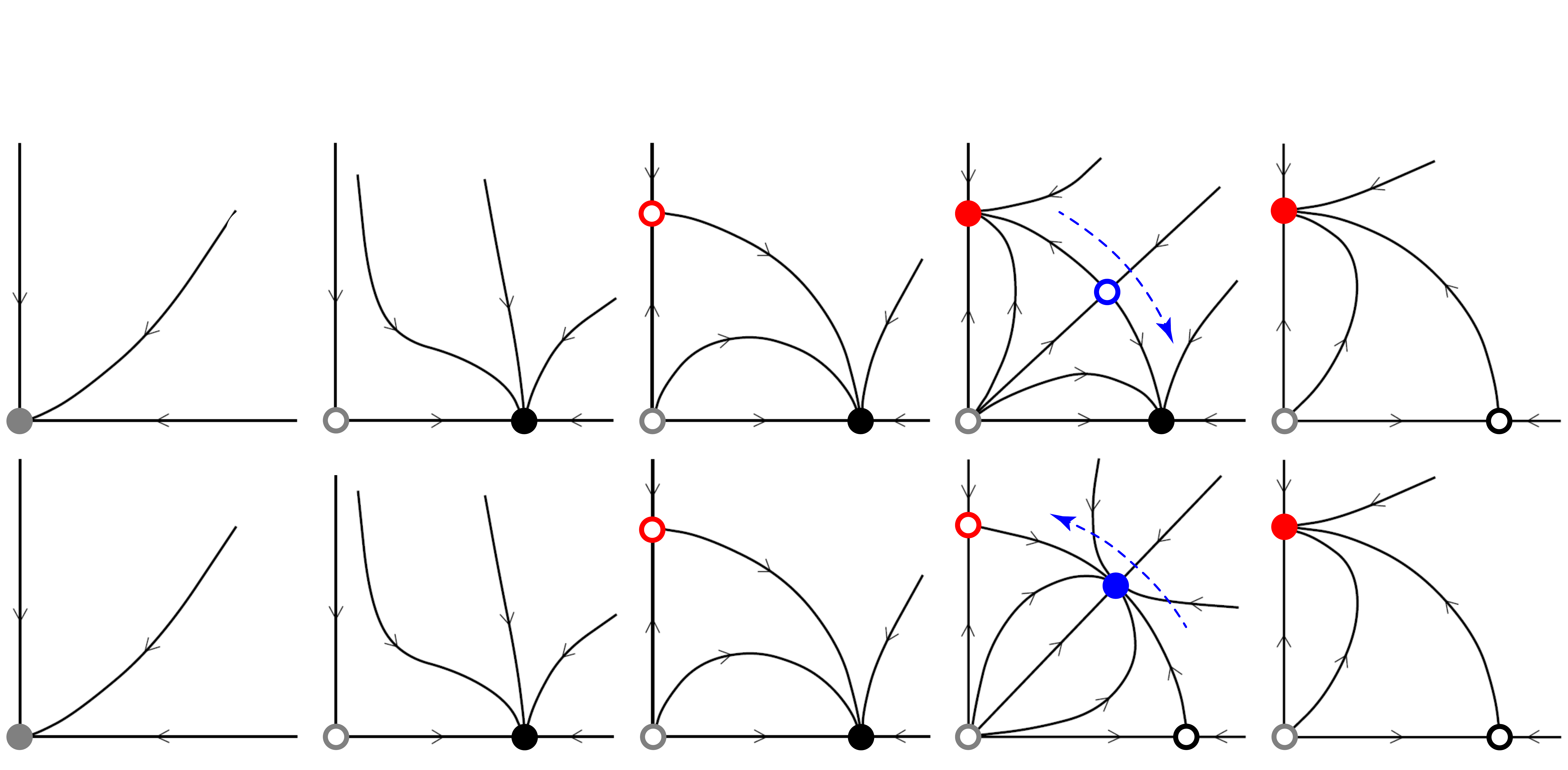}   
      \put(-15,42){$(c)$}
      \put(-15,30){Region I}
      \put(-15,10){Region II}
      \put(-2,17.5){$B$}
      \put(16.5,-1){$A$}
      \put( 9,  42){\scriptsize $\circled{1}$}
      \put(29.2,42){\scriptsize $\circled{2}$}
      \put(49.4,42){\scriptsize $\circled{3}$}
      \put(69.6,42){\scriptsize $\circled{4}$}
      \put(89.8,42){\scriptsize $\circled{5}$}
 	\end{overpic}
}
\caption{Schematic bifurcations diagrams in $(a)$~region I and $(b)$~region II of the $\theta$--$\delta$ plane of Fig.~\ref{fig:sketch_Kuznetsov}.
Stable and unstable states are shown with solid and dashed lines, respectively.
$(c)$~Corresponding phase portraits in the $A$--$B$ amplitude plane.
Stable and unstable states are shown with filled and open symbols, respectively.
Dashed arrows in stage 4 show how the mixed state $(A,B)$ moves with $Re$. 
}
\label{fig:sketch_BD_regions_I_II}
\end{figure}

A convenient way to summarise the results is to consider the different possible bifurcation diagrams for this type of amplitude equations.
As detailed in \citet{Kuznetsov}, there are five topologically different bifurcation diagrams, corresponding to different regions of the  $\theta$--$\delta$ plane, where $\theta$ and $\delta$ are mixed ratios of the non-linear coefficients of (\ref{eq:A})-(\ref{eq:B}),
\begin{align}
\theta = \frac{\eta_A}{\chi_B},
\quad
\delta = \frac{\eta_B}{\chi_A}.
\end{align}
Figure~\ref{fig:sketch_Kuznetsov}$(a)$ shows the five regions, labelled I to V, for the case $\theta \geq \delta$ relevant here.
As will soon become clear, regions I ($\theta>0$, $\delta>0$, $\theta\delta>1$) and II ($\theta>0$, $\delta>0$, $\theta\delta<1$) are of particular interest to our study.
In region I, the bifurcation diagram is as sketched in figure~\ref{fig:sketch_BD_regions_I_II}$(a)$, and is topologically equivalent to that obtained in  section~\ref{sec:WNL-bifurc_diag}.
We briefly recall the bifurcation sequence for increasing $Re$:
\begin{enumerate}
\item  
Only the trivial state $(0,0)$ is stable;

\item  
Mode $A$ bifurcates and the stable state $(A,0)$ appears;

\item  
Mode $B$ bifurcates and the unstable state $(0,B)$ appears;

\item  
State $(B,0)$ becomes stable and the unstable mixed state $(A,B)$ appears;

\item  
State $(A,B)$ disappears and state $(A,0)$ becomes unstable; only $(0,B)$ remains stable.
\end{enumerate}
In the neighbouring region II, the bifurcation diagram is as sketched in figure~\ref{fig:sketch_BD_regions_I_II}$(b)$, and corresponds to another bifurcations sequence:
%In region I, the following sequence is observed  as $Re$ increases: 
%the stable state $(A,0)$ appears;
%the state $(0,B)$ appears and is initially unstable;
%the state $(0,B)$ becomes stable  and the double-symmetry-breaking state $(A,B)$ appears but is unstable;
%the  state $(A,0)$ becomes unstable and the state $(A,B)$  disappears, leaving $(0,B)$ as the only stable state.
%
%By contrast, the following sequence is observed  in region II:
\begin{enumerate}
\item  
Only the trivial state $(0,0)$ is stable;

\item  
Mode $A$ bifurcates and the stable state $(A,0)$ appears;

\item  
Mode $B$ bifurcates and the unstable state $(0,B)$ appears;

\item  
State $(A,0)$ becomes unstable and the stable mixed state $(A,B)$ appears;

\item  
State $(A,B)$ disappears and  state $(0,B)$ becomes stable.
\end{enumerate}
Although the initial stages (1-3) and the final stage (5) are identical in regions I and II, what happens in between (stage 4, shaded region in figure~\ref{fig:sketch_BD_regions_I_II}$(a)$) is fundamentally different: 
in region I the pure states $(A,0)$ and $(B,0)$ are  simultaneously stable (bistability) while the mixed state $(A,B)$ cannot be observed;
by contrast, in region II the pure states $(A,0)$ and $(B,0)$ are unstable and only the mixed state $(A,B)$ can be observed.

This is also apparent in the phase portraits of figure~\ref{fig:sketch_BD_regions_I_II}$(c)$.
From stages 3 to 5, the stability of the pure states $(A,0)$ and $(0,B)$ is exchanged.
In region I, this happens as the unstable mixed state $(A,B)$ is born in $(0,B)$ (making it stable)
and dies when colliding with $(A,0)$ (making it unstable).
In region II, the stable mixed state $(A,B)$ is born in $(A,0)$ (making it unstable) and dies when colliding with $(0,B)$ (making it stable).

Figure~\ref{fig:sketch_Kuznetsov}$(b)$ shows that all the Ahmed bodies considered in this study fall in region I of the $\theta$--$\delta$ plane.
Our reference Ahmed body corresponds to $\theta=     2.68$, $\delta=0.554$ (filled symbol).
The effect of $L$ is very limited, with $\theta$ and $\delta$ barely varying compared to their reference values (inset).
The effect of $W$ is much more significant, with $\theta$ and $\delta$ varying substantially:
the point $(\theta,\delta)$ moves away from the boundary between regions I and II as the body becomes wider, and  vice-versa.
By symmetry, the limiting case $W=1$ (body as wide as tall, seldom used in practice for Ahmed bodies) lies exactly on the boundary.
We can conclude that, in the WNL framework, the bifurcation sequence obtained in section~\ref{sec:WNL-bifurc_diag} is robust to width and length variations commonly encountered for Ahmed bodies.

One must keep in mind that the WNL analysis is rigorously valid in the vicinity of $Re_c$ only. 
Therefore, one cannot completely rule out a fully non-linear bifurcation diagram that differs from the weakly non-linear one, e.g. an incursion in region II. 
More generally, it is  likely that the deflected wake undergoes one or several secondary instabilities at sufficiently large Reynolds number.
Nonetheless, the fact that stages 1--3 and 5 are identical in regions I and II suggests that the transition from state $(A,0)$ to state $(0,B)$ is, in any case, a robust feature of this flow.
Furthermore, the fully non-linear DNS bifurcation sequence (section~\ref{sec:WNL-DNS}) is in very good agreement with the WNL prediction for the reference Ahmed body, which attests to the reliability of the WNL analysis.

%
%\begin{figure}
%  \centerline{
%  \fbox{
%    \begin{overpic}[width=8cm, trim=30mm 90mm 40mm 90mm, clip=true]{C:/Users/boujo/Documents/_Projects_MSc/2021_Adrien_Gimonnet/results/quarter-other-ARs-fillet/sketch_Kuznetsov_varying_W_L.pdf}
% 	\end{overpic}
% 	}
%  }
%\caption{Variation of $\theta$ and $\delta$ with  $(a)$ $W$ and $(b)$ $L$. 
%The fillet radius is fixed to $R=0.3472$.
%Dashed lines are a guide to the eye.
%In the considered range of width $1 \leq W \leq 1.35$ and length $2.6 \leq L \leq 3.8$, representative of most Ahmed bodies, $\theta$ and $\delta$ stay inside region I,  i.e. the bifurcation diagram remains as in figures \ref{fig:bifurc_diagram} and 
%\ref{fig:sketch_BD_regions_I_II}$(a)$.
%}
%\label{fig:W_L_effect}
%\end{figure}

%-----------------------------------------------------
%-----------------------------------------------------
%-----------------------------------------------------
\section{Conclusion}

In this study, we have investigated the stability of 3D rectangular prism wakes of width-to-height ratio $W/H=1.2$. 
Linear stability analysis showed that two stationary modes become unstable via pitchfork bifurcations, at critical Reynolds numbers close to one another. 
Mode $A$ breaks the top/down planar symmetry, and mode $B$ the left/right planar symmetry.
At larger $Re$, two oscillatory modes become unstable via Hopf bifurcations, each mode breaking either planar symmetry. 
The critical Reynolds numbers all increase with the body length $L$, similar to the pitchfork and Hopf bifurcations of axisymmetric wakes.
The effect of the leading edge fillet radius $R$ is limited, and depends on $L$.

Next, a weakly non-linear analysis performed in the vicinity of the critical $Re$ of modes $A$ and $B$ yielded a set of two coupled amplitude equations. 
For Ahmed bodies, the bifurcation diagram revealed that the flow first undergoes a stationary bifurcation leading to a wake deflection in the top/down direction, or state $(A,0)$. Then, another stationary state $(0,B)$ with the opposite symmetry breaking, i.e. in the left/right direction, becomes stable. 
Simultaneously, a double symmetry-breaking state $(A,B)$ appears but remains unstable at all $Re$. The two single-symmetry-breaking states thus coexist in a range of Reynolds number. 
Finally, state $(A,0)$ becomes unstable, leaving state $(0,B)$ as the only stable state. 
The corresponding wake deflection, i.e. along the larger dimension of the body, is the same as the static deflection observed in the turbulent wakes of Ahmed bodies, with or without ground proximity.

Fully non-linear DNS confirmed the whole bifurcation sequence, including bistability-induced hysteresis, and the final $(0,B)$ state. 
We also demonstrated that the bifurcation sequence was robust to  variations  in body width $W$ and body length $L$ in the range of common Ahmed body geometries.

Natural extensions of this work may include: 
(i)~sensitivity analysis, in order to determine optimal control strategies and locations, 
(ii)~analysis of possible secondary instabilities  at larger $Re$, especially along the $(0,B)$ branch, and 
(iii)~analysis of the linear and weakly-non linear stability of Ahmed bodies with modifications  suppressing one of the two planar symmetries, such as ground proximity, yaw/pitch angles, and asymmetric body geometries.

\begin{small}

%%-----------------------------------------------------
%%-----------------------------------------------------
%%-----------------------------------------------------
%%\section*{Acknowledgements}
%\bigskip
%
%\noindent 
%\textbf{Acknowledgements.}
%E.B. thanks 
%Olivier Cadot for stimulating discussions   about Ahmed bodies and symmetry breaking,
%Eunok Yim for initial versions of the base flow and linear stability codes,
%Adrien Gimmonet for his work in the initial stage of this study,
%and Scitas for support with EPFL's high-performance computing clusters. 
%%
%Useful comments from Alessandro Bongarzone and  François Gallaire about the weakly non-linear analysis are also acknowledged.

%-----------------------------------------------------
%-----------------------------------------------------
%-----------------------------------------------------
\bigskip
\noindent 
\textbf{Declaration of Interests.}
The authors report no conflict of interest.

%-----------------------------------------------------
%-----------------------------------------------------
%-----------------------------------------------------
\bigskip
\noindent 
\textbf{Author ORCID.}

\noindent 
G. A. Zampogna, https://orcid.org/0000-0001-7570-9135;
\\
E. Boujo,    https://orcid.org/0000-0002-4448-6140.

%-----------------------------------------------------
%-----------------------------------------------------
%-----------------------------------------------------
\bigskip
\noindent 
\textbf{Author contributions.}
G.A.Z. performed the DNS and wrote the paper.
E.B. conceived the study, performed the linear stability and weakly non-linear analyses, and wrote the paper.

\end{small}

%-----------------------------------------------------
%-----------------------------------------------------
%-----------------------------------------------------
\section*{Appendix A. Numerical domain and mesh}

%\begin{tabular}{cccccccccc} 
%Mesh & $n$ & $N_{elmts}$ & $N_{DOF}$ & $C_D$  & $L_r$ & $Re_c^A$ & $Re_c^B$ & $\sigma^A$ & $\sigma^B$ \\ 
%\\
%M1   & 40  & 1'005'762   & xxxxxxxx  & 0.771 & •     & •         &          & 0.00560 & -0.00242 \\ 
%M2   & 60  & 1'382'998   & xxxxxxxx  & 0.778 & •     & 293	     & 304      & 0.00578 & -0.00253 \\ 
%M3   & 80  & 1'936'779   & xxxxxxxx  & 0.781 & •     & •         & •        & 0.00589 & -0.00264 \\ 
%\end{tabular} 
%
%\bigskip
%

% Obs: drag = 2/W * 4 * (FF(1,1)+FF(1,2))

\begin{table}
\begin{tabular}{c c ccc} 
                 & Mesh              & M1        & M2        & M3 \\ \\

                 & $n$               & 40        & 60        & 80 \\
                 & $N_{\mathrm{elmts}}$       & 1.01$\times10^6$    & 1.38$\times10^6$    & 1.94$\times10^6$ \\ 
                 & $N_{\mathrm{DOF}}$        & 3.73$\times10^6$  & 5.13$\times10^6$  & 7.18$\times10^6$ \\ \\

Base flow        & $C_D$             & 0.771     & 0.778     & 0.781 \\ 
                 & $L_r$             & 2.25      & 2.24      & 2.24  \\ \\

Linear stability & $Re_c^A$          & 293       & 293       & 292 \\ 
                 & $Re_c^B$          & 304       & 304       & 304 \\ 
                 & $\sigma^A$        &  0.00560  & 0.00578   &  0.00589 \\ 
                 & $\sigma^B$        & -0.00242  & -0.00253  & -0.00264 \\ \\

WNL              & $\tilde\lambda_A$ & $ 0.00560+68.3\tilde\alpha$ & $ 0.00578+68.5\tilde\alpha$ & $ 0.00589+68.3\tilde\alpha$ \\
                 & $\tilde\lambda_B$ & $-0.00242+57.1\tilde\alpha$ & $-0.00253+57.1\tilde\alpha$ & $-0.00264+56.8\tilde\alpha$ \\
                 & $\tilde\eta_A$    & 0.467                       & 0.466                       & 0.467           \\
                 & $\tilde\eta_B$    & 0.556                       & 0.557                       & 0.558           \\
                 & $\tilde\chi_A$    & 1.01                        & 1.01                        & 1.01            \\
                 & $\tilde\chi_B$    & 0.183                       & 0.174                       & 0.171           \\
                 & $\theta$          & 2.56                        & 2.68                        & 2.74            \\
                 & $\delta$          & 0.552                       & 0.554                       & 0.552           \\
\end{tabular} 
\caption{
Dependence on the mesh of several properties of the base flow, linear stability analysis and weakly non-linear analysis,
for the reference Ahmed body ($W=1.2$, $L=3$, $R=0.3472$).
The drag coefficient $C_D$, recirculation length $L_r$ and eigenvalues  $\sigma^A$, $\sigma^B$ are calculated at $Re=300$.
The WNL coefficients are computed with the reference Reynolds number $Re_c=300$.
}
\label{tab:convergence}
\end{table}

All FreeFEM calculations (base flow, linear stability, weakly-non linear analysis) are performed on the quarter-space domain  $\Omega' = \{ x,y,z \, | \,  -10 \leq x \leq 20, \, 0 \leq y,z \leq 10 \}$.
Mesh points are strongly clustered near the body, with density 1 on the outermost boundaries,
10 on the boundaries of the sub-domain $\{ x,y,z \, | \,  -5 \leq x \leq 15, \, 0 \leq y,z \leq 2 \}$, and $n$ on the body surface.
Table~\ref{tab:convergence} reports the variation with $n$ of several quantities: base drag coefficient and recirculation length at $Re=300$,
growth rates of modes $A$ and $B$ at $Re=300$, critical Reynolds number of modes $A$ and $B$, and WNL coefficients for $Re_c=300$.
Increasing $n$ from 40 to 80, thus doubling the number of elements $N_{\mathrm{elmts}}$ and degrees of freedom $N_{\mathrm{DOF}}$, leads to reasonably small variations of all the quantities of interest.
Throughout the study we have used $n=60$.

%-----------------------------------------------------
%-----------------------------------------------------
%-----------------------------------------------------
\section*{Appendix B. Effect of the reference Reynolds number on the bifurcation diagram}

\begin{figure}
  \centerline{
    \begin{overpic}[width=8cm, trim=35mm 90mm 40mm 95mm, clip=true]{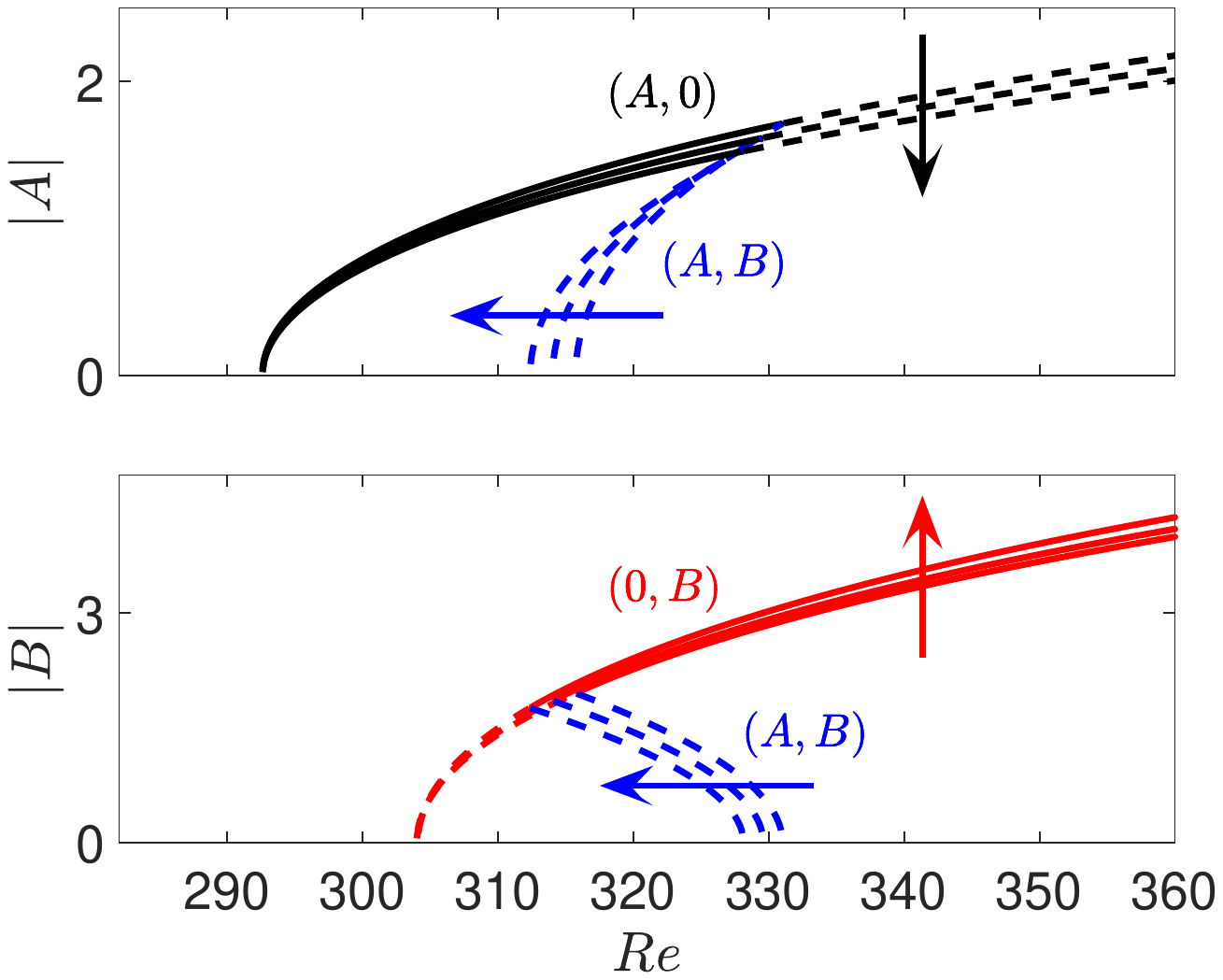}
 	\end{overpic}
  }
\caption{
Effect of the reference Reynolds number on the bifurcation diagram, for the reference Ahmed body ($W=1.2$, $L=3$, $R=0.3472$): arrows show increasing values $Re_c=295$, 300, 305.
}
\label{fig:effect_of_Rec}
\end{figure}

Figure~\ref{fig:effect_of_Rec} shows the bifurcation diagram obtained with different choices of the reference Reynolds number,  $Re_c=295$, 300 and 305, close to the first two pitchfork bifurcations ($Re_c^A=293$ and $Re_c^B=304$).
We observe no significant effect on the bifurcation sequence, i.e. on the symmetry-breaking states, their stability and  amplitudes.
The onset of existence of states $(A,0)$ and $(0,B)$ is not modified either.
Only the extent of the bistability region is slightly modified as $Re_c$ increases: the lower bound varies between $Re=312$ and 316 and the upper bound between $Re=328$ and 331.

\bibliographystyle{apalike}
\bibliography{../ALL2}
%%\printbibliography

\end{document}